\documentclass[11pt]{aastex}
\usepackage{emulateapj5,apjfonts}
\usepackage{apjfonts}
\usepackage{epsf}
\slugcomment{{\sc Accepted to ApJ:} April, 2003}
\received{XXth XXXXXXXX 2002}

\newcommand{\begit}{\begin{itemize}}
\newcommand{\enit}{\end{itemize}}
\newcommand{\begen}{\begin{enumerate}}
\newcommand{\enen}{\end{enumerate}}
\newcommand{\beq}{\begin{equation}} 
\newcommand{\eeq}{\end{equation}} 
\newcommand{\beqa}{\begin{eqnarray}} 
\newcommand{\eeqa}{\end{eqnarray}} 
\newcommand{\p}{\partial}    
\newcommand{\pr}{^\prime}

\newcommand{\aenu}{\bar{\nu}_e}
\newcommand{\enu}{\nu_e}
\newcommand{\unu}{\nu_\mu}
\newcommand{\aunu}{\bar{\nu}_\mu}

\newcommand{\cms}{cm s$^{-1}$}

\newcommand{\modot}{M$_\odot$}

\def\sles{\lower2pt\hbox{$\buildrel {\scriptstyle <}
   \over {\scriptstyle\sim}$}}

\def\sgreat{\lower2pt\hbox{$\buildrel {\scriptstyle >}
   \over {\scriptstyle\sim}$}}
\def\lesssim{\sles}

\begin{document} 

\title{Shock Breakout in Core-Collapse Supernovae and its Neutrino Signature}

\author{Todd A.~Thompson\altaffilmark{1}}
\altaffiltext{1}{Hubble Fellow}
\affil{Astronomy Department and Theoretical Astrophysics Center, 601 Campbell Hall, \\
The University of California, Berkeley, CA 94720;
thomp@astro.berkeley.edu}
\author{Adam Burrows}
\affil{Steward Observatory, 
The University of Arizona, Tucson, AZ 85721;
burrows@zenith.as.arizona.edu}
\author{Philip A.~Pinto}
\affil{Steward Observatory, 
The University of Arizona, Tucson, AZ 85721;
ppinto@as.arizona.edu}

\vspace{.5cm}

\begin{abstract}

We present results from dynamical models of core-collapse supernovae
in one spatial dimension, employing a newly-developed Boltzmann neutrino radiation 
transport algorithm, coupled to Newtonian Lagrangean hydrodynamics and a consistent high-density 
nuclear equation of state.  The transport method is multi-group, employs the Feautrier technique,
uses the tangent-ray approach to resolve angles, is implicit in time, and is second-order
accurate in space.  We focus on shock breakout and follow the dynamical evolution 
of the cores of 11\,M$_\odot$, 15\,M$_\odot$, and 20\,M$_\odot$ progenitors
through collapse and the first 250 milliseconds after bounce.
The shock breakout burst is the signal event 
in core-collapse evolution, is the brightest phenomenon in astrophysics,
and is largely responsible for the initial debilitation and stagnation
of the bounce shock.  As such, its detection and characterization could 
test fundamental aspects of the current collapse/supernova paradigm. 
We examine the effects on the emergent neutrino spectra, light curves, and 
mix of species (particularly in the early post-bounce epoch) 
of artificial opacity changes, the number of energy groups, the weak magnetism/recoil corrections,
nucleon-nucleon bremsstrahlung, neutrino-electron scattering, and the compressibility
of nuclear matter.  Furthermore, we present the first high-resolution
look at the angular distribution of the neutrino radiation field both in the semi-transparent
regime and at large radii and explore the accuracy with which our tangent-ray
method tracks the free propagation of a pulse of radiation in a near vacuum.
Finally, we fold the emergent neutrino spectra with the 
efficiencies and detection processes for a selection of modern underground 
neutrino observatories and argue that the prompt electron-neutrino breakout
burst from the next galactic supernova 
is in principle observable and usefully diagnostic of 
fundamental collapse/supernova behavior.  
Though we are not in this study focusing on the supernova mechanism 
per se, our simulations support the theoretical conclusion (already 
reached by others) that spherical (1D) supernovae 
do not explode when good physics and transport methods are employed.

\end{abstract}

\keywords{supernovae, neutrinos, radiative transfer, spectra}

\section{Introduction}

The most dynamical phase of core-collapse supernova evolution
is the epoch of core bounce, shock wave formation, and early shock propagation.
Within a few milliseconds (ms) of birth, the shock wave propagates down the
density gradient from where it is formed at $\sim 10^{14}$ g\,cm$^{-3}$ and $\sim$10 kilometers 
to where it dies into an accretion shock at $\sim 10^{10}$ g\,cm$^{-3}$ and $\sim$100--120 kilometers.
Between birth and death, the shock wave, born in the neutrino-opaque region,
emerges into the neutrino-transparent region.  In the process, it ``breaks out''
through the neutrinospheres of the various neutrino species ($\nu_{e}$, $\bar{\nu}_{e}$, 
and ``$\nu_{\mu}$''s),
and, thereby, produces an ultra-luminous burst of neutrinos brighter 
than any other phenomenon in astrophysics.  This short transient 
is responsible for sapping the shock of energy and aborting the
supernova explosion within 10-20\,ms of its birth and is 
a primary reason the supernova theorist has shifted focus
to the ``long-term'' mechanism of shock revival in its various forms (Bethe \& Wilson 1985; 
Herant et al. 1994;
Burrows, Hayes, \& Fryxell 1995; Fryer et al. 1999; Fryer \& Heger 2000; 
Rampp \& Janka 2000; Janka 2001; Liebend\"orfer et al. 2001; 
Liebend\"{o}rfer, Mezzacappa, \& Thielemann 2001).  

Given the central importance of the breakout phenomenon to the dynamics
of core collapse (as we think we understand it), and given the possibility that
the next galactic supernova will be detected by neutrino telescopes far more
sensitive than the underground detectors IMB and Kamioka II which caught SN1987A,
we have made a detailed study of the breakout phenomenon, its neutrino signature,
and its dependence on various parameters, progenitors, and neutrino microphysics.  
Neutrino breakout is the signature characteristic of the theory of collapse and supernovae,
that if measured would speak volumes concerning the dynamics of a dense core otherwise
obscured from view.  

In \S\ref{section:ingredients}, we describe our multi-group, 
multi-angle radiation/hydrodynamics code 
SESAME\footnote{{\bf S}pherical {\bf E}xplicit/Implicit {\bf S}upernova 
{\bf A}lgorithm for {\bf M}ulti-Group/Multi-Angle {\bf E}xplosion Simulations}
and the microphysics we use in this survey of the breakout phenomenon.
We include a discussion of our implementation of the Lattimer-Swesty high-density nuclear
equation of state (LSEOS) in Appendix \ref{appendix.eos}.
In \S\ref{section:baseline}, we present our baseline core-collapse simulation 
using an 11\,M$_\odot$ progenitor from Woosley \& Weaver (1995).
We follow the discussion of this reference model with studies of
the dependence on spectral resolution (\S\ref{section:eres}), progenitor mass 
(\S\ref{section:progenitor}),
the nuclear equation of state (\S\ref{section:comp}), the weak magnetism/recoil 
correction (\S\ref{weak:recoil}), and the role of nucleon-nucleon
bremsstrahlung (\S\ref{section:dyn_brem}).  In \S\ref{section:inelasticnue}, using 
a new explicit algorithm for handling it (see Appendix \ref{section:inelastic}),
we explore the effect of inelastic neutrino-electron scattering.
In \S\ref{section:detection}, event rates and signatures in the underground 
telescopes Super Kamiokande, SNO, and ICARUS
are calculated and described.  It is shown that the breakout burst is in principle detectable
in all three of these neutrino observatories.  Changes in the observed
neutrino signal due to modifications in the microphysics employed in the  
simulations are also presented.
We wrap up in \S\ref{summary} with a summary 
and general conclusions
on the breakout phenomenon and its model dependencies. 
Our focus in this paper is on the breakout phenomenon, 
not the supernova itself. Hence, our simulations are restricted (for the most part)
to the first 250\,ms after bounce.

\section{Numerical Methods}
\label{section:ingredients}

\subsection{Radiation Transport}

The mass densities, temperatures, and compositions that obtain during core collapse
conspire to produce regions in the deep core of a supernova where the mean-free-path for neutrinos
is just meters and the transport could be handled using simple diffusion theory.
Simultaneously, there are regions at larger radius where each neutrino species
decouples from the matter and free-streams to infinity.  In between, at modest optical
depth, where the neutrino mean-free path is a modest fraction of the size of the object
as a whole, the full transport problem must be solved in order to obtain
accurate values of the local
neutrino flux and energy density - both critical in determining the 
energy deposition profile and, hence, the subsequent evolution.

Our Boltzmann transport solver (Burrows et al. 2000) is due primarily to Eastman \& Pinto (1993),
who developed an algorithm for solving the comoving transport equation
using the Feautrier variables and the tangent-ray method for angular discretization.
The transport solver is implicit in time and second-order accurate in space.
The zeroth- and first-moment equations are iterated with the Boltzmann/transport
equation using Accelerated Lambda Iteration (ALI) to speed convergence
of the radiative transfer solution (see e.g., Mihalas \& Mihalas 1984).
Only the diagonal and the adjacent off-diagonal terms of the $\Lambda$ matrix
are used. This method is akin to the standard variable Eddington
factor (VEF) approach, but no ad hoc flux limiters or
artificial closures are necessary.  The iteration scheme 
generally converges to a part in 10$^6$ in 2 to 10 steps, 
and automatically conserves energy in the transport sector.
The Feautrier scheme can transition to the diffusion
limit seamlessly and accurately and the tangent-ray method automatically adapts with
the Lagrangean hydrodynamic grid as it moves.  In constructing the tangent rays, we cast
them from every outer zone to every inner zone.  Hence, if there are 300 radial zones,
the outer zone has 299 angular groups in each quadrant of the unit circle. 
Because of the spherical nature of the core-collapse
problem and the need to accurately reproduce the angular distribution of the radiation
field that transitions from the opaque (inner) to the transparent (outer) regions,
such fine angular resolution is useful, though computationally
demanding, as radiation becomes more and more forward-peaked.

The coupling of the neutrino radiation to matter is done implicitly in 
operator-split fashion, using the Accelerated Lambda operator to obtain the 
derivative `$\p J/\p S$' (where $S$ is the source function) 
after each hydrodynamic update.  This derivative is required for the  
implicit updates of temperature ($T$) and electron fraction ($Y_e$) (Burrows et al.~2000).

We solve the transport equation
for $\nu_e$ neutrinos, $\bar{\nu}_e$ neutrinos, and $\nu_\mu$ neutrinos, where
`$\nu_\mu$' stands for $\nu_\mu$, $\bar{\nu}_\mu$, $\nu_\tau$, and $\bar{\nu}_\tau$.
Although in principle the neutrino microphysics is different between
$\nu_\mu$, $\bar{\nu}_\mu$, $\nu_\tau$, and $\bar{\nu}_\tau$ we make
the assumption standard in the field that they can be treated identically.
The energy grouping is different for each species.  For $\nu_e$ neutrinos,
$1\leq\varepsilon_{\nu_e}\,\,{\rm (MeV)}\leq320$. For $\bar{\nu}_e$ neutrinos,
$1\leq\varepsilon_{\bar{\nu}_e}\,\,{\rm (MeV)}\leq100$.  Both are grouped
logarithmically.  For $\nu_\mu$ neutrinos, $1\leq\varepsilon_{\nu_\mu}\,\,{\rm (MeV)}\leq100$
with uniform grouping in energy. Our baseline models employ 40 energy groups for each species.
In \S\ref{section:eres}, we compare our results with calculations employing just 10 and 20 energy
groups. Our standard set of microphysics includes all of the emission/absorption
and scattering processes described in Burrows (2001) plus weak-magnetism and
recoil modifications to the cross sections for neutrino-nucleon absorption and scattering,
as per Horowitz (1997,2002).  In addition, we include inelastic neutrino-electron scattering as
described in Appendix \ref{section:inelastic}.  In our baseline models
for this paper we do not include inelastic neutrino-nucleon scattering, but include it
only as a purely elastic scattering opacity source.

\subsection{Hydrodynamics}

During most epochs in supernova evolution, the radiation pressure term is negligible.
However, during breakout when electron neutrino luminosities approach
10$^{54}$ erg s$^{-1}$ locally, this term can be important.
Since there is as yet no natural way in PPM-like hydro codes
to include the radiation pressure term in the solution to the Riemann problem,
we have opted to couple our transport algorithm to a Newtonian and Lagrangean predictor/corrector
hydrodynamics scheme that uses artificial viscosity for shock resolution.
During and just after shock breakout, since the neutrinospheres
and the regions of spectrum formation are found at 30
to 100 kilometers, general-relativistic effects are small to modest, ${GM}/{rc^2}$ 
being just 3 to 10\%.
Nevertheless, particularly at later times, general-relativistic
effects can introduce $\sim$10\% corrections (Bruenn, De Nisco,
\& Mezzacappa 2001; Liebend\"{o}rfer et al.~2001).

In the core at bounce the zone size in our
calculations is $\sim$200 meters.  The sound speed at nuclear densities approaches
$\sim c/3$. Hence, our Courant-limited explicit timestep is $\sim10^{-6}$ seconds.   
This implies that to
evolve a calculation for $\sim$1 second would require $\sim10^6$ timesteps.
The solution to the transport equation for all species is computationally expensive
and on current single-processor machines $10^6$ timesteps is large.
In order to circumvent this problem we sub-cycle the hydrodynamics after shock
breakout and stall.  After this most dynamical phase is over
we take $2-4$ hydro steps per transport step, thus cutting the total computational
requirements significantly.  Introducing sub-cycling $\sim$20 milliseconds
after bounce seems to introduce errors of less than 1\% in all radiation and hydrodynamic
quantities.

For all of our simulations, we employ the tabular equation of state
described in Appendix \S\ref{appendix.eos}. 

\section{The Baseline Model Results}
\label{section:baseline}

We take as a reference model to establish the general context of the study
an 11\,M$_\odot$ progenitor from Woosley \& Weaver (1995).
The model employs 300 mass zones out to M = 1.43 \modot, where
M is the interior mass; 
in this progenitor, at M = 1.43 \modot, $r$ equals 2500\,km.
For the simulations presented here, which focus on collapse and the early ($\sim100-200$ ms)
post-bounce evolution, we find this zoning sufficient.  
In \S\ref{section:comp}, we explore the effects of changes in the
nuclear compressibility modulus, but for our baseline model we use $\kappa=220$ MeV in the LSEOS 
(Appendix \ref{appendix.eos}).  

\begin{figure} 
\vspace*{6.0in}
\hbox to\hsize{\hfill\includegraphics{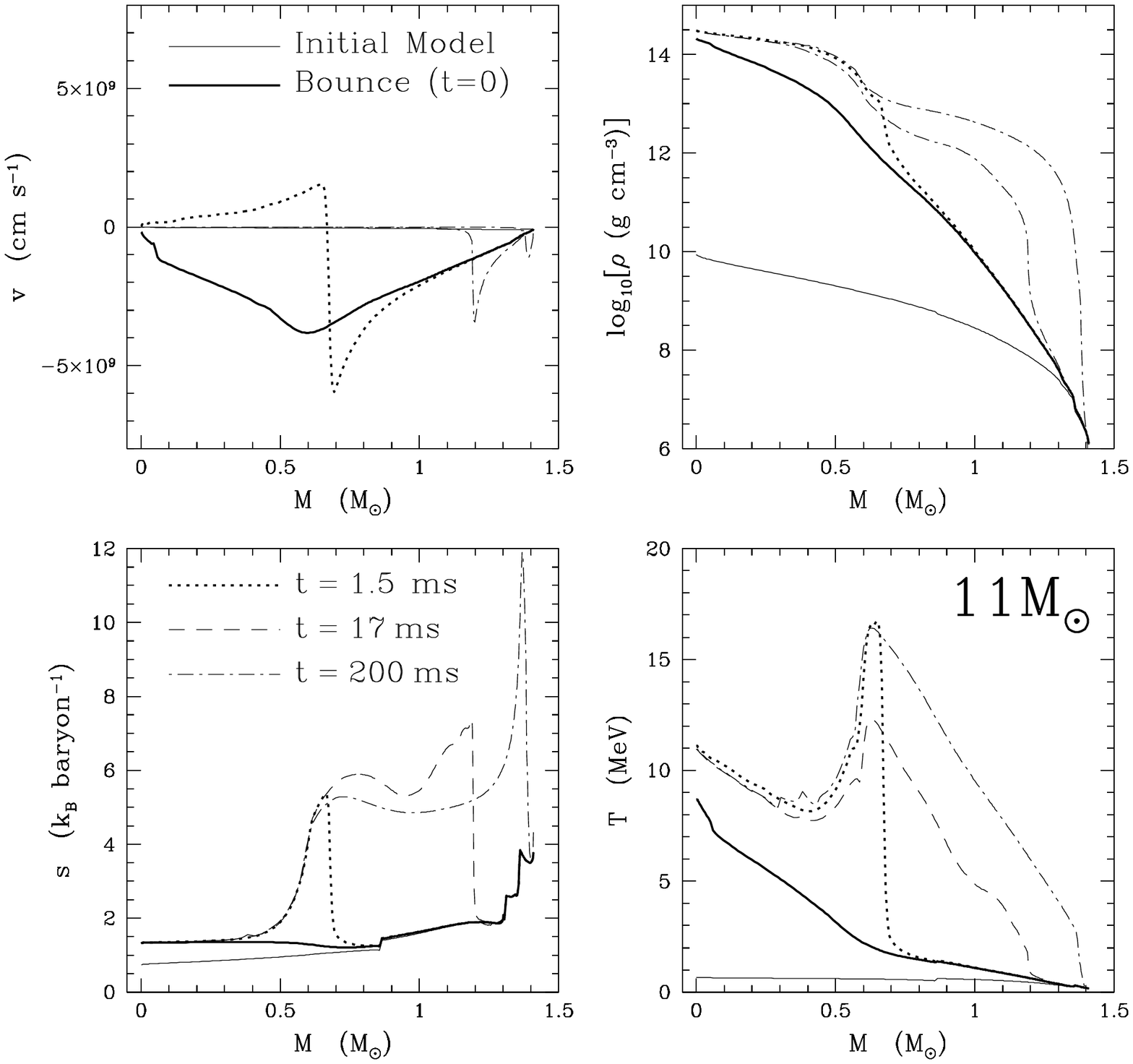}\kern+6in\hfill}
\caption[Snapshots of $v$, $\rho$, $T$, and $s$ vs.~$M$ for fiducial 11\,M$_\odot$ model]
{Velocity $v$ (in cm s$^{-1}$, upper left panel),
$\log_{10}[\rho]$ (in g cm$^{-3}$, upper right panel),
entropy $s$ (in k$_B$ baryon$^{-1}$, lower left panel),
and temperature $T$ (in MeV, lower right panel) in the 11\modot\,
progenitor, as a function
of mass coordinate in \modot\, for five snapshots in time.
The thin solid line shows the initial configuration and the thick
solid line shows the model at bounce.
The dotted, dashed, and dot-dashed lines are snapshots 
at 1.5\,ms, 17\,ms, and 200\,ms after bounce, respectively. 
}
\label{tots11}
\end{figure}

\begin{figure} 
\vspace*{6.0in}
\hbox to\hsize{\hfill\includegraphics{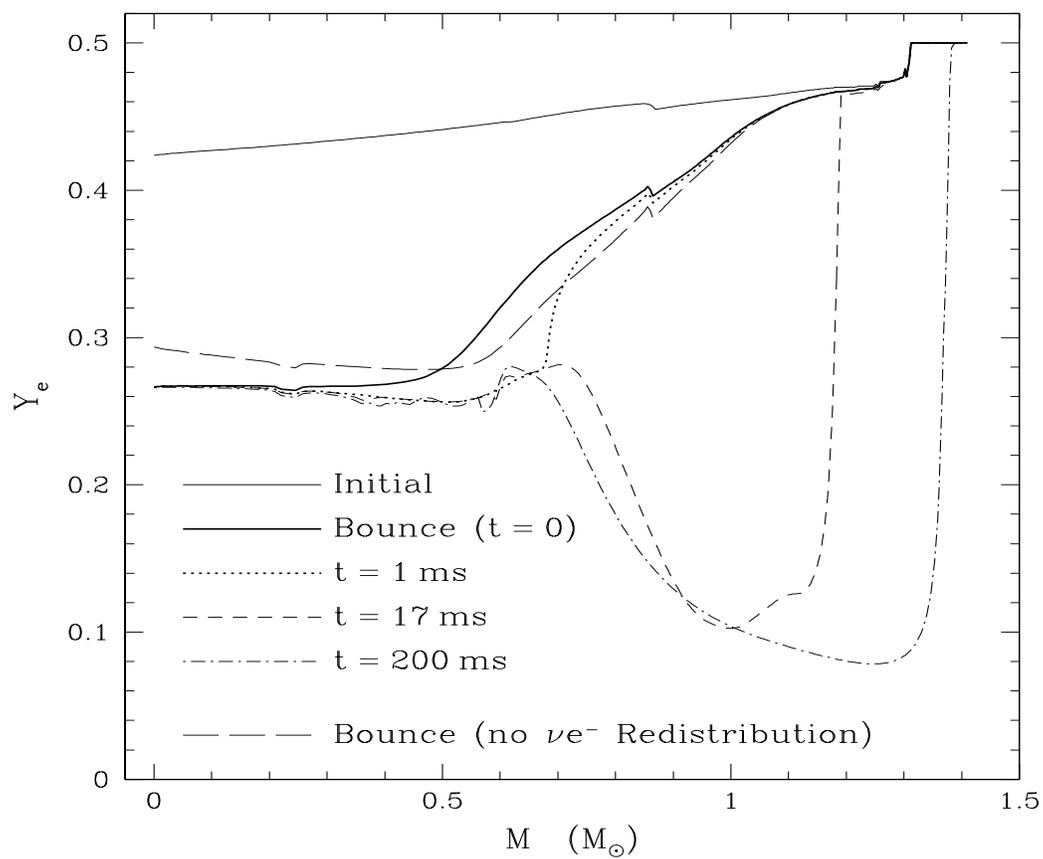}\kern+6in\hfill}
\caption[Snapshots of $Y_e$ vs.~$M$ for fiducial 11\,M$_\odot$ model]
{The electron fraction $Y_e$ in the 11\modot\,
progenitor, as a function
of mass coordinate in \modot\, for five snapshots in time.
The thin solid line shows the initial configuration and the thick
solid line shows the model at bounce.
The dotted, short-dashed, and dot-dashed lines are snapshots 
at 1\,ms, 17\,ms, and 200\,ms after bounce, respectively. 
For comparison, we include the bounce $Y_e$ profile with 
inelastic neutrino-electron redistribution turned off (long dashed line).
Compare this figure with Fig.~\ref{tots11}, which shows the basic
hydrodynamical evolution for this same model.
}
\label{yes11}
\end{figure}

Figure \ref{tots11} shows velocity $v$ (in cm s$^{-1}$, upper left panel),
$\log_{10}[\rho]$ (in g cm$^{-3}$, upper right panel),
entropy $s$ (in k$_B$ baryon$^{-1}$, lower left panel),
and temperature $T$ (in MeV, lower right panel), as a function
of mass coordinate in \modot\, at five snapshots in time 
in our 11\,\modot\, baseline model.  Fig.~\ref{tots11} is
to be compared with Fig.~\ref{yes11}, which shows the
electron fraction ($Y_e$) as a function of mass in the same model.
The thin solid line in Figs.~\ref{tots11} and \ref{yes11} is the initial configuration.  
The thick solid line is at 
hydrodynamical bounce (approximately 200\,ms after we start the calculation).
We generally quote times relative to bounce.
Bounce coincides with the core density reaching $\sim2-3\times10^{14}$ g cm$^{-3}$,
where the EOS stiffens dramatically. 
Note that $Y_e$ in the very core reaches its minimum value {\it before}
bounce.  At core densities of $\sim2\times10^{12}$ g cm$^{-3}$ the
electron-neutrino outward diffusion speed becomes smaller than the
the inward collapse speed and the neutrinos are effectively trapped.
Soon afterward, the process primarily responsible
for the decrease in $Y_e$, $e^- p\rightarrow n\nu_e$, is 
balanced by its inverse and beta equilibrium is achieved 
(Bruenn 1985; Mezzacappa \& Bruenn 1993abc).
Although we have included inelastic neutrino-electron scattering in
our baseline model, the long dashed line in Fig.~\ref{yes11} shows
the bounce $Y_e$ profile if one turns off this equilibration process.  
Because small changes in the $Y_e$ profile at bounce can affect
the shock energetics significantly (Burrows \& Lattimer 1983) and because 
inelastic neutrino-electron scattering affects the approach to beta equilibrium by modifying 
the neutrino phase space occupancy, inelastic neutrino-electron scattering
should be included in a full treatment (Bruenn 1985; Mezzacappa \& Bruenn 1993abc).  
We discuss this process and its effects on breakout
more fully in \S\ref{section:inelasticnue}.

\begin{figure} 
\vspace*{6.0in}
\hbox to\hsize{\hfill\includegraphics{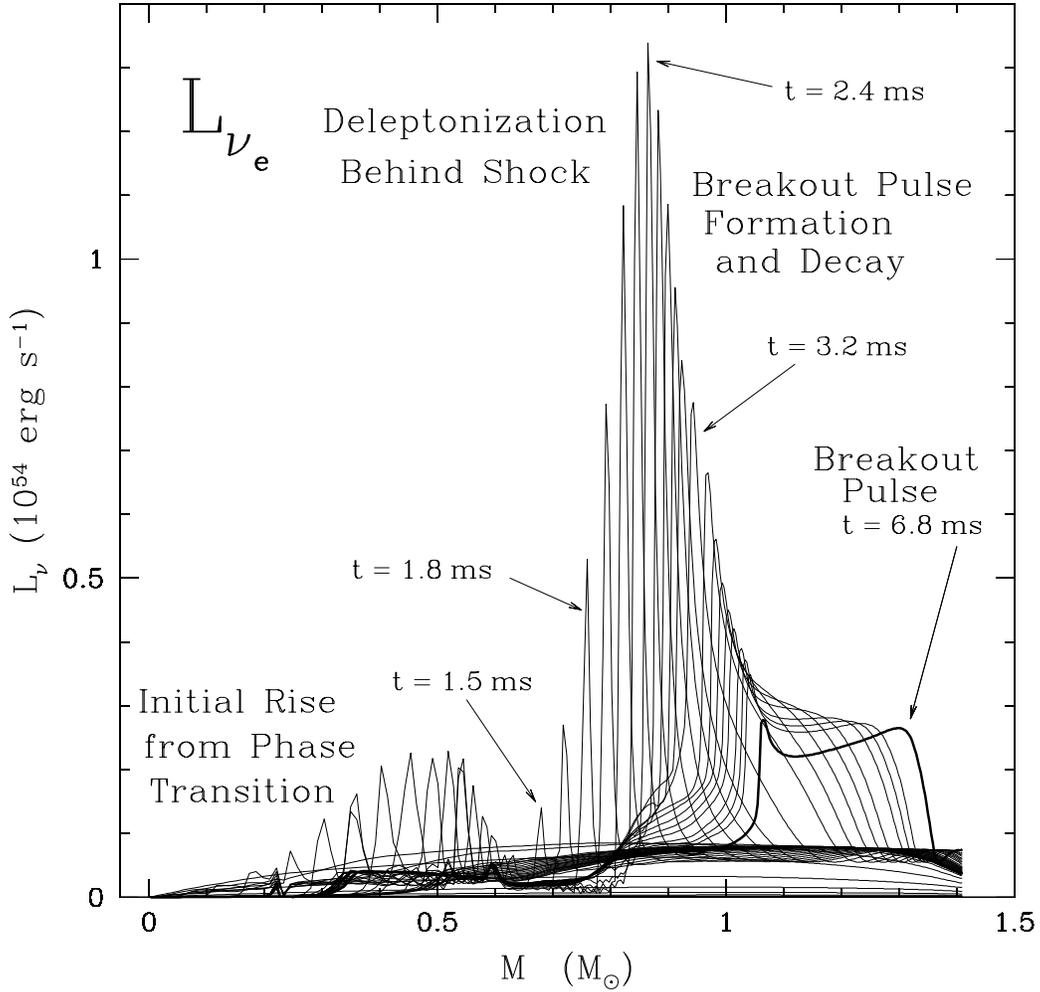}\kern+6in\hfill}
\caption[Snapshots of $L_{\nu_e}$ vs.~$M$ at breakout in 11\,M$_\odot$ model]
{Snapshots of $L_{\nu_e}$ in 10$^{54}$ erg s$^{-1}$ as a function of mass
for the fiducial 11\,M$_\odot$ progenitor, showing 
the whole envelope of 
luminosities realized in this progenitor from bounce through 
electron-neutrino breakout.
Times re relative to hydrodynamical bounce.
The initial rise in $L_{\nu_e}$ at $\sim0.3$\,M$_\odot$ comes from
electron capture on newly liberated free protons before the shock forms,
but just after bounce.  The second, larger peak
forms after shock formation. The temperature and density increase
across the shock dissociates nuclei into free nucleons.  Subsequent 
electron capture on free protons generates the breakout
pulse, which reaches $L_{\nu_e}\sim1.5\times10^{54}$ erg s$^{-1}$ locally.
}
\label{lms113}
\end{figure}

At bounce, a strong sound wave is formed deep in the core, which 
propagates down the density gradient set up by the infalling outer
stellar material.  The sound wave steepens into a shock near M$\simeq$0.6\modot, 
just $\sim$1\,ms after bounce.
The dotted line in Fig.~\ref{tots11} shows the shock fully formed and at maximum positive
velocity ($\sim$$1.9\times10^9$ cm s$^{-1}$).  Of course, 
the shock is associated with the generation of
entropy and an increase in temperature, as seen in the lower two panels of Fig.~\ref{tots11}.
As the shock moves outward in mass and radius it dissociates nuclei
into free nucleons.  Although the matter is hot in this region, the electrons
are still quite degenerate and the process $e^- p\rightarrow n\nu_e$ builds
up a sea of electron neutrinos that are trapped and advected with the matter.  
Figure \ref{lms113} shows the luminosity of electron neutrinos as a function of mass and time
from bounce all the way through breakout. 
Although the shock does not
form until M$\simeq$0.6\modot, a phase-transition front which liberates free nucleons
from nuclei moves in tandem with the sound wave generated
at bounce.  This causes the initial luminosity peak seen in Fig.~\ref{lms113}
at M$\simeq$0.5\modot, or, in this model, at a radius of just 11\,km (labeled `Initial
rise due to phase transition').
The sudden drop in $L_{\nu_e}$ at M$\simeq0.6-0.7$\,M$_\odot$ comes from a sharp decrease
in $X_p$, coincidentally nearly simultaneous with shock formation.  Just as the 
shock is forming, but before the temperature rises dramatically, as shown in the
lower right panel of Fig.~\ref{tots11} (dotted line), a region of heavy nuclei
exists and, due to the fact that the absorption cross section is not continuous
across the phase transition, there is a decrease in $L_{\nu_e}$ at M$\simeq0.6-0.7$\,M$_\odot$.
In Fig.~\ref{lms113}, the specific snapshots in time 
between the 1\,ms (dotted) line and the 17\,ms (short dashed) line
in Figs.~\ref{tots11} and \ref{yes11} are labeled.  At 2.4\,ms after bounce 
the spike reaches a local maximum at M$\simeq0.85$\modot, with $L_{\nu_e}$
exceeding  $1.4\times10^{54}$ erg s$^{-1}$.  This spike 
roughly denotes the position of the shock.  It is at this moment
that the shock crosses the $\nu_e$ neutrinosphere ($R_{\nu_e}$).
The neutrinospheric radius ($R_\nu$) for any neutrino species 
of energy $\varepsilon_{\nu}$ is set approximately by the following 
integral criterion:
\begin{equation}
\tau_\nu(R_\nu,\varepsilon_\nu)=
\int_{R_\nu}^\infty\,\kappa_\nu(\varepsilon_\nu,r)\rho(r)\, dr=\frac{2}{3},
\label{nusphere}
\end{equation}
where $\kappa_\nu$ is the total opacity.
At just 3.2\,ms after bounce, one
can see clearly that the breakout pulse of electron neutrinos 
is beginning to move ahead of the shock.  Having moved into a region
where the optical depth is below $\sim$2/3, the neutrinos begin to free-stream.

\begin{figure} 
\vspace*{6.0in}
\hbox to\hsize{\hfill\includegraphics{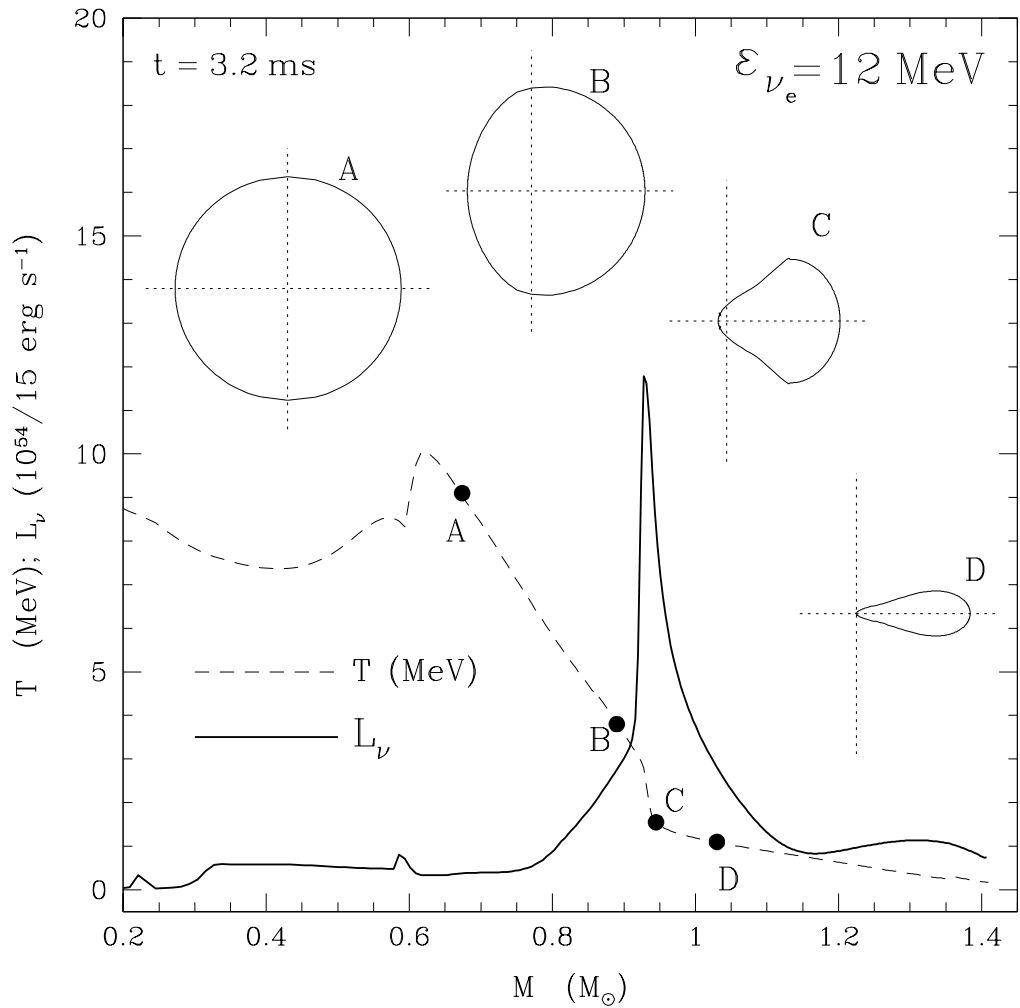}\kern+6in\hfill}
\caption[plot]
{
Temperature (MeV, dashed line) and electron neutrino luminosity
($10^{54}/15$ erg s$^{-1}$, thick solid line) as a function of enclosed
mass 3.2\,ms after bounce.  
The four small insets (thin solid lines on dotted axes)
are polar plots of the angular distribution of the electron-neutrino specific 
intensity (i.e., $I_\nu(\theta)$) for $\varepsilon_{\nu_e}=12$\,MeV 
at four mass points:  (A) 0.674\,M$_\odot$,
 (B) 0.89\,M$_\odot$, (C) 0.945\,M$_\odot$, and (D) 1.03\,M$_\odot$,
each point indicated specifically by large dots on the temperature profile.
At (A), the radiation field is isotropic, indicating that the flux is zero and
that the neutrinos are trapped.
At (D), the radiation field is nearly decoupled from the matter,
as indicated by the forward-peaked angular profile.
}
\label{p11r}
\end{figure}

The decoupling of the radiation during breakout is perhaps better seen in 
Fig.~\ref{p11r}, which shows the temperature profile
(dashed line) with the $\nu_e$ luminosity 3.2\,ms after bounce (compare with
Fig.~\ref{lms113}).  The four insets (thin solid lines on dotted axes) show polar plots of the
specific intensity, $I_{\nu_e}(\theta)$, for $\varepsilon_{\nu_e}=12$\,MeV.
The large dots on the temperature profile show the position, in units of enclosed
mass, at which the specific intensity has been sampled.  At point `A' 
($M\simeq$0.674\,M$_\odot$, $r\simeq21.8$\,km), $I_{\nu_e}(\theta)$ is virtually a circle,
indicating that the radiation field is isotropic, that the net flux is nearly zero, and that the
neutrinos at that position and energy are trapped.  In contrast, at point `D'
($M\simeq$1.03\,M$_\odot$, $r\simeq124.0$\,km), $I_{\nu_e}(\theta)$ is very forward-peaked
and well-resolved by our tangent-ray algorithm with 160 angles in each quadrant of the unit circle.
Insets `B' and `C' show  $I_{\nu_e}(\theta)$ at intermediate positions, on either
side of the shock.  The small increases in $I_{\nu_e}(\theta)$ in the forward quadrants 
of `C' are consistent with limb-brightening effects from the shock, never before resolved.  

Figure \ref{lms113} shows a marked decrease in the
breakout $\nu_e$ luminosity pulse just as it begins to propagate ahead of the
shock, between the lines labeled `t=2.4\,ms' and `t=6.8\,ms'.
The local maximum of $L_{\nu_e}$ at 2.4\,ms occurs at a radius of approximately 
$r_{\rm peak}\simeq41$\,km.
We expect the peak luminosity ($L^{\rm peak}_{\nu_e}$) to be attenuated approximately by an 
amount $\sim\exp(-\tau_a)$ in propagating to infinity ($L^{\infty}_{\nu_e}$),
where $\tau_a$ is the total absorptive optical depth exterior to $r_{\rm peak}$.
>From the results presented here, we find $\tau_a(r_{\rm peak})\sim1.9$ 
at $\varepsilon_{\nu_e}\simeq12$ MeV.  This simple estimate shows that
we should expect $L^{\rm peak}_{\nu_e}$ to decay by $\sim85$\%.  
With $L^{\rm peak}_{\nu_e}\simeq1.45\times10^{54}$ erg s$^{-1}$
we expect $L^{\infty}_{\nu_e}\sim0.22\times10^{54}$ erg s$^{-1}$ and
find from the simulation that $L^{\infty}_{\nu_e}\simeq0.25\times10^{54}$ erg s$^{-1}$,
in good agreement with our estimate.  This asymptotic luminosity is shown
clearly as the bump in the last line in Fig.~\ref{lms113}, labeled `Breakout Pulse',
which has moved out to 1.35\,\modot.
Behind the pulse, at M$\simeq1.08$\,\modot\,, is a sharper peak
in the luminosity that denotes the shock position.
During this breakout phase of shock evolution, much of the shock energy
is sapped by the neutrino losses that attend electron capture on newly-liberated free protons.
This process also creates 
a characteristic deleptonization trough manifest by the marked decrease in 
$Y_e$ in Fig.~\ref{yes11} near M$\simeq$0.8\,\modot.  The shock stalls quickly and in just 
a few milliseconds after bounce all velocities are $\leq0$.  Note the negative
entropy gradient between M$\simeq0.75$\,\modot\, and M$\simeq0.95$\,\modot.
Depending upon the compositional gradients in this region, in a multi-dimensional
simulation this region might be convectively unstable 
(Burrows and Fryxell 1992; Keil, Janka, and M\"uller 1996).

\begin{figure} 
\vspace*{6.0in}
\hbox to\hsize{\hfill\includegraphics{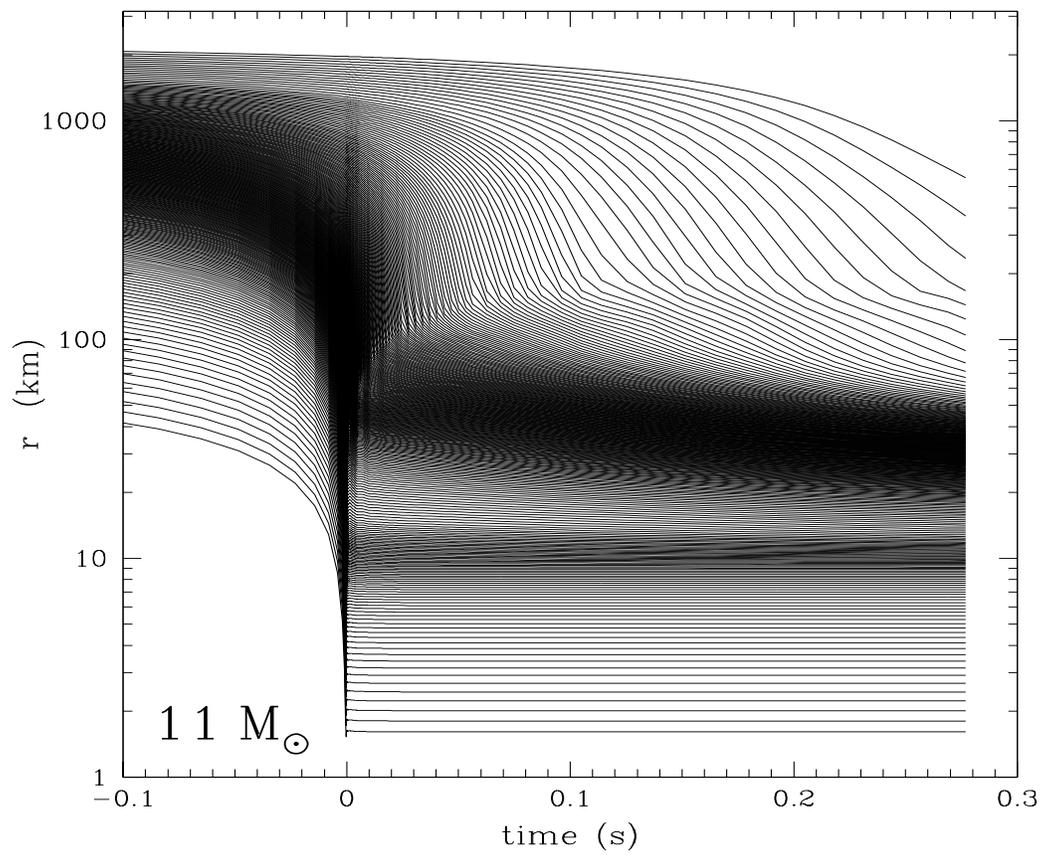}\kern+6in\hfill}
\caption[Radial position of mass zones vs.~$t$ for fiducial 11\,M$_\odot$ model]
{The radial position (in km) of selected mass shells as a function
of time in our fiducial 11\modot\, model.
}
\label{massts11}
\end{figure}

Although the shock continues to move outward in mass as matter flows through
it from the free-falling outer core, all matter velocities are negative by the time the shock reaches
$r\sim80-90$\,km.   The short dashed lines in Figs.~\ref{tots11} and
\ref{yes11} show the configuration of the model 17\,ms after bounce and $\sim$10\,ms
after the shock has stalled.  The dot-dashed line shows these basic
hydrodynamical quantities 200\,ms post-bounce.  The shock has consumed another
0.2\,\modot\, of infalling material and the entropy behind the shock 
has increased dramatically.  The temperature over the whole post-shock profile
has increased as a result of the compressional work done on these zones by the 
infalling stellar material.  In particular, at the peak in the temperature 
profile at M$\sim$0.7\,M$_\odot$, $T$ increased by $\sim$4 MeV between 
$t=17$\,ms  and $t=200$\,ms.
In these hundreds of milliseconds after bounce, the shock moves outward in radius
to $\sim175$\,km.  Although it is in this epoch that neutrino heating behind the shock
is thought to revive it, no explosion is seen in our simulations at this time.  
Figure \ref{massts11} shows the evolution of selected mass zones in our 11\,\modot\,
model as a function of time.  Small oscillations are visible just after bounce 
- the `ringing' of the core - and
the shock position is made clear by the change in the infalling matter velocity
between 100\,km and 200\,km during the 280\,ms of post-bounce evolution shown here.  
Although the neutrino-driven mechanism is not the focus 
of this work, from this figure and from many similar calculations 
we see no evidence of explosion.

\begin{figure} 
\vspace*{6.0in}
\hbox to\hsize{\hfill\includegraphics{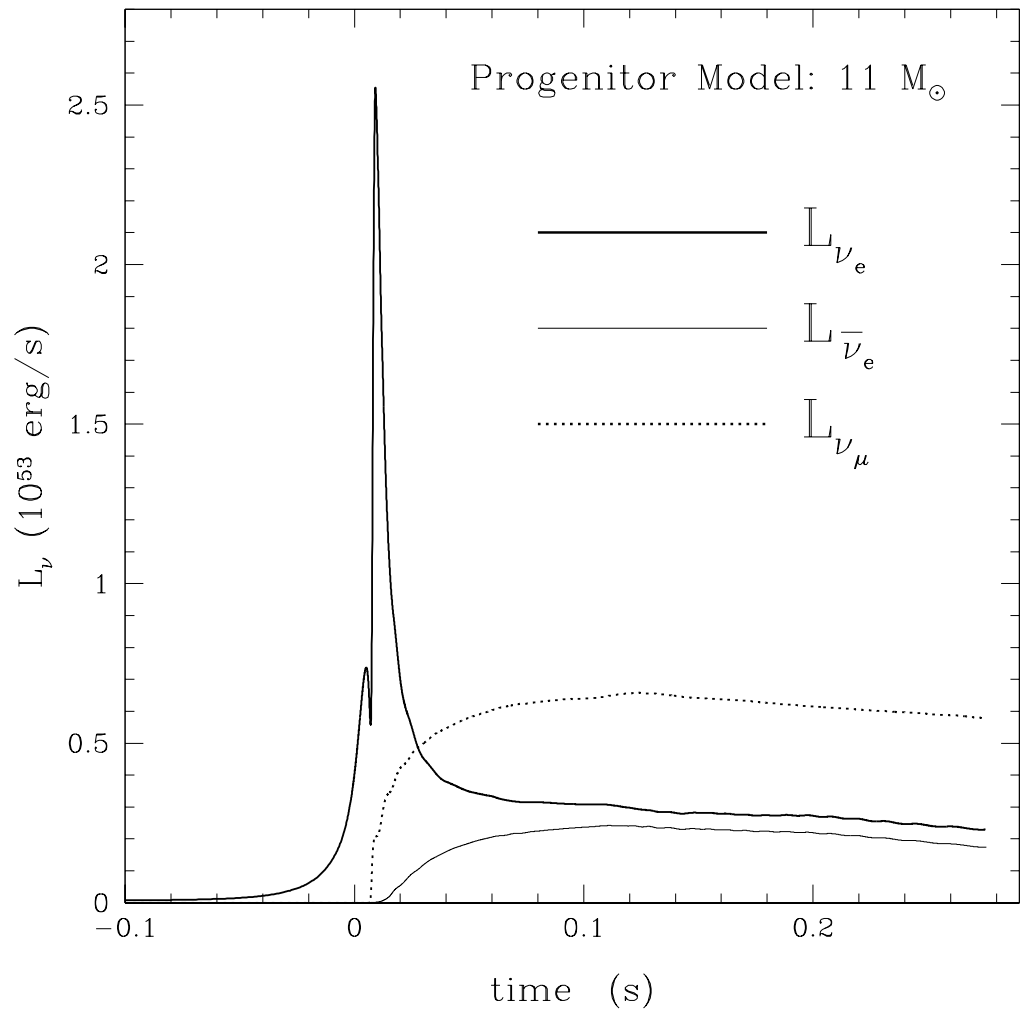}\kern+6in\hfill}
\caption[$L_{\nu_e}$, $L_{\bar{\nu}_e}$, and $L_{\nu_\mu}$ vs.~$t$ 
at infinity for 11\,M$_\odot$ progenitor]
{$L_{\nu_e}$ (thick solid line), $L_{\bar{\nu}_e}$ (thin solid line), 
and $L_{\nu_\mu}$ (dotted line) measured at the outer edge of the grid
in erg s$^{-1}$ as a function
of time for the fiducial M$=11$ \modot\, progenitor.
Time is measured relative to bounce.  Note that we define $t=0$
as the time of hydrodynamical bounce.  The finite light travel-time 
to the edge of the grid creates a $\sim7$\,ms offset between
hydrodynamical bounce and the initial dip before the large
$\nu_e$ breakout pulse.
}
\label{lts11}
\end{figure}

In Fig.~\ref{lms113}, we showed $L_{\nu_e}$ as a function of mass at various
snapshots in time.  The last snapshot shows the $\nu_e$ breakout pulse 
at 1.35\modot\, and propagating at the speed of light to the edge of the
grid.  Complementary to Fig.~\ref{lms113}, the time evolution of $L_{\nu_e}$ 
at infinity is shown in Fig.~\ref{lts11} (thick solid line).  
The luminosity of anti-electron neutrinos ($L_{\bar{\nu}_e}$, thin solid line) and
the combined luminosity of $\nu_\mu$, $\bar{\nu}_\mu$, $\nu_\tau$, and 
$\bar{\nu}_\tau$ (collectively, $L_{\nu_\mu}$, dotted line) are also shown
for the first 250\,ms of post-bounce evolution.  
Note that just before the major breakout pulse of electron neutrinos, there is
another small peak that reaches $\sim0.72\times10^{53}$ erg s$^{-1}$.  This
peak results from the deleptonization of the core as the $Y_e$ drops 
in Fig.~\ref{yes11} from that of the initial model to $Y_e\simeq0.27$.  There
is a small dip in $L_{\nu_e}$ just after this initial rise.  This dip
in the asymptotic luminosity signals the end of the infall/collapse phase 
of the supernova.  After the initial increase in $L_{\nu_e}$ due to deleptonization,
$L_{\nu_e}$ decreases because of the increased opacity in the dense
collapsing core.  The high opacity decreases the local luminosity at small 
radii and isolates the precursor peak in luminosity at larger radii caused by 
early deleptonization. 
If it were not for shock formation, $L_{\nu_e}$ at infinity would continue to decline.
The main breakout pulse just after the small
downturn in  $L_{\nu_e}$ results from the dissociation of nuclei and
subsequent electron capture on free protons,
as described above and shown in Fig.~\ref{lms113}.

\begin{figure} 
\vspace*{6.0in}
\hbox to\hsize{\hfill\includegraphics{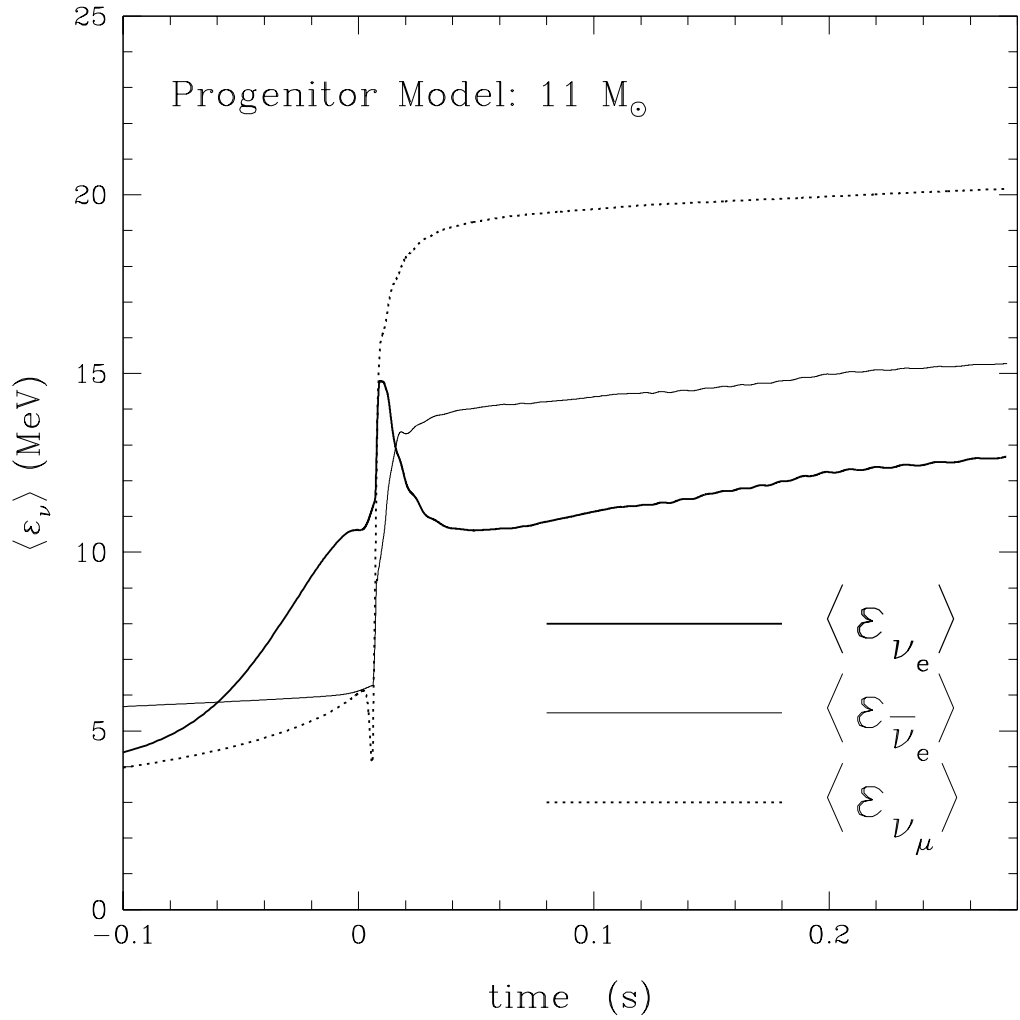}\kern+6in\hfill}
\caption[$\langle\varepsilon_{\nu_e}\rangle$, $\langle\varepsilon_{\bar{\nu}_e}\rangle$,
and $\langle\varepsilon_{\nu_\mu}\rangle$ at infinity for 11\,M$_\odot$ progenitor]
{$\langle\varepsilon_{\nu_e}\rangle$ (thick solid line), 
$\langle\varepsilon_{\bar{\nu}_e}\rangle$ (thin solid line), 
and $\langle\varepsilon_{\nu_\mu}\rangle$ (dotted line) 
at the outer edge of the grid in MeV as a function
of time for the fiducial M$=11$ \modot\, progenitor.
Averages are computed using eq.~(\ref{averagee}).
Compare with
Fig.~\ref{lts11}, which shows the corresponding luminosities:
$L_{\nu_e}$, $L_{\bar{\nu}_e}$, and $L_{\nu_\mu}$.
}
\label{ets11}
\end{figure}

In Fig.~\ref{ets11} we
show the corresponding average neutrino energy for each species over the 
same time interval as Fig.~\ref{lts11}. 
There are many ways one might define the average neutrino energy.  
The literature  does not consistently employ a unique prescription and considerable
confusion exists.  The average energy we present in our figures 
is the $rms$ average energy, defined by:
\begin{equation}
\langle\varepsilon_\nu\rangle=\left[
\frac{\int\,d\varepsilon_\nu\varepsilon_\nu^2J_\nu(\varepsilon_\nu)}
{\int\,d\varepsilon_\nu J_\nu(\varepsilon_\nu)}\right]^{1/2},
\label{averagee}
\end{equation}
where $J_{\nu}(\varepsilon_\nu)$ is the zeroth moment of the specific intensity.
We use this definition in order to make comparisons with other recent work
(Liebend\"{o}rfer et al.~2001; Rampp et al.~2002).
After approximately 50\,ms, the average energies show a nearly linear increase
with time.  Mezzacappa et al.~(2001)
and Liebend\"{o}rfer et al.~(2001) find similar behavior.  In addition,
although we employ a different progenitor, the values for the 
average neutrino energies they obtain compare well with those
presented in  Fig.~\ref{ets11}.  For example, 250\,ms after bounce in their
13\,$M_\odot$ progenitor,
Mezzacappa et al.~(2001) obtain $\langle\varepsilon_{\nu_\mu}\rangle\simeq21$ MeV,
$\langle\varepsilon_{\bar{\nu}_e}\rangle\simeq16$ MeV, and 
$\langle\varepsilon_{\nu_e}\rangle\sim14$ MeV.
We find that $\langle\varepsilon_{\nu_\mu}\rangle\simeq20$ MeV,
$\langle\varepsilon_{\bar{\nu}_e}\rangle\simeq15.5$ MeV, and 
$\langle\varepsilon_{\nu_e}\rangle\sim13$ MeV.

\begin{figure} 
\vspace*{6.0in}
\hbox to\hsize{\hfill\includegraphics{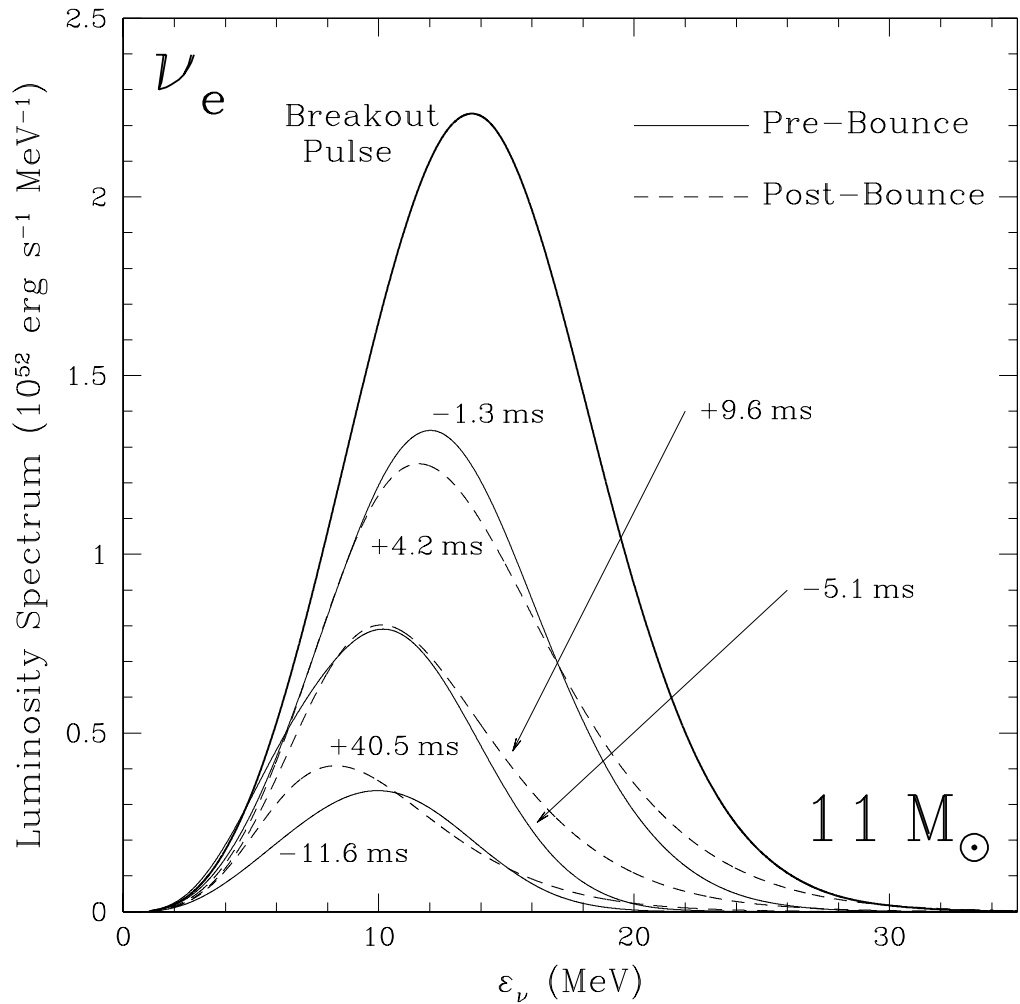}\kern+6in\hfill}
\caption[Snapshots of $\nu_e$ spectrum near breakout] 
{Luminosity spectrum of $\nu_e$ neutrinos at infinity at various 
pre- (thin solid lines) and post-breakout (dashed lines) times. 
In this figure, 
time is measured relative to the peak breakout spectrum (thick solid line). 
The thin solid lines correspond to 
11.6, 5.1, and 1.3\,ms before the peak
and the dashed lines denote the $\nu_e$ spectrum 4.2, 9.6, and 40.5\,ms after 
the peak.
}
\label{lepres11}
\end{figure}

\begin{figure} 
\vspace*{6.0in}
\hbox to\hsize{\hfill\includegraphics{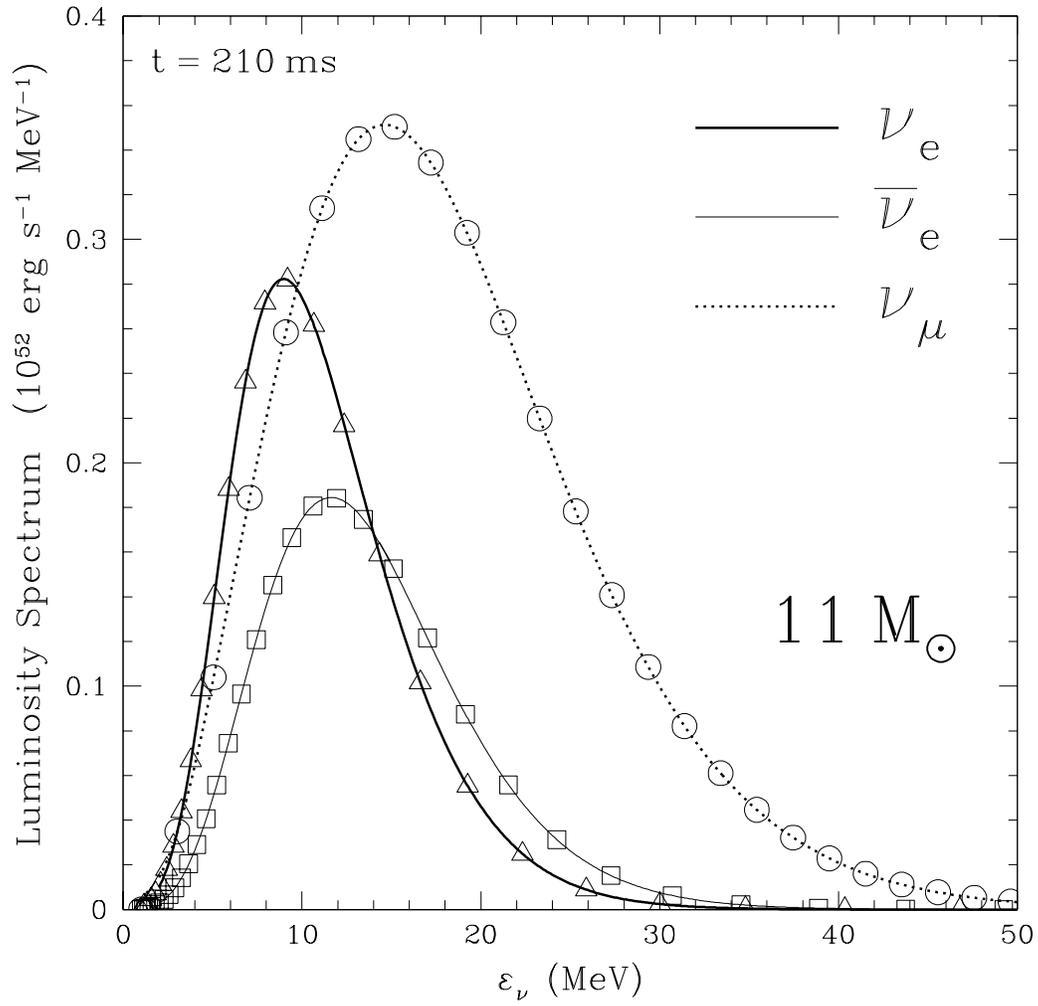}\kern+6in\hfill}
\caption[Snapshot of $\nu_e$, $\bar{\nu}_e$, $\nu_\mu$ spectra post-breakout]
{Luminosity spectrum of $\nu_e$ (thick solid line),
$\bar{\nu}_e$ (thin solid line), and $\nu_\mu$ neutrinos (dotted line)
at infinity at $t=210$\,ms after bounce.  The actual energy groups
are denoted by triangles, squares, and circles for $\nu_e$, $\bar{\nu}_e$,
and $\nu_\mu$ neutrinos, respectively.
}
\label{alems11}
\end{figure}

Figure \ref{lepres11} shows the luminosity spectrum of electron neutrinos 
at infinity in units of 10$^{52}$ erg s$^{-1}$ MeV$^{-1}$.  The thick solid line denotes the
maximum peak breakout pulse luminosity and corresponds to the 
$2.55\times10^{53}$ erg s$^{-1}$ peak in $L_{\nu_e}$ in Fig.~\ref{lts11}.
The thin solid lines in this plot show the rise up to the peak
and the simultaneous shift in $\langle\varepsilon_{\nu_e}\rangle$ depicted
in Fig.~\ref{ets11}.  The dashed lines show the evolution of
the luminosity spectrum after the breakout pulse.  
The pre-breakout lines show the spectrum at 11.6, 5.1, and 1.3\,ms before
the peak in the breakout pulse.  The dashed, post-breakout, lines
show the spectrum at 4.2, 9.6, and 40.5\,ms after the $\nu_e$ breakout pulse.
Although in this figure the peak energy
is dropping as the luminosity decays, the high energy tail
becomes broader and with our definition of the average energy (eq.~\ref{averagee})
$\langle\varepsilon_{\nu_e}\rangle$ increases.  Figure \ref{alems11} shows
the emergent spectrum much later (at 210\,ms after bounce) and for all
three neutrino species.  The thick solid line shows the $\nu_e$ spectrum,
the thin solid line shows the $\bar{\nu}_e$ spectrum, and the dotted line shows
the combined $\nu_\mu$ spectrum. The post-bounce energy hierarchy one expects
from simple considerations of the opacity sources for the various species,
$\langle\varepsilon_{\nu_e}\rangle<\langle\varepsilon_{\bar{\nu}_e}\rangle
<\langle\varepsilon_{\nu_\mu}\rangle$, is evident in both Figs.~\ref{ets11} 
and \ref{alems11}.
 
Although this exact progenitor model has not been used in any of the recent 
supernova modeling done by groups employing sophisticated neutrino
transport (Rampp 2000; Rampp \& Janka 2000, 2002; 
Liebend\"{o}rfer et al. 2001; Liebend\"{o}rfer, Mezzacappa, \& Thielemann 2001; 
Mezzacappa et al.~2001), we find that the basic structures and systematics are similar
between groups.  In particular, our core entropy ($s_{\rm core}$) drops slightly 
at the beginning of the calculation
and then climbs during bounce to $s_{\rm core}\simeq1.32$ $k_B$ baryon$^{-1}$.  
We observe similar behavior for our 15\,M$_\odot$ progenitor with
$s_{\rm core}\simeq1.33$ $k_B$ baryon$^{-1}$.
Without inelastic neutrino-electron scattering we obtain 
$s_{\rm core}\simeq1.16$ $k_B$ baryon$^{-1}$.
The absolute magnitude of these results and the effect of neutrino-electron scattering 
on $s_{\rm core}$ are duplicated in the work of Rampp (2000), whose Figs.~5.3a and 5.6b
shows that with and without neutrino-electron scattering in their 15\,M$_\odot$ model
they obtain $s_{\rm core}\simeq1.3$ and 1.14 $k_B$ baryon$^{-1}$, respectively.
A recent collapse calculation by Liebend\"{o}rfer et al.~(2002b) of a 13\,M$_\odot$
progenitor also shows $s_{\rm core}\sim1.3$ $k_B$ baryon$^{-1}$.
Core $Y_e$ ($Y^{\rm core}_e$) in our 11\,M$_\odot$ model drops
to 0.266 and 0.293 at bounce in our models with and without 
neutrino-electron scattering, respectively.
Both numbers from Rampp (2000) are slightly higher; his Fig.~5.3b
gives the corresponding numbers as approximately 0.28 and $0.30-0.31$, respectively.
Liebend\"{o}rfer et al.~(2002b) have $Y^{\rm core}_e\simeq0.30$, including neutrino-electron
scattering. Peak positive velocities in all our calculations reach $\sim2\times10^9$ cm s$^{-1}$
at M$\simeq0.7$\modot\,- in good agreement with previous work.

One important difference bears mention.
Our peak breakout electron-neutrino luminosity is slightly lower than that obtained
in other work.  Both Liebend\"{o}rfer et al.~(2001) and Rampp \& Janka (2000)
obtain $L_{\nu_e}^{\rm\,peak}$ of $\sim3.5\times10^{53}$ erg s$^{-1}$
with comparable progenitors.  Indeed,  Liebend\"{o}rfer et al.~(2002a) present
the breakout pulse for several different progenitors and find
virtually identical $L_{\nu_e}^{\rm\,peak}$ for all models.
We obtain $L_{\nu_e}^{\rm\,peak}\simeq2.55\times10^{53}$ erg s$^{-1}$
in our 11\modot\, and virtually the same number with the 15 \modot\, and 20 \modot\,
models we present in \S\ref{section:progenitor}.  Our results
consistently give lower $L_{\nu_e}^{\rm\,peak}$ by about 30\%.
This luminosity pulse is a signature of the most dynamical phase
in supernova modeling and it is important to achieve adequate angular,
energy, and spatial zoning to resolve it.  For comparison,
Mezzacappa et al.~(2001) and Liebend\"{o}rfer et al.~(2001) 
typically employ eight angular bins using the S$_n$ method at each radial point and twelve 
energy groups.  In particular, poor angular resolution at large radii may
compromise the propagation of the breakout pulse.
Liebend\"{o}rfer et al.~(2002b) state that they
experience significant ``numerical diffusion'' of their breakout pulse
as it propagates from the point of decoupling to the edge of their 
computational grid.  
The peak asymptotic luminosity they quote, 
$L_{\nu_e}^{\rm\,peak}\sim$3.5$\times10^{53}$ erg s$^{-1}$,
is sampled at 500\,km in their calculations.  This
peak drops $\sim$30\% to $\sim$2.25$\times10^{53}$ erg s$^{-1}$ in propagating
to $\sim$4200\,km (see their Fig.~14 and Fig.~20).
Plotting the radial coordinate of the peak in the luminosity 
pulse as a function of $r-ct$, Liebend\"{o}rfer et al.~(2002b)
find that $L_{\nu_e}^{\rm\,peak}$ decreases, shifts by $\sim500-1000$\,km,
and broadens due to numerical diffusion.  
Constructing a similar plot with our collapse calculations,
over a similar range of radii, we find a shift in the pulse
peak of just $\lesssim10$\,km, or a fractional deviation 
from $c$ of $\sim$0.3\%.  We also find a decrease in $L_{\nu_e}^{\rm\,peak}$
between 500\,km and 4200\,km of 8.5\%.  In short,
we obtain a lower asymptotic $\nu_e$ breakout pulse
than Liebend\"{o}rfer et al.~(2002b), that cannot be
explained by the considerable numerical diffusion in their
first-order upwind differencing scheme.  
Using the tangent-ray method, our $L_{\nu_e}^{\rm\,peak}$ 
does not experience the same degradation and maintains 
an on-grid speed remarkably close to $c$ after fully decoupling
from the matter.

Although they describe it as diffusion, 
the temporal dispersion and decay Liebend\"{o}rfer et al.~(2002b) 
experience in the electron-neutrino breakout pulse might be better interpreted as a consequence
of the progressive loss of angular resolution as the luminosity
pulse propagates outward and, thereby, becomes more and more forward-peaked.
Though a fixed angular grid with eight points can not resolve a sharply-pointed pulse,
it should be adequate for calculating the flux factors of relevance to
neutrino energy deposition in the gain region and for gauging the
viability of the neutrino mechanism of supernova explosions.

\begin{figure} 
\vspace*{6.0in}
\hbox to\hsize{\hfill\includegraphics{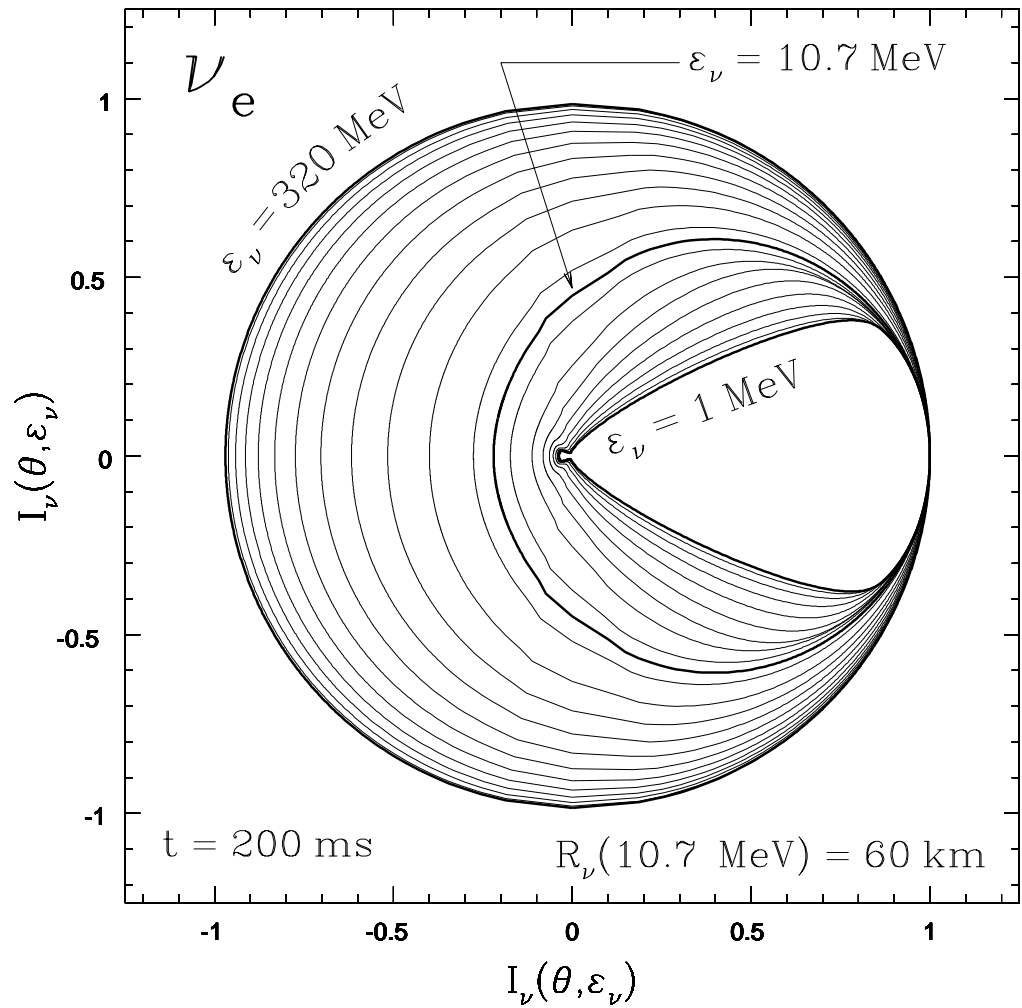}\kern+6in\hfill}
\caption[$I_\nu(\theta,\varepsilon_\nu)$ for $\nu_e$ neutrinos at $R_{\nu_e}$]
{Polar plot of the normalized specific intensity,
constructed from the Feautrier variables at the neutrinosphere
(as defined in eq. \ref{nusphere}) for $\varepsilon_\nu\simeq10.7$ MeV for $\nu_e$ neutrinos.
This plot shows only every other energy grid point (solid lines).
There are 269 angular bins in each quadrant.   At the largest
energies, the radiation field is nearly isotropic and the
flux is quite small.  At the lowest energies,
the radiation field is beginning to decouple from the matter and
$I_\nu$ is forward-peaked.
}
\label{polare}
\end{figure}

Two hundred milliseconds after bounce, the shock has moved
out to about 200\,km in radius.  A gain region of net heating
has formed between the gain radius $R_g\simeq$100\,km  
(where heating balances cooling) and the shock. 
For $\varepsilon_{\nu_e}=10.7$ MeV, the $\nu_e$ neutrinosphere
($R_{\nu_e}$, as defined in eq.~\ref{nusphere}) occurs at $\simeq60$\,km.
Between $R_{\nu_e}$ and $R_g$ is a region of net cooling.
The solution to the Boltzmann equation yields the specific intensity
of the neutrino radiation field, $I_\nu$, as a function of both energy
and angle, at every radial coordinate and timestep.  
When neutrinos of a given energy are trapped and diffusive, their
radiation field is isotropic in angle.  Beyond the neutrinosphere,
where they approach the free-streaming limit, the neutrino radiation
field becomes continuously more forward-peaked.
In an effort to understand the angular distribution of $I_{\nu_e}$, 
Fig.~\ref{polare} shows a polar plot of $I_{\nu_e}(\theta)$ for many $\nu_e$ energies
at $R_{\nu_e}(\varepsilon_{\nu_e}=10.7\,{\rm MeV})=60$\,km. 
All curves in this figure have been normalized to the same maximum specific intensity.
The outer near-circle (thick solid line) in
this figure shows $I_{\nu_e}(\theta)$ for $\varepsilon_{\nu_e}=320$\,MeV.
Fig.~\ref{polare} shows that although 60\,km is $R_{\nu_e}$ for neutrinos 
with $\varepsilon_{\nu_e}=10.7$ MeV, 320\,MeV neutrinos are completely trapped 
at this radius; $I_{\nu_e}(\theta)$ is isotropic and the net flux is nearly
zero for $\varepsilon_{\nu_e}=320$\,MeV.  This is a consequence of the fact 
that the neutrino opacity is proportional to $\varepsilon_\nu^2$.  
Also shown is $I_\nu(\theta)$ for other neutrino energies 
down to 1\,MeV (also a thick solid line).
Only every other neutrino energy calculated is shown.
Going from high to low energies, one sees that the radiation field
becomes less isotropic, the lowest energies being the most forward-peaked.  
The $\varepsilon_{\nu_e}=10.7$\,MeV point is highlighted (middle thick
solid line) to show the angular distribution of the radiation field
at the neutrinosphere for that energy. 
A glance at Fig.~\ref{polare} shows how forward-peaked the radiation
field is at the point of decoupling (the neutrinosphere), for
a given energy (in this case, 10.7\,MeV).
Note that in this calculation, $R_{\nu_e}$ occurs at our
270th radial zone, so that in each quadrant of the polar plot there 
are 269 angular bins.  

\begin{figure} 
\vspace*{6.0in}
\hbox to\hsize{\hfill\includegraphics{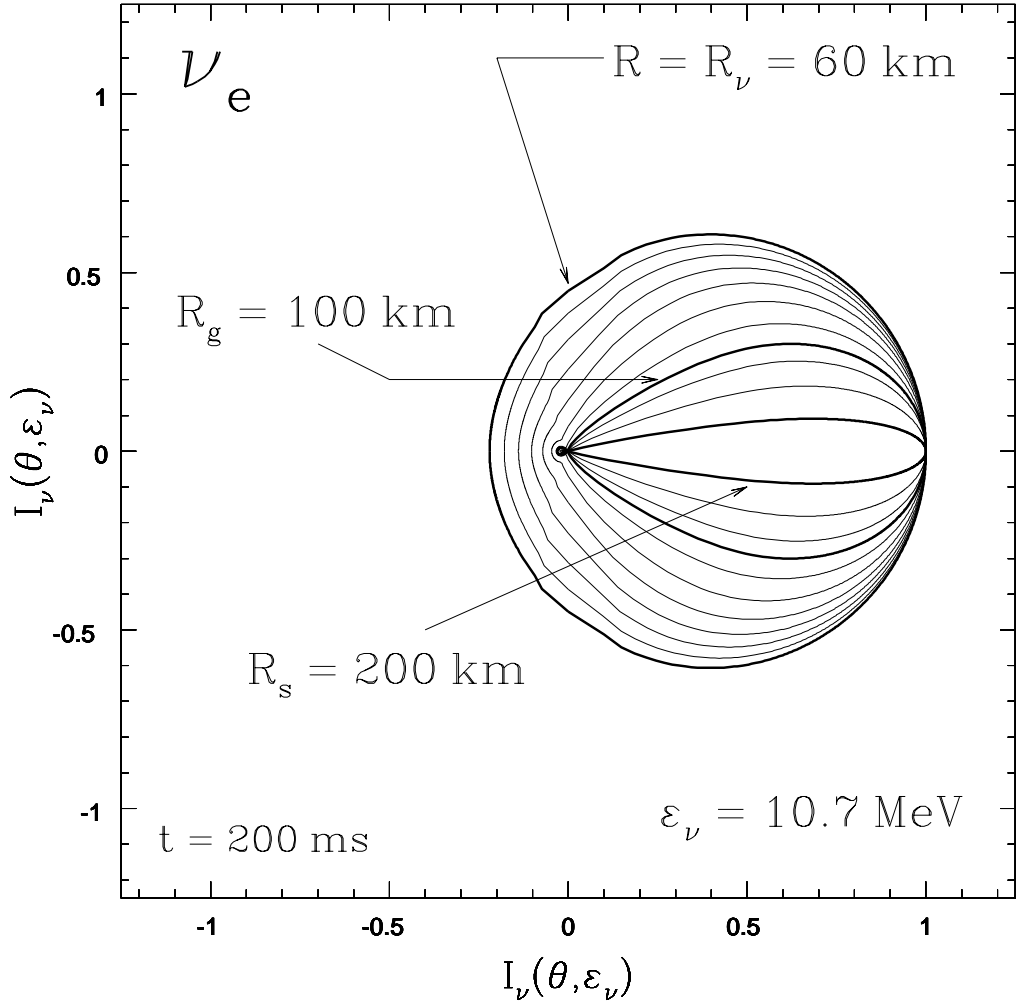}\kern+6in\hfill}
\caption[$I_\nu(\theta,\varepsilon_\nu)$ for $\nu_e$ neutrinos for radii from $R_g$ to $R_s$]
{Polar plot of the specific intensity ($I_{\nu_e}$) 
as a function of $\theta$ for 
$\nu_e$ neutrinos with $\varepsilon_\nu=10.7$ MeV at selected radii. 
We maintain the same scale as in Fig.~\ref{polare} for the line labeled `60 km'
and normalize the absolute value  $I_\nu$ for all other radii to that reference $I_\nu$.
Thick solid
lines denote $I_{\nu_e}(\theta,\varepsilon_\nu=10.7\,{\rm MeV})$ 
at the neutrinosphere ($R_\nu\simeq60$\,km), 
at the gain radius ($R_g\simeq100$\,km), 
and at the shock radius ($R_s\simeq200$\,km) 200\,ms after bounce.  
Thin solid lines show the specific intensity at 
various intermediate radii.
}
\label{polarr}
\end{figure}

\begin{figure} 
\vspace*{6.0in}
\hbox to\hsize{\hfill\includegraphics{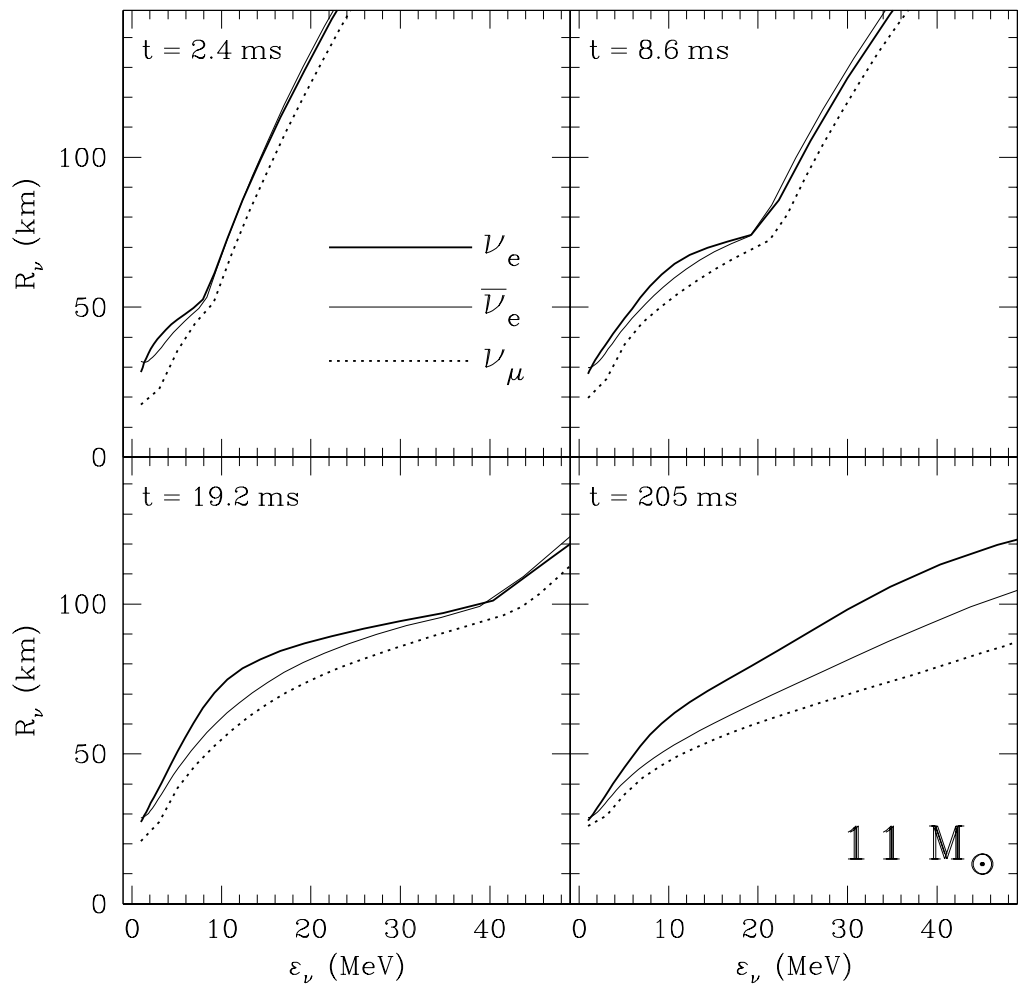}\kern+6in\hfill}
\caption[neutrinospheres]
{The neutrinospheres (as defined by eq.~\ref{nusphere})
as a function of neutrino energy for each neutrino species
($\nu_e$, thick solid line; $\bar{\nu}_e$, thin solid line;
$\nu_\mu$, dotted line) for four post-bounce times in the
baseline 11\,M$_\odot$ model.  
}
\label{rnu2}
\end{figure}

Also of interest is the dependence of the angular distribution 
of the radiation field on radius for a given energy.  Figure \ref{polarr} shows 
how $I_{\nu_e}(\theta,\varepsilon_{\nu_e}=10.7\,\,{\rm MeV})$ evolves from the 
neutrinosphere radius ($\sim$60\,km, outer thick solid line) 
to the shock radius ($R_s\simeq200$\,km, inner thick solid line).
Also shown is $I_{\nu_e}(\theta)$ at $R_g$, where heating balances cooling,
and $I_{\nu_e}(\theta)$ at larger radii, where heating dominates.
This figure shows that the angular distribution of the radiation field 
is quite forward-peaked in the heating region.  As noted in 
Liebend\"{o}rfer et al.~(2002b), insufficient angular resolution in this region
can lead to artificial enhancements in the energy deposition profile and, hence,
could affect the subsequent dynamics. In Fig.~\ref{rnu2}, we show the 
neutrinosphere radius (as defined in eq.~\ref{nusphere}) 
as a function of energy for all neutrino species at four different snapshots in time from just after 
breakout to 200\,ms after bounce.  $R_{\nu_e}$, $R_{\bar{\nu}_e}$, and 
$R_{\nu_\mu}$ are the thick, thin, and dotted lines, respectively.
The break in the neutrinosphere radius profiles at approximately 50\,km, 70\,km, 
and 100\,km in the first three panels roughly denotes the position of the
shock.  The bottom-right panel corresponds roughly in time with the polar plots
of the radiation field in Figs.~\ref{polare} and \ref{polarr}.
Note that the energy hierarchy, $\langle\varepsilon_{\nu_e}\rangle<
\langle\varepsilon_{\bar{\nu}_e}\rangle<\langle\varepsilon_{\nu_\mu}\rangle$, 
is reflected here: $R_{\nu_\mu}<R_{\bar{\nu}_e}<R_{\nu_e}$.

\section{Spectral Resolution}
\label{section:eres}

Our baseline model employs 40 energy groups for each neutrino species,
with the actual grouping described in \S\ref{section:baseline}.  In an effort
to understand the effects of degrading the spectral resolution, we have run
the same 11 \modot\, model, with the same spatial zoning and the same physics as in the
baseline model, but with 20 and 10 energy groups.  All three calculations
were carried out to 250 milliseconds after bounce.

We observe only small quantitative differences and virtually no qualitative differences
in the overall evolution of all quantities in comparing the run with 40 energy groups to that
with only 20.  Over the whole post-bounce evolution, $\langle\varepsilon_{\nu_e}\rangle$
is $\sim$0.05 MeV higher in the 20-group calculation, a difference of just 0.5\%
at 250 ms.  However, $L_{\nu_e}$ is higher by  2\% in the model with just 20 energy groups.
Differences in the average energy and luminosity of $\bar{\nu}_e$ and $\nu_\mu$
neutrinos amount to less than $\sim1$\% overall, with each of these quantites
lower in the 20-group calculation than in the model run with 40 energy groups.
The basic hydrodynamical quantities were similarly unaffected.  The core entropy 
was reproduced in the the 20-group model to within 0.5\%, with the higher resolution model
having slightly lower entropy.  The core $Y_e$ at bounce was just 0.5\% lower
in the model with 20 groups.  The only qualitative difference
observed between the two models was a slight oscillation in the velocity versus
mass profile inside 0.5 \modot\ just before bounce.

We find significant qualitative and quantitative differences between our 10-group
model and our models with higher spectral resolution.  The small oscillations
observed in the matter velocity in the 20-group model are much more
pronounced on infall with just 10 groups.  There are similar and seemingly correlated
oscillations in $Y_e$.  Oscillations very similar to these have been observed in 
Rampp (2000) and Mezzacappa \& Bruenn (1993c).  The oscillations result
from insufficient spectral resolution of the Fermi surface of electron neutrinos.
In addition to the oscillations, many components of the hydrodynamic and
neutrino spectral evolution are affected by the use of only 10 energy groups
in our calculation.
At bounce, the core entropy, which had come within 5\% of
that obtained in the 20- and 40-group calculations ($s_{\rm core}\sim1.35$ $k_B$ baryon$^{-1}$), 
dropped to 0.8 $k_B$ baryon$^{-1}$.  This, of course, also manifests itself in the temperature profile.
The core temperature after bounce reaches only 7.5 MeV, as compared with the 11 MeV
obtained in the calculations with 20 and 40 energy groups.  Despite the fact that the core $Y_e$
was slightly higher in the 10-group calculation, the peak positive shock velocity 
just after bounce was 15\% lower.  The $\nu_e$ breakout pulse was 10\% lower
in the 10-group model than in the 40- or 20-group models. 
$L_{\nu_e}$ was $\sim0.03\times10^{53}$ erg s$^{-1}$ higher in the low-resolution model
throughout the post-bounce evolution, this difference amounting to $\sim8$\% at 250 ms.
In addition, we find that $L_{\bar{\nu}_e}(t)$ 
in the 10-group model peaks faster and higher than the 40-group model, but drops off
more quickly in time.  Although $L_{\bar{\nu}_e}$ at 250 ms is just 9\% lower 
in the 10-group model than the 40-group model, the slope of the luminosity
is steeper and that difference grows with time.  The average energy for all
species is higher in the 10-group calculation than in the 40-group calculation,
but not by more than a few percent over the whole post-bounce evolution.

In sum, we conclude that 10 neutrino energy groups are insufficient for resolving
the neutrino radiation field.  We see the largest differences in the emergent
spectrum for the electron neutrinos.  This is likely due to the fact that
the $\varepsilon_{\nu_e}$ grouping must extend to $\sim300$ MeV in order to 
resolve the Fermi energy of the electron neutrinos within the core.  Because $\nu_e$
transport, through the charged-current interaction,
affects the dynamics so significantly, failure to resolve the $\nu_e$
radiation field affects most observables.  Our model employing 20 energy
groups comes quite close to reproducing our results with 40 energy groups.
Differences are no larger than 2\% and, for most quantities, the fractional differences
are closer to 0.5$-$1\%.  Considering the large decrease in computation time 
(a factor of 2), 20 groups are probably desirable for all but the most detailed models. 
In comparison, with 21 energy groups Rampp (2000) finds that the $\nu_e$ Fermi surface 
is under-resolved, as the large oscillations he sees in the trapped lepton fraction demonstrate.
Typical recent calculations by Rampp \& Janka (2002)
employ $20-30$ energy groups spaced geometrically.
Mezzacappa et al.~(2001) and Liebend\"{o}rfer et al.~(2001)
employ 12 energy groups.

\section{Comparison Between Progenitors: \\ 15\,M$_\odot$ and 20\,M$_\odot$ Models}
\label{section:progenitor}

A complete theory of core-collapse supernovae must hope to understand the detailed
dynamics and neutrino signatures of all possible supernova progenitors.  If 
successful neutrino-driven supernova models were obtained, one might hope
to find systematic trends between, for example, the progenitor mass and the peak 
neutrino luminosity, final supernova energy, or mass of the nascent protoneutron
star.  We begin such an investigation by first considering the dynamics and emergent
neutrino characteristics of two more massive progenitors.

\begin{figure} 
\vspace*{6.0in}
\hbox to\hsize{\hfill\includegraphics{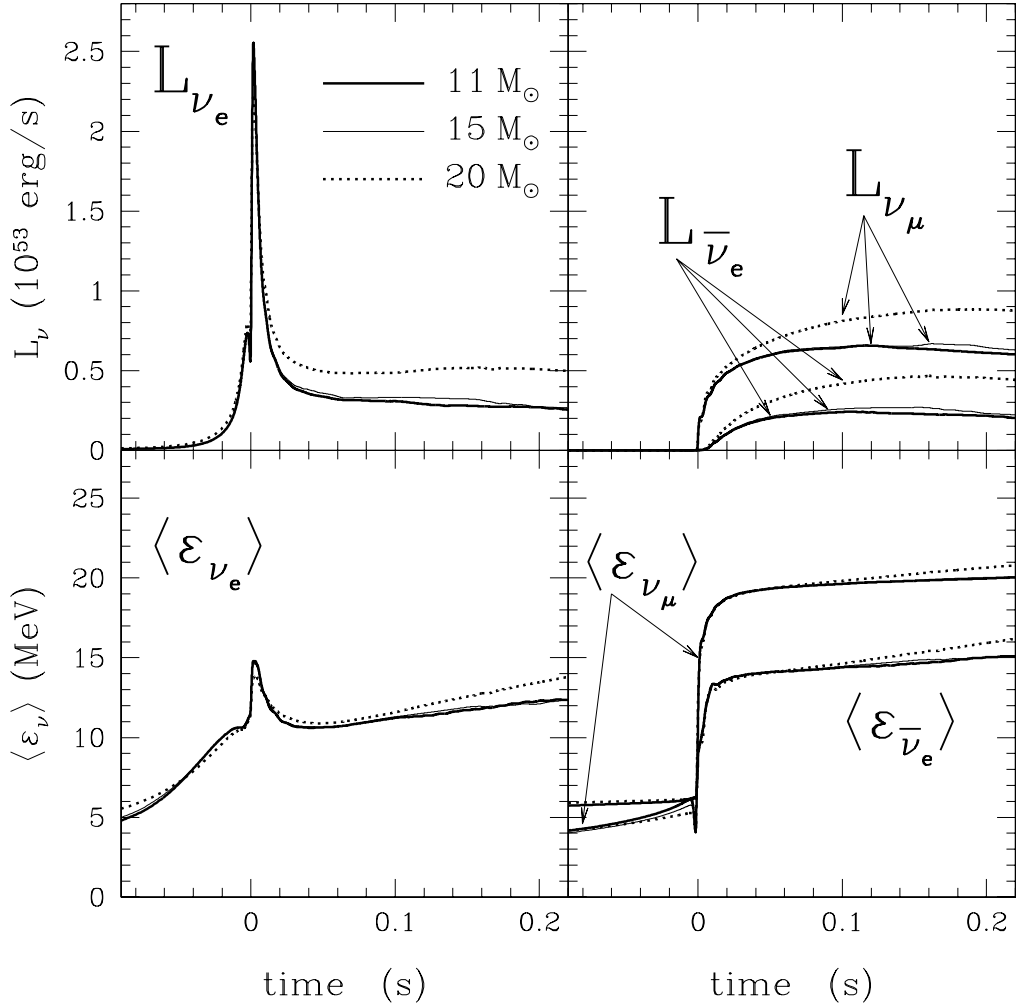}\kern+6in\hfill}
\caption[plot]
{$L_{\nu_e}$ (10$^{53}$\,erg\,s$^{-1}$, upper left-hand panel), $L_{\bar{\nu}_e}$ 
and $L_{\nu_\mu}$ (upper right-hand panel), $\langle\varepsilon_{\nu_e}\rangle$ 
(MeV, lower left-hand panel), and $\langle\varepsilon_{\bar{\nu}_e}\rangle$ and
$\langle\varepsilon_{\nu_\mu}\rangle$ (lower right-hand panel) as a function of time
at infinity for three different progenitors: 11\,M$_\odot$ (the baseline model, thick
solid line), 15\,M$_\odot$ (thin solid line), and 20\,M$_\odot$ (dotted line).
}
\label{pro}
\end{figure}

The top two panels of Fig.~\ref{pro} show the $L_{\nu_e}$ (10$^{54}$\,erg\,s$^{-1}$, 
upper left-hand panel) and 
$L_{\bar{\nu}_e}$ and $L_{\nu_\mu}$ (upper right-hand panel), 
at infinity as a function of time for three different progenitor masses:
11\,M$_\odot$ (the baseline model, thick solid line), 15\,M$_\odot$ (thin solid line), 
and 20\,M$_\odot$ (dotted line).  Note that the $\nu_e$ breakout pulse for each
progenitor is remarkably similar, a result also recently obtained in the work of 
Liebend\"{o}rfer et al.~(2002a).
Interestingly, as discussed in \S\ref{section:baseline}, 
our $\nu_e$ breakout pulse reaches $\sim2.55\times10^{53}$ erg s$^{-1}$, whereas
Liebend\"{o}rfer et al.~(2002a) obtain a characteristic peak $L_{\nu_e}$ of 
$\sim3.5\times10^{53}$ erg s$^{-1}$.  Note that our 20 \modot\, model,
which also employs 300 spatial zones, 40 neutrino energy groups,  
and the tangent-ray algorithm for angular binning, reaches a somewhat
lower peak luminosity at $\sim2.225\times10^{53}$ erg s$^{-1}$.

$L_{\bar{\nu}_e}$ and $L_{\nu_\mu}$ are virtually identical in both
the 11\,M$_\odot$ and 15\,M$_\odot$ models over the first 100 ms of post-bounce evolution
and only later develop small differences.  In marked contrast, the 20\,M$_\odot$ model exhibits
higher neutrino luminosities for all species after electron-neutrino breakout.  
For $L_{\nu_e}$ and $L_{\bar{\nu}_e}$, 
the difference between the 11\,M$_\odot$ model and the 20\,M$_\odot$
model is a factor of two
at 200 milliseconds after bounce.  For $L_{\nu_\mu}$, the difference is a factor of 1.5.
In addition, $L_{\nu_\mu}$  and $L_{\bar{\nu}_e}$ peak approximately 100 ms later
in the 20\,M$_\odot$ model than in the 11\,M$_\odot$ progenitor.

\begin{figure} 
\vspace*{6.0in}
\hbox to\hsize{\hfill\includegraphics{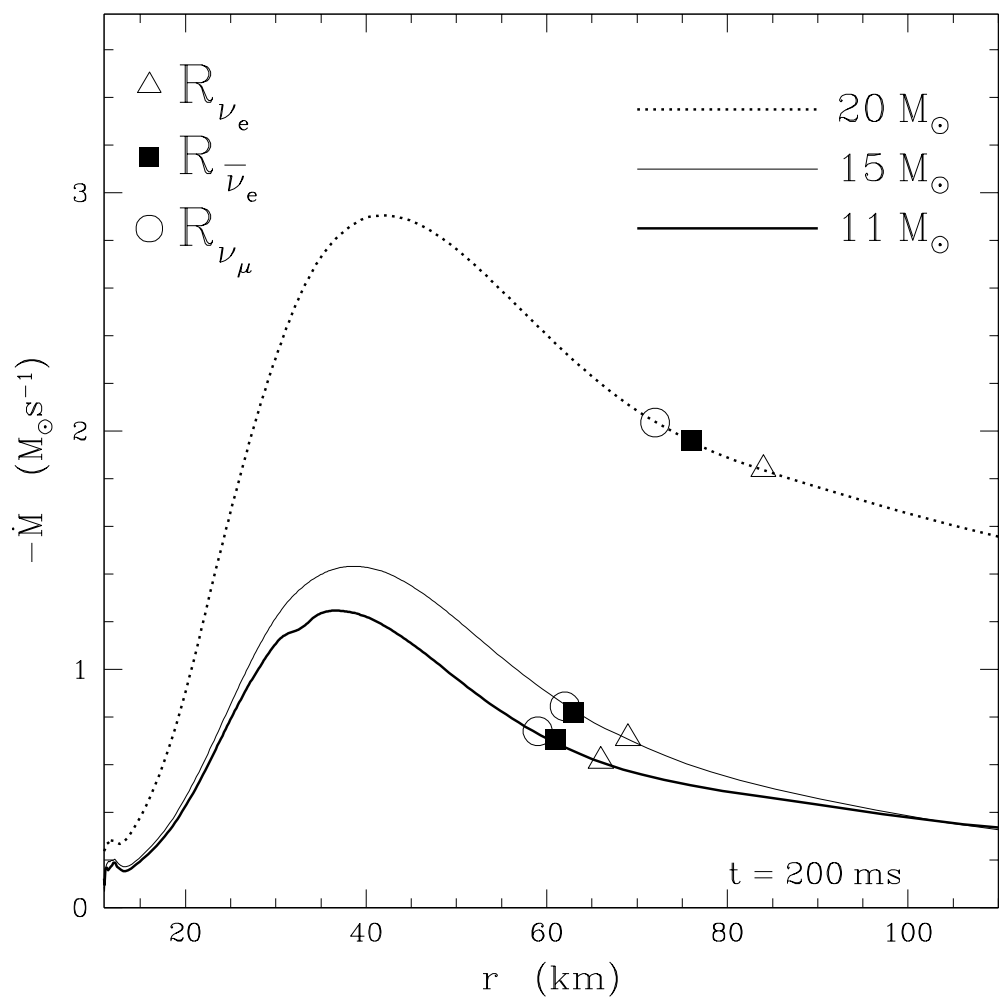}\kern+6in\hfill}
\caption
[$\dot{{\rm M}}$ vs.~$r$ \& $R_{\nu_e}$, $R_{\bar{\nu}_e}$, \& $R_{\nu_\mu}$
for 11\,M$_\odot$, 15\,M$_\odot$, 
\& 20\,M$_\odot$ progenitors at 200\,ms post-bounce]
{Mass flux $\dot{{\rm M}}$ in M$_\odot$ s$^{-1}$ as a function of radius
for our 11 \modot\, (thick solid line), 15 \modot\, (thin solid line), 
and 20 \modot\, (dotted line) models, all at 200\,ms after bounce.
Open triangles mark the $\nu_e$ neutrinosphere ($R_{\nu_e}$) at 
$\langle\varepsilon_{\nu_e}\rangle\simeq12$, 12, and 13.5\,MeV for the 11\modot,
15\modot, and 20\modot\, models, respectively. Filled squares and
open circles mark 
$R_{\bar{\nu}_e}(\langle\varepsilon_{\bar{\nu}_e}\rangle\simeq15,\,15,\,16\,{\rm MeV})$ 
and $R_{\nu_\mu}(\langle\varepsilon_{\nu_\mu}\rangle\simeq20,\,20,\,21\,{\rm MeV})$
for the  11 \modot,
15 \modot, and 20 \modot\, models, respectively.
}
\label{massfpro}
\end{figure}

Also shown in Fig.~\ref{pro} (lower two panels) are the corresponding average
neutrino energies (defined by eq.~\ref{averagee}) of the emergent spectra.
The lower left-hand panel shows the time evolution of $\langle\varepsilon_{\nu_e}\rangle$ 
and the lower right-hand panel shows the same for 
$\langle\varepsilon_{\bar{\nu}_e}\rangle$ and $\langle\varepsilon_{\nu_\mu}\rangle$.
One hundred milliseconds after bounce the differences between the spectra 
from each model become significant.  Not only are the spectra for
the 20\,M$_\odot$ model harder, but the average energies 
are increasing faster than those for the 11\,M$_\odot$ and 15\,M$_\odot$ models.  
Some of these systematics can be understood by inspecting Fig.~\ref{massfpro},
which shows the mass flux ($\dot{M}$) as a function of radius 
in units of \,M$_\odot$ s$^{-1}$ for the 11\,M$_\odot$ (thick solid line),
15\,M$_\odot$ (thin solid line), and 20\,M$_\odot$ (dotted line) models.  Also shown
are the positions of the neutrinosphere for the average emerging neutrino energy.
Open triangles correspond to $R_{\nu_e}$, whereas filled squares and open circles
denote $R_{\bar{\nu}_e}$ and $R_{\nu_\mu}$, respectively. 
In the steady-state, the total luminosity of a given neutrino species
should be given approximately 
by the accretion luminosity, written in terms of $\dot{M}$ and $R_\nu$:
$L_\nu^{\rm acc}\sim GM_\nu\dot{M}/R_\nu$, where $M_\nu$ is the mass
enclosed by $R_\nu$.  This reflects the role of the neutrino luminosity
in carrying away the energy of infall. 
To zeroth order, the higher mass flux in the 20\,$M_\odot$ model
in Fig.~\ref{massfpro} explains the higher luminosity at that epoch in 
the top two panels of Fig.~\ref{pro}.  This conclusion is modified slightly
by the actual radial positions  of the various neutrinospheres for
each progenitor in this epoch and the mass enclosed within $R_\nu$.
Given the simple expression for $L_\nu^{\rm acc}$, the ratio of the total 
neutrino luminosity of the 20\,$M_\odot$ model
to that of the 11\,$M_\odot$ model should be approximately 
$$\frac{L^{\rm 20\,M_\odot}_{\nu}}{L^{\rm 11\,M_\odot}_{\nu}}\,=\,
\left(\frac{M_{\nu}^{\rm \,20\,M_\odot}}{M_{\nu}^{\rm \,11\,M_\odot}}\right)\,
\left(\frac{\dot{M}^{\rm \,20\,M_\odot}}{\dot{M}^{\rm \,11\,M_\odot}}\right)\,
\left(\frac{R_{\nu}^{\rm \,11\,M_\odot}}{R_{\nu}^{\rm \,20\,M_\odot}}\right)\sim
2.5.$$
We find that the  actual ratio, as obtained by solving the full problem, is $\sim1.9$.
These significantly higher neutrino luminosities in the post-bounce phase and slightly harder
average energies affect the number of neutrino events we expect from the next galactic supernova.
In \S\ref{section:detection} we present detection results for each progenitor.

Although we obtain higher average neutrino luminosities and energies for
the 20\,M$_\odot$ progenitor, there is no sign of explosion or developing
explosion in this model.  In fact, approximately 200\,ms after bounce the
shock in this progenitor has reached $r\sim190$\,km and begins to recede;
at approximately 250\,ms, it has dropped to $r\sim150$\,km.  In contrast,
both the 11\,M$_\odot$ and 15\,M$_\odot$ progenitors maintain shock
radii of 180-190\,km throughout this epoch.  Although none of these simulations
yields an explosion, the lower mass models seem more promising.

\section{Dependence on the Nuclear Equation of State}
\label{section:comp}

One might expect that large modifications to the nuclear equation of state 
could have a significant impact on the collapse and bounce dynamics, as well as
on the emergent neutrino spectra.  We have not yet studied nuclear equations
of state based on finite-temperature mean-field theory, nor have we incorporated
exotic particle species and phase transitions.  
Such equations of
state are of considerable interest, but are not always easily incorporated into
existing dynamical codes. 
In addition, the neutrino microphysics would have to be made consistent
with the presence of exotic constituents (e.g., neutrino-meson and neutrino-quark 
interactions) (Reddy, Prakash, \& Lattimer 1998).

Performing an albeit limited exploration of the parameter space 
available to high-density equations of state,
we have constructed three tabular versions of the 
LSEOS (see Appendix \S\ref{appendix.eos}) with different nuclear compressibilities:
$\kappa=180$, 220, and 375 MeV.  For all of our baseline models we took $\kappa=220$ MeV.
Note that, since Lattimer and Swesty (1991) attempted to construct
a thermodynamically-consistent EOS with self-consistent Maxwell constructions
at the phase transition to nuclear matter, LSEOS models with different 
compressibilities also have different $\hat{\mu}$s, compositions, and energies,
not just pressures.

In order to illustrate the role of the ``nuclear compressibility," 
we have run collapse and bounce simulations 
with all three $\kappa$s, evolving each to 200\,ms after bounce.
Figure \ref{tcomp} shows the 
temperature profile in an 11\modot\, progenitor 
for calculations with $\kappa=180$ (dashed lines) and $\kappa=375$ (solid lines) 
$\sim$2\,ms and $\sim$100\,ms after bounce.  
The model with $\kappa=180$ reaches a higher core temperature, but has a steeper 
negative temperature gradient,
reaching a minimum at M$\simeq0.4$ \modot.  The peak temperature at M$\simeq0.65$ \modot\, is
also higher for the lower compressibility model.  Overall, however, the two models are
remarkably similar. Peak velocities in both calculations reach $\sim$$2\times10^9$ \cms\,
at M$\sim0.7$ \modot. In addition, the entropy and $Y_e$ profiles are qualitatively identical for
all three $\kappa$s.
Although the $\nu_e$ breakout luminosity pulse for the model with $\kappa=375$ is delayed by 1\,ms 
compared with the corresponding pulse for both $\kappa=180$ and $\kappa=220$, the shape and 
magnitude of the pulses are very similar. 

Figure \ref{ltcomp} shows that for both EOS models  
the luminosities at the edge of the grid for all neutrino species
(apart from the 1 ms offset for $\kappa=375$) evolve similarly during the first $\sim30$\,ms 
after bounce.  The average neutrino energies for both EOS models
are also similar during this earlier phase.
After 50\,ms, the differences between
the models with $\kappa=180$ and $\kappa=375$ in 
both the neutrino luminosities and spectra become discernible.
The post-bounce neutrino luminosity for all species is lower in the $\kappa=375$ model.  
At 200\,ms after bounce the fractional difference is approximately 4\%, 5\%, and 9\% for $L_{\nu_e}$,
$L_{\bar{\nu}_e}$, and $L_{\nu_\mu}$, respectively.  The model with $\kappa=375$
also produces characteristically softer spectra with the average energy for each
species being $\sim$3\% lower.  Although $\langle\varepsilon_{\nu_e}\rangle$ and
$\langle\varepsilon_{\bar{\nu}_e}\rangle$ are lower by a few percent
in the model with $\kappa=375$, $\langle\varepsilon_{\nu_\mu}\rangle$ is not evolving
to higher energies as quickly as in the model with $\kappa=180$. For all models,
$\langle\varepsilon_{\nu_\mu}\rangle$ evolves approximately linearly with time, but the model
with $\kappa=180$ (the softer EOS) evolves with a larger slope.
In \S\ref{section:microdetect} we quantify the importance of the
nuclear compressibility modulus in shaping the detected neutrino
signal from supernovae.

These differences in spectra and some of the basic hydrodynamical quantities are not
large enough to affect the post-bounce dynamics significantly on 100\,ms timescales.
In fact, 200\,ms after bounce, there is very little difference between the
shock position, gain radius, or neutrinosphere positions in these two models.
Perhaps larger differences between models with different compressibilities
develop on one- and ten-second timescales in the post-explosion protoneutron
star cooling epoch.  This might be important both for detectability of the
neutrino signature at late times and the evolution of the neutrino-driven protoneutron
star wind (Thompson, Burrows, and Meyer 2001).

\section{Modifications to the Standard Cross Sections and Sources}
\label{section:opacity}

The work of Burrows \& Sawyer (1998,1999), Raffelt \& Seckel (1998),
Janka et al.~(1996), and Reddy, Prakash, \& Lattimer (1998) all point to a reduction
at high density of the dominant neutral- and charged-current opacity sources for neutrinos 
due to nucleon-nucleon correlations caused by the strong interaction and Fermi statistics.
Such modifications to the standard neutrino opacities  
depend on the model for strong interactions employed.
Ideally, they should be included consistently in constructing the nuclear 
equation of state (Reddy, Prakash, \& Lattimer 1998).
Although they did not take into account these changes to the nuclear EOS,
Rampp et al.~(2002) have recently investigated the role of nuclear correlations
in their dynamical models.  They found modest enhancements in $L_{\nu_e}$ and
$L_{\nu_\mu}$ and considerable enhancements in $L_{\bar{\nu}_e}$ ($\sim20$\%).
Although inclusion of these processes did not lead to explosions, they did affect
the dynamics and it seems they should not be absent from a full treatment of the
problem.

In a future work we will incorporate the dynamical structure function formalism of
Burrows \& Sawyer (1998,1999) and Reddy, Prakash, \& Lattimer (1998).  For now, we provide
a set of simple tests, varying some of our cross sections and reaction rates
in order to test the sensitivity of the observables and the dynamics.

\subsection{Artificial Opacity Reduction}
\label{subsub:artificial}

\begin{figure} 
\vspace*{6.0in}
\hbox to\hsize{\hfill\includegraphics{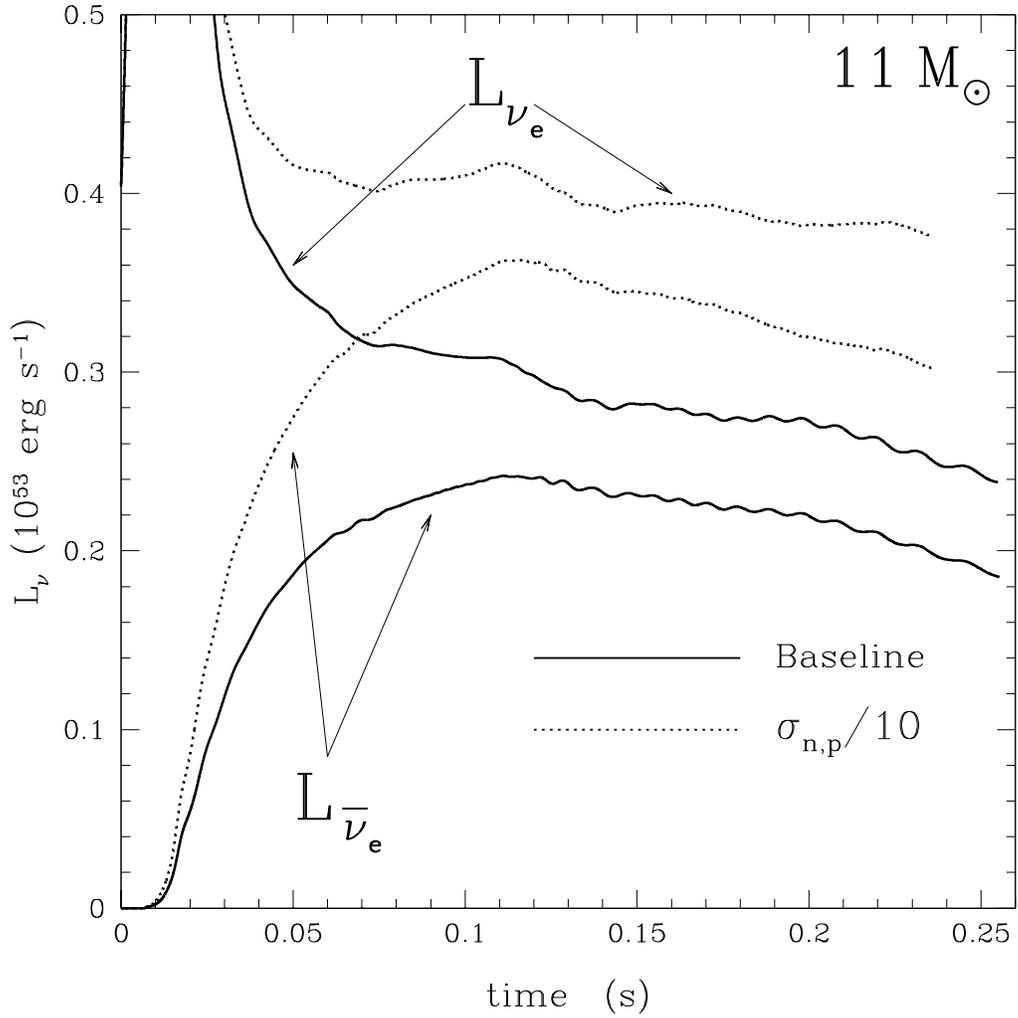}\kern+6in\hfill}
\caption[The effects $\sigma_{n,p}$ on $L_{\nu_e}$ \& $L_{\bar{\nu}_e}$ vs.~$t$]
{$L_{\nu_e}$ and $L_{\bar{\nu}_e}$ in erg s$^{-1}$ 
at infinity as a function of time, for the fiducial model
and for the model with artificially decreased neutral-current
neutrino-neutron and neutrino-proton cross sections.
}
\label{ltsige}
\end{figure}

\begin{figure} 
\vspace*{6.0in}
\hbox to\hsize{\hfill\includegraphics{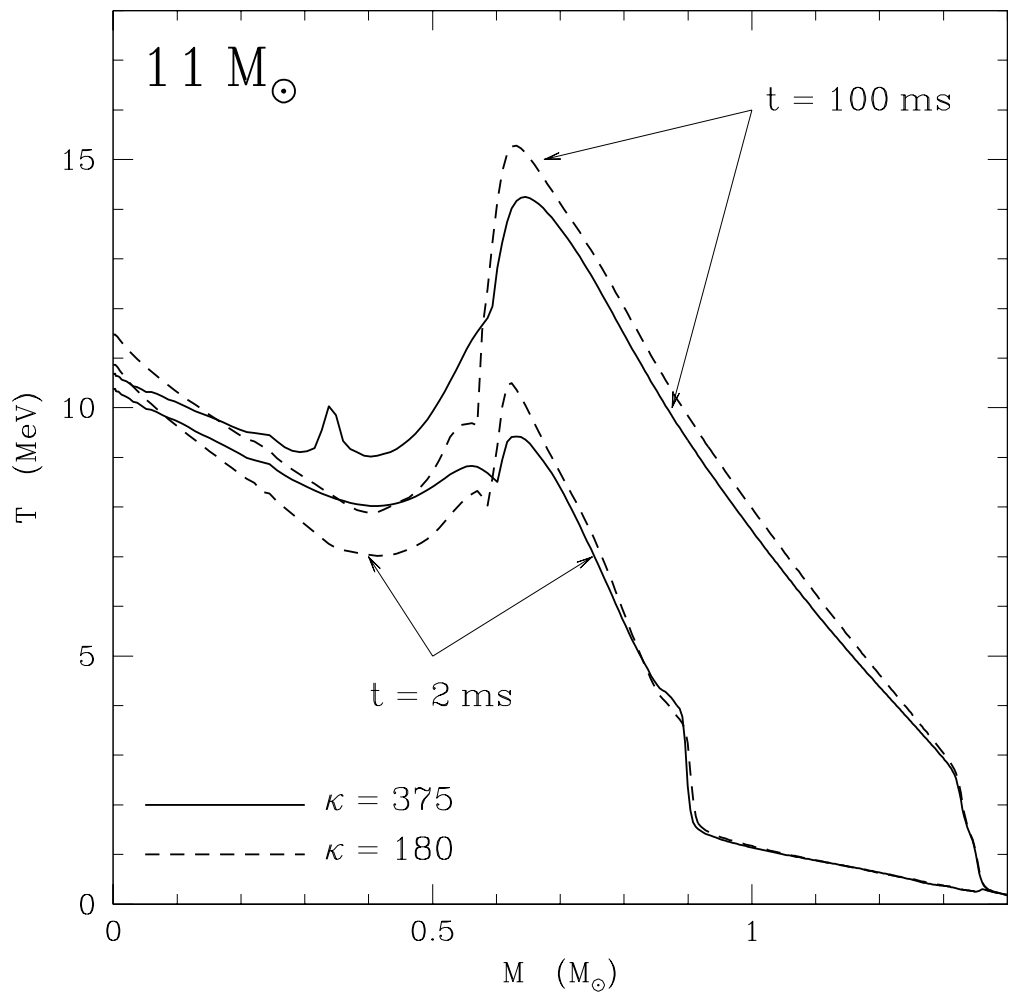}\kern+6in\hfill}
\caption[The effects of $\kappa$ on bounce \& post-bounce $T$ profile]
{Temperature ($T$) in MeV as a function of mass coordinate in units
of M$_\odot$ at two snapshots in time ($\sim2$\,ms and $\sim100$\,ms post-bounce)
for the 11 M$_\odot$ progenitor, using two different nuclear compressibilities,
$\kappa=180$ (dashed lines) and $\kappa=375$ (solid lines).}
\label{tcomp}
\end{figure}

\begin{figure} 
\vspace*{6.0in}
\hbox to\hsize{\hfill\includegraphics{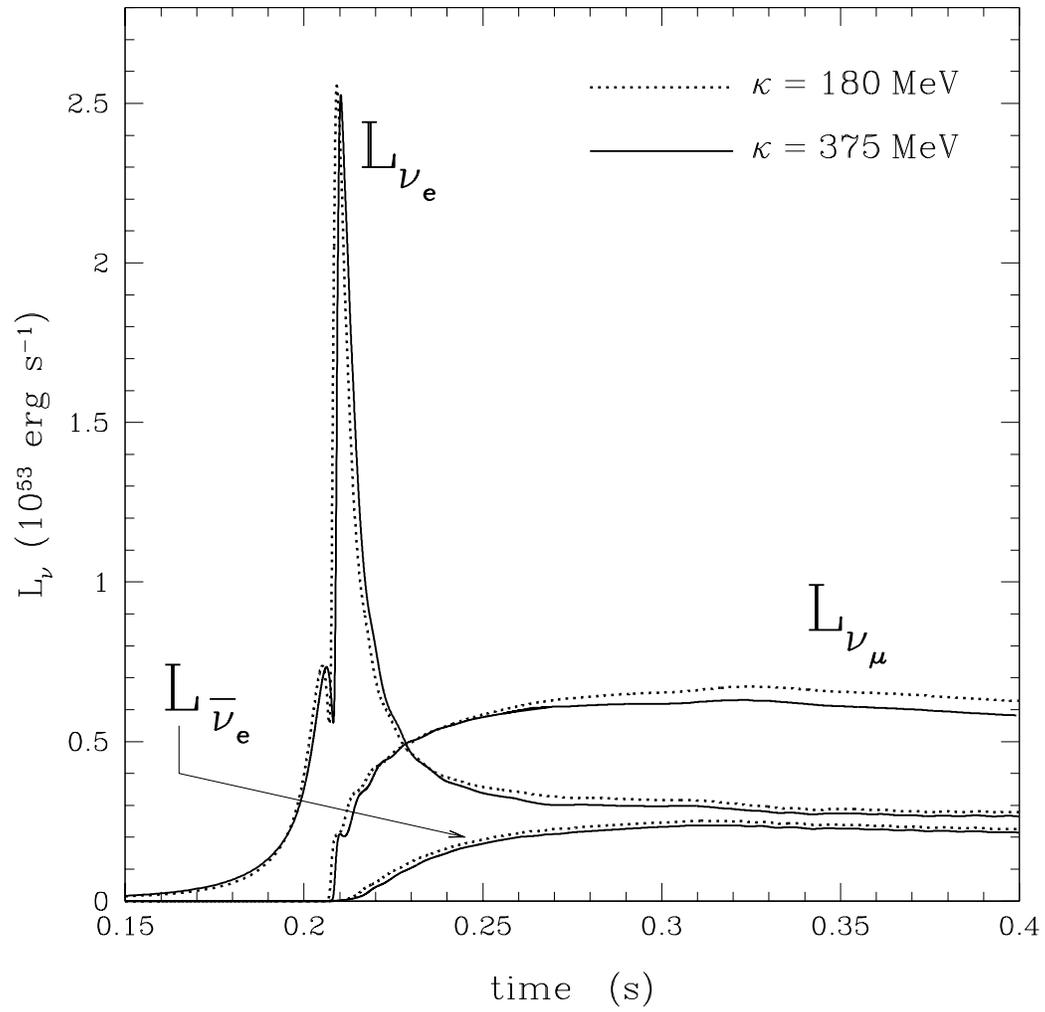}\kern+6in\hfill}
\caption[The effects of $\kappa$ on $L_{\nu_e}$, $L_{\bar{\nu}_e}$, 
and $L_{\nu_\mu}$]
{$L_{\nu_e}$, $L_{\bar{\nu}_e}$, 
and $L_{\nu_\mu}$ in units of $10^{53}$ erg s$^{-1}$ at infinity
for $\kappa=180$ (dotted lines) and $\kappa=375$ (solid lines).}
\label{ltcomp}
\end{figure}

In Liebend\"{o}rfer (2000), erroneous explosions were obtained in one-dimensional
models of supernovae because of an artificially decreased neutral-current neutrino-nucleon
opacity.  As a simple test, we have decreased the neutrino-neutron ($\sigma_n$) and 
neutrino-proton ($\sigma_p$) scattering
cross sections by a factor of 10 everywhere, for all neutrino energies.
Despite such a change, because the dynamics during collapse and bounce
are dominated by neutral-current scattering
on nuclei and the charged-current process $\nu_e n\leftrightarrow e^-p$, the
breakout phenomenon is only modestly affected;  
the peak $\nu_e$ breakout pulse is just 10\% higher than in our baseline model
and reaches $2.85\times10^{53}$ erg s$^{-1}$.  Two hundred and fifty milliseconds after breakout,
$L_{\nu_e}\simeq0.25\times10^{53}$ erg s$^{-1}$ in our baseline model.  In our
model with $\sigma_{n,p}/10$, we obtain $L_{\nu_e}\simeq0.35\times10^{53}$ erg s$^{-1}$.
Figure \ref{ltsige} shows $L_{\nu_e}$ and $L_{\bar{\nu}_e}$ at infinity
as a function of time for the baseline model and the model presented here.
Both $L_{\nu_e}$ and $L_{\bar{\nu}_e}$ are increased by $\sim$40$-$50\%  
in the calculation with lower opacity.
These changes, however, are modest compared with those for $L_{\nu_\mu}$
and $\langle\varepsilon_{\nu_\mu}\rangle$. Two hundred and fifty milliseconds after bounce,
$L_{\nu_\mu}$ equals $2.4\times10^{53}$ erg s$^{-1}$, a factor of four larger 
than in the baseline model.  For $\nu_\mu$ neutrinos, there is no
charged-current absorption process on free nucleons to contribute to the
total opacity, as there is for the $\nu_e$ and $\bar{\nu}_e$ neutrinos.
For the $\nu_\mu$s, the neutral-current scattering dominates the opacity and so 
changing that opacity source so dramatically has significant consequences for
$L_{\nu_\mu}$ and $\langle\varepsilon_{\nu_\mu}\rangle$.  

Importantly, because of the increase in the $L_{\nu_e}$ and $L_{\bar{\nu}_e}$ 
we obtain a larger gain region in this calculation, higher entropy behind the shock,
and a larger shock radius 250\,ms after bounce.  In the baseline model,
at 250\,ms post-bounce the shock is sitting at 180\,km, whereas in the model with reduced opacities,
$R_s\simeq210$\,km.  Still, although the dynamics are affected by this drastic
decrease in neutrino opacity, we do not obtain an explosion in this first 250\,ms
of post-bounce evolution.

\subsection{The Weak Magnetism/Recoil Correction}
\label{weak:recoil}

Our baseline 11\,\modot\, model includes the weak magnetism/recoil correction 
to the charged-current opacities for $\nu_e$ and $\bar{\nu}_e$
and to the neutral-current scattering
opacities off of nucleons for all neutrino species
(Vogel 1984; Horowitz 1997,2002).  
The weak magnetism/recoil correction lowers the cross section
for all processes; the decrease is largest for anti-neutrinos ($\bar{\nu}_e$ and $\bar{\nu}_\mu$).
We ran our baseline model (\S\ref{section:baseline})
without these corrections and 
found only small differences in the first 250 milliseconds of post-bounce evolution.
$L_{{\nu}_e}$, $L_{{\nu}_\mu}$, $\langle\varepsilon_{\nu_e}\rangle$,and 
$\langle\varepsilon_{\nu_\mu}\rangle$ are all lower at the 0.5\% level without
the weak-magnetism correction. $\langle\varepsilon_{\bar{\nu}_e}\rangle$
and $L_{\bar{\nu}_e}$ are less by 4\% in this representative 11 \modot\, calculation.
This slight softening and dimming of the emergent spectra can be understood
simply as the result of increasing the total opacity for all species
by removing the small weak magnetism/recoil correction.  Hence, though
the corrections themselves for given temperatures, densities, and
neutrino energies can be greater than 10--15\%, due to feedbacks on the dynamics
and thermodynamics, the self-consistent inclusion of weak-magnetism/recoil
effects in a core-collapse calculation has only very limited consequences.

\subsection{The Role of Nucleon-Nucleon Bremsstrahlung}
\label{section:dyn_brem}

\begin{figure} 
\vspace*{6.0in}
\hbox to\hsize{\hfill\includegraphics{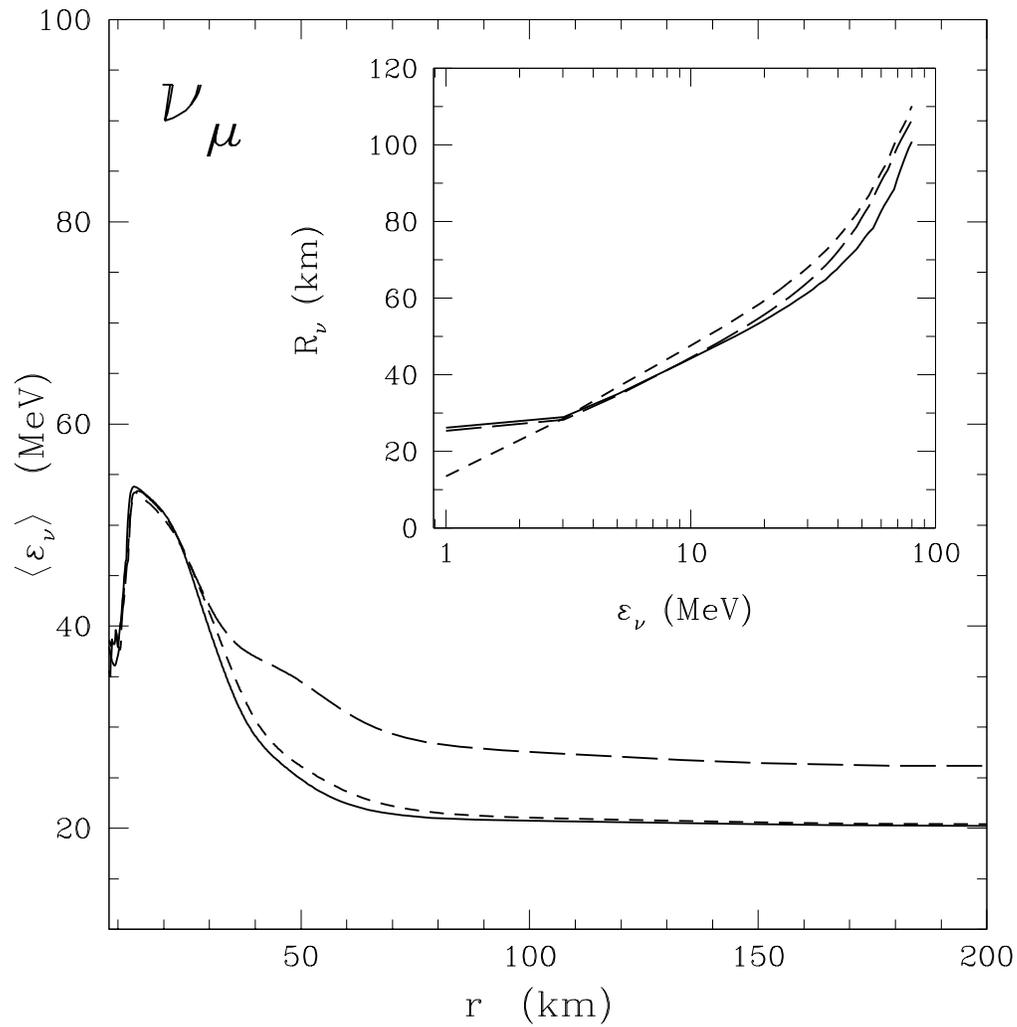}\kern+6in\hfill}
\caption[$\langle\varepsilon_{\nu_\mu}\rangle(r)$
with and without inelastic $\nu_\mu e^-$ scattering]
{$\langle\varepsilon_{\nu_\mu}\rangle$ as a function of radius 
for 11\,M$_\odot$ models with bremsstrahlung and inelastic neutrino-electron
scattering (solid line), without bremsstrahlung and with inelastic neutrino-electron
scattering (short dashed line), and with bremsstrahlung and without 
inelastic neutrino-electron scattering (long dashed line),
approximately 220\,ms after bounce.
The inset shows the neutrinosphere $R_{\nu_\mu}$, defined by
eq.~(\ref{nusphere}) as a function of energy in the same models.
}
\label{aem}
\end{figure}

The effects of bremsstrahlung on the $\nu_e$ and $\bar{\nu}_e$ 
spectra are negligible at virtually all times because for them
the charged-current processes $\nu_e n\leftrightarrow p e^-$ and 
$\bar{\nu}_e p\leftrightarrow n e^+$ dominate.  
Thompson, Burrows, and Horvath (2000) 
and Raffelt (2001) showed that nucleon-nucleon bremsstrahlung might
have important consequences for the emergent $\nu_\mu$ spectra and that it dominates
$e^+e^-\rightarrow\nu_\mu\bar{\nu}_\mu$ as a production process.
Figure \ref{aem}
shows the effect of bremsstrahlung on the average energy of $\nu_\mu$s.
The plot itself shows the average energy (in MeV) as a function of radius 
approximately 230 ms after bounce.  The inset shows
$R_{\nu_\mu}(\varepsilon_{\nu_\mu})$, the neutrinosphere radius, defined
by eq.~(\ref{nusphere}), as a function of energy.
The solid line is the
baseline model, with neutrino-electron scattering redistribution and nucleon-nucleon
bremsstrahlung.  The short dashed line is the same model run through
the full evolution, but without nucleon-nucleon bremsstrahlung.  The long dashed
line shows the same model without neutrino-electron scattering energy redistribution,
but with nucleon-nucleon bremsstrahlung.  One can see immediately from
the inset that at low energies, bremsstrahlung increases the opacity
significantly.  It is also clear from the main figure that inelastic 
neutrino-electron scattering profoundly effects the emergent $\nu_\mu$ 
spectrum.

\begin{figure} 
\vspace*{6.0in}
\hbox to\hsize{\hfill\includegraphics{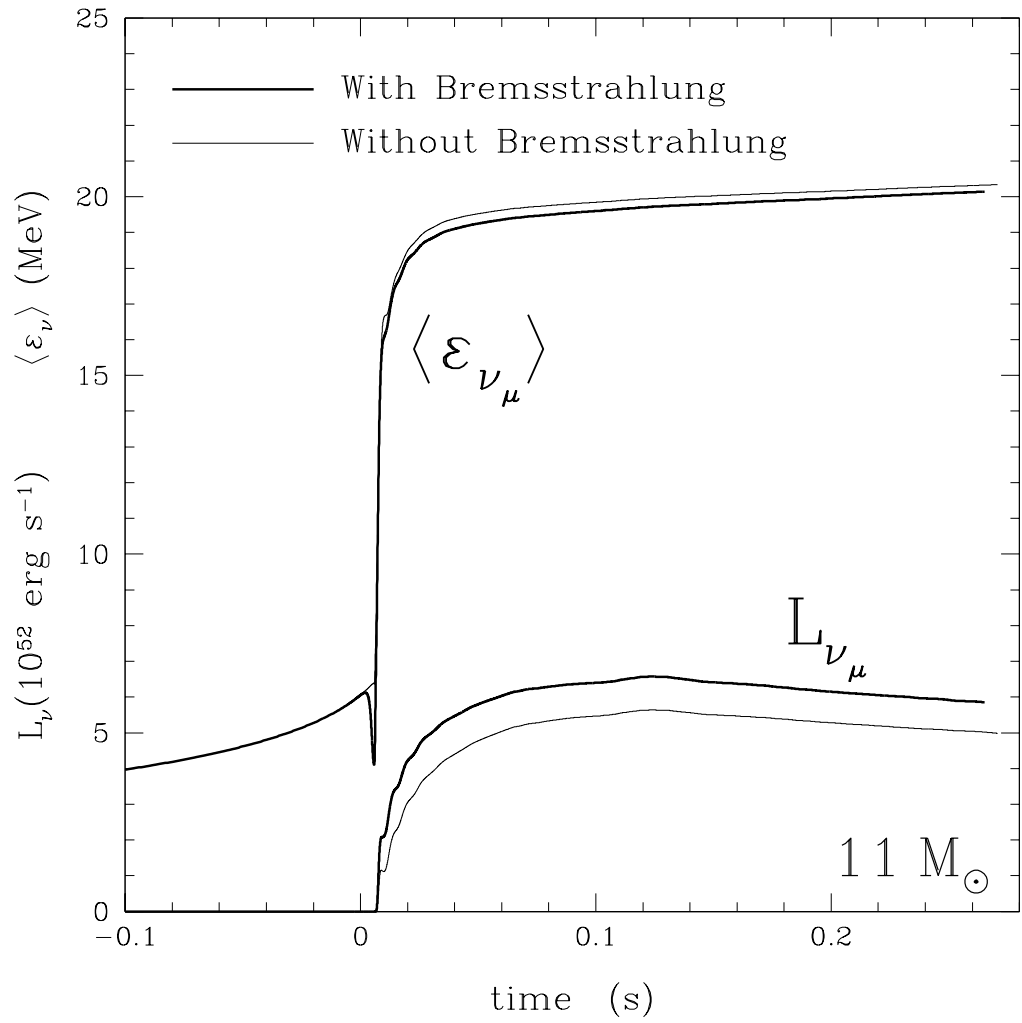}\kern+6in\hfill}
\caption[$\langle\varepsilon_{\nu_\mu}\rangle$ and $L_{\nu_\mu}$
with and without bremsstrahlung]
{$\langle\varepsilon_{\nu_\mu}\rangle$ and $L_{\nu_\mu}$
of the emergent spectrum at infinity as a function of time, for 11\,M$_\odot$ models
with (thick solid line) and without (thin solid line) nucleon-nucleon bremsstrahlung.
}
\label{etm}
\end{figure}

Figure \ref{etm} shows $L_{\nu_\mu}$ and $\langle\varepsilon_{\nu_\mu}\rangle$
at infinity as a function of time for the baseline model (thick solid line), 
which includes nucleon-nucleon bremsstrahlung, and the same model computed 
without bremsstrahlung (thin solid line).  With bremsstrahlung, the average energies
of the $\nu_{\mu}$ neutrinos 
are lower and the luminosity is significantly higher.  
The results of Thompson, Burrows, and Horvath (2000) 
anticipated these results based on comparisons between
the bremsstrahlung and $e^+e^-$-annihilation neutrino production rates.  
Near the $\nu_\mu$ neutrinosphere, the bremsstrahlung
emission spectrum is peaked at low neutrino energies ($<10$\,MeV) compared with
$e^+e^-$ annihilation.  For this reason, Thompson, Burrows, and Horvath (2000),
Burrows et al.~(2000), and Raffelt (2001) concluded that the emergent $\nu_\mu$
spectra have both higher luminosity and lower average energy.  
As evidenced by Fig.~\ref{etm}, this 
prediction is borne out in our dynamical models.

\subsection{The Effects of Inelastic Neutrino-Electron Scattering}
\label{section:inelasticnue}

\begin{figure} 
\vspace*{6.0in}
\hbox to\hsize{\hfill\includegraphics{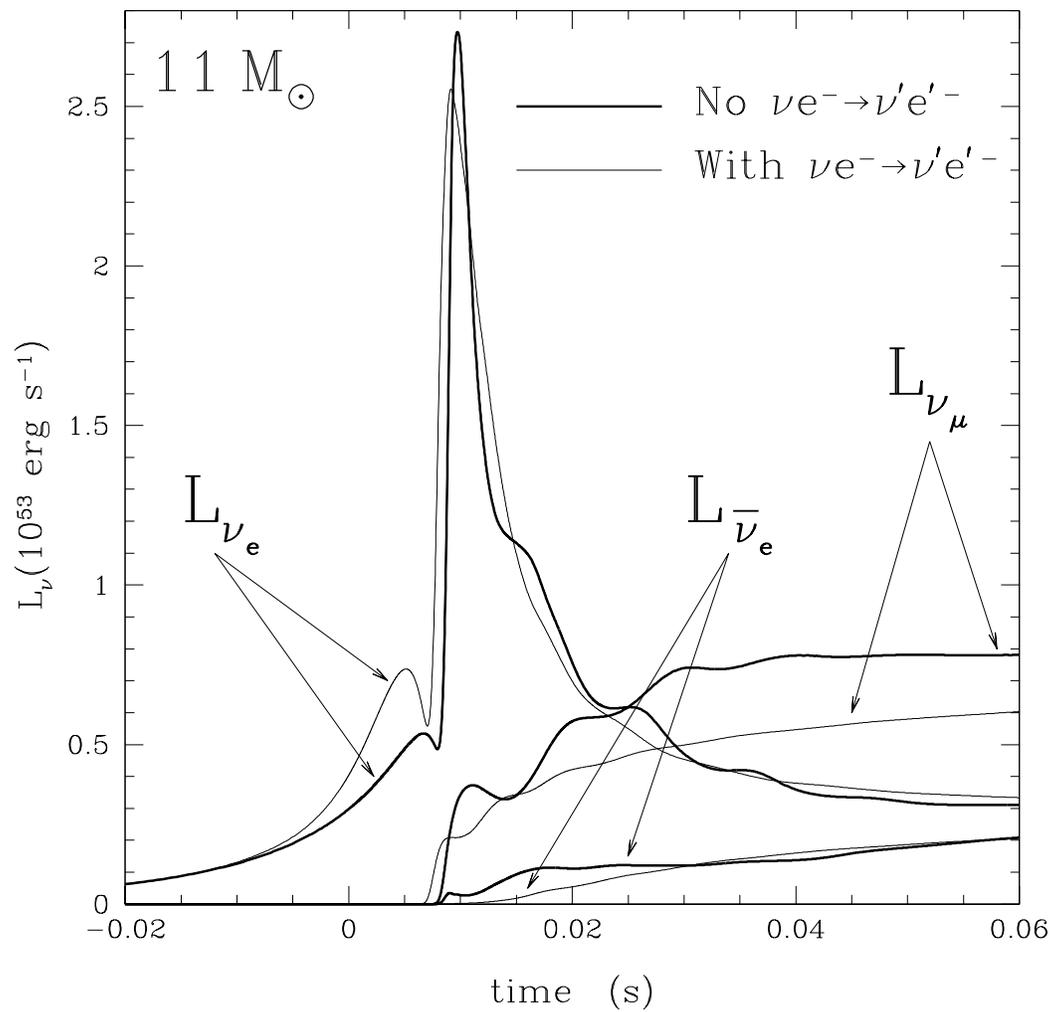}\kern+6in\hfill}
\caption[$L_{\nu_i}$ with and without inelastic $\nu_i e^-$ scattering]
{$L_{\nu_e}$, $L_{\bar{\nu}_e}$, and  $L_{\nu_\mu}$
at infinity for the 11\,M$_\odot$ progenitor with (thick solid lines) and without
(thin solid lines)
inelastic neutrino-electron scattering as described in Appendix \S\ref{app:escatt}.
}
\label{ltnr}
\end{figure}

Figure \ref{ltnr} shows the luminosity at infinity as a function of
time for all species in the 11\,M$_\odot$ models with (thin solid line)
and without (thick solid line) inelastic neutrino-electron scattering,
using the prescription described in detail in Appendix \ref{app:escatt}.
Figure \ref{etnr} shows the $rms$ neutrino energy for each species
for the model without inelastic $\nu_ie^-$ scattering and should 
be compared with Fig.~\ref{ets11}.  
The main $\nu_e$ breakout pulse
is enhanced slightly and reaches $2.73\times10^{53}$ erg s$^{-1}$,
compared with $2.55\times10^{53}$ erg s$^{-1}$ in the baseline model.
Also noticeable in $L_{\nu_e}$ is the decrease in the pre-breakout
pulse associated with the deleptonization of the core on infall as described 
in \S\ref{section:baseline}.
Shown here at $t\sim0.007$\,s, the pre-breakout peak reaches $0.75\times10^{53}$ erg s$^{-1}$
with neutrino-electron scattering and only $0.52\times10^{53}$ erg s$^{-1}$
without.  By downscattering neutrinos during collapse and bounce via $\nu_e e^-$ scattering, 
more neutrinos are able to escape before the core becomes opaque. 
Thus, there is a trade-off.  Including inelastic scattering
off electrons makes the {\it collapse} luminosity peak higher, but the main 
{\it breakout} luminosity peak lower.  
Although the breakout
phenomenon is slightly affected, the spectra of $\nu_e$ and $\bar{\nu}_e$ neutrinos
at 200\,ms are dominated by the charged-current interactions and so the
effects of $\nu e^-$ scattering on $L_{\nu_e}$, $L_{\bar{\nu}_e}$, 
$\langle\varepsilon_{\nu_e}\rangle$, and $\langle\varepsilon_{\bar{\nu}_e}\rangle$ are modest.  
$\nu_\mu$ neutrinos
are more significantly affected.  Throughout the evolution, $L_{\nu_\mu}$ is decreased
by $\sim$30\% by including $\nu_\mu e^-$ redistribution.  $\langle\varepsilon_{\nu_\mu}\rangle$
is also profoundly altered.  Comparing Fig.~\ref{etnr} with Fig.~\ref{ets11},
we see that $\langle\varepsilon_{\nu_\mu}\rangle$ is $\sim6$ MeV higher without
$\nu_\mu e^-$ redistribution. 

Figure \ref{yes11} shows that there is significantly 
more deleptonization of the core at bounce with inelastic neutrino-electron scattering 
(thick solid line) than there is without it (long dashed line).  
The magnitude of the difference in core $Y_e$ at bounce is in keeping
with other recent models (cf. Rampp 2000, \S\ref{section:baseline}).
The fact that
$Y^{\rm core}_e$ is higher at bounce in the latter case has important consequences
for the shock energy and the early post-bounce neutrino characteristics.  In
the model without neutrino-electron scattering, the peak positive velocity after 
bounce is higher ($\sim$2.5$\times10^9$ cm s$^{-1}$, compared with 
$\sim$2.0$\times10^9$ cm s$^{-1}$)
and peaks at larger interior mass (0.85\,M$_\odot$, compared with 0.7\,M$_\odot$).
In both models, as the shock stalls and the briefly positive velocities achieved
become negative, the matter behind the shock (the protoneutron star) 
oscillates on $5-10$ millisecond timescales.
As the protoneutron star pulsates, pressure waves move through the neutrinosphere
of each species.  These waves act to modulate the emergent luminosity and average neutrino energy.
The amplitude of these near-periodic variations in neutrino luminosity and energy 
are damped on $\sim$10\,ms timescales.
They are smaller 
for the model with neutrino-electron scattering because the shock in this model has less energy 
and the corresponding post-bounce oscillatory mass motions (`ringing') are thereby smaller
at the neutrinosphere for each species.  The smaller the local changes in temperature and density
near the neutrinosphere, the smaller the changes in the local
emission and absorption.  Hence, we should expect smaller amplitude temporal modulations in the 
emergent neutrino spectral characteristics. 

Also potentially important is the fact that $L_{\bar{\nu}_e}$ rises
later in the model with neutrino-electron energy redistribution than in the model without
such redistribution.  Twenty milliseconds after bounce, 
Fig.~\ref{ltnr} shows that $L_{\bar{\nu}_e}$
with $\nu_ie^-$ scattering is approximately half that for the model without
this thermalization process.  This has important implications for the detection of
the $\nu_e$ breakout pulse in terrestrial light-water \v{C}erenkov neutrino detectors
like SuperKamiokande.  In \S\ref{section:sk}, we show that in such detectors,
if one is to identify the $\nu_e$ breakout signature uniquely, one must do
so within  $5-20$ milliseconds of bounce, but before the $\bar{\nu}_e$ signal 
becomes appreciable.  Hence, because neutrino-electron scattering suppresses
the  $\bar{\nu}_e$ signal during this epoch, it has consequences for the 
observation of the next galactic supernova.

\section{Neutrino Detector Signatures}
\label{section:detection}

A total of just $19$ neutrino events were observed 
from supernova SN1987A in the IMB (Irvine-Michigan-Brookhaven) 
and Kamioka II detectors.  The signal lasted $\sim$10 seconds and 
confirmed the basics of our theoretical understanding of core collapse; the amount of energy 
radiated in neutrinos was comparable to that expected in protoneutron star cooling models at 
the time (Burrows \& Lattimer 1986) and the inferred $\bar{\nu}_e$ neutrino spectra had 
average energies of $\sim$10$-$15\,MeV (compare with Fig.~\ref{ets11}).

Far surpassing the sensitivity of Kamioka II and IMB, the current generation of neutrino 
detectors might collectively see thousands of neutrino events from the next galactic 
supernova (Burrows, Klein, \& Gandhi 1992).  Such an event would provide exquisite 
neutrino spectral information and would put strong constraints on the physics of 
the core during collapse and explosion.  Besides elucidating supernova physics, 
much information on the mass hierarchy of neutrinos could also be collected.  
For example, the neutrino signature might clearly reveal whether a black hole or a 
neutron star had been created (Burrows 1988) and if the former is born, direct eV-scale
measurements of $\nu_{\mu}$- and $\nu_{\tau}$-neutrino masses could be made 
by the current generation of detectors (Beacom, Boyd, \& Vogel 2000).

Using the models presented in the preceding sections, we have made predictions
about the dynamics and emergent neutrino spectra expected from a class of supernova progenitors,
with reasonable variations in microphysics and the models.
In this section, we go a step further and fold our neutrino spectra
with the thresholds and sensitivities of a subset of modern neutrino detectors.
The focus is on the detectability of the breakout phenomenon itself.

In underground light-water detectors, such as a fully-repaired Super Kamiokande (SK), we can
expect $\sim$5000 neutrino events from the next galactic supernova at a distance of 10\,kpc.
In such detectors the primary mechanism for neutrino detection
is the $\bar{\nu}_e p\rightarrow n e^+$ process.  The positron secondary
emits \v{C}erenkov radiation. For an electron or positron secondary, 
the differential number detected at distance $D$ can be written (Burrows, Klein, \& Gandhi 1992):
\beqa
dN&=&\frac{N_T}{4\pi D^2}L^{\rm num}_\nu(t^\prime,\varepsilon_\nu)
\frac{d\sigma(\varepsilon_\nu,\varepsilon_e)}{d\varepsilon_e}\nonumber \\
&\times&E(\varepsilon_e)
\delta(t-t^\prime-\Delta t)\,dt^\prime\,d\varepsilon_e\,d\varepsilon_\nu\,dt,
\label{gendetect}
\eeqa
where $\varepsilon_e$ is the electron or positron energy, $E(\varepsilon_e)$ is
the detection efficiency, $N_T$ is the number of targets, $L^{\rm num}_\nu$
is the number luminosity of neutrinos, 
$t^\prime$ is the source time, $t$ is the detector time with $D/c$ subtracted,
$\Delta t(\simeq D/2c\,(m_\nu c^2/\varepsilon_\nu)^2)$ is the time delay of a 
neutrino with mass $m_\nu$,
and $d\sigma/d\varepsilon_e$ is the differential cross section for neutrino capture
and electron or positron production.  For most purposes 
$d\sigma/d\varepsilon_e\simeq
\sigma(\varepsilon_\nu)\delta(\varepsilon_\nu-\varepsilon_e-\Delta)$, where
$\Delta$ is the reaction threshold (in the case of the reaction 
$\bar{\nu}_e p\rightarrow n e^+$, $\Delta=m_n-m_p+m_e\simeq1.804$ MeV).
Taking this approximation for $d\sigma/d\varepsilon_e$, $E(\varepsilon_e)=1$,
and $m_\nu=0$, we obtain 
\begin{equation}
\frac{dN}{dt\,d\varepsilon_\nu}=\frac{N_T}{4\pi D^2}
L^{\rm num}_\nu(t,\varepsilon_\nu)\sigma(\varepsilon_\nu).
\label{detect}
\end{equation}
Using eqs.~(\ref{gendetect}) and (\ref{detect}), it is a relatively 
simple matter to fold our time-dependent spectra with 
terrestrial detector characteristics, including efficiency and threshold corrections,
in order to obtain the observed neutrino signal.  We consider
three different detectors, focusing on the $\nu_e$-rich 
breakout phase and the early neutrino signal that might 
be observed from the next galactic supernova.
Although eq.~(\ref{gendetect}) includes the effects of time-delay 
due to a finite neutrino mass, in what follows we assume massless 
neutrinos and employ eq.~(\ref{detect}).

\subsection{SuperKamiokande}
\label{section:sk}

The SuperKamiokande (SK) neutrino observatory is a light-water
\v{C}erenkov detector whose baseline volume for supernova neutrino
detection is $\sim$32 ktonne (Totani et al.~1998).  We take as a threshold $\sim$5 MeV 
(Beacom \& Vogel 1998).  The background is typically of order 
several $\times$0.1 s$^{-1}$ if the full detector volume is
considered (Beacom \& Vogel 1998) and will not figure into our discussion here.

Light-water \v{C}erenkov detectors offer several different channels for detection of the
various neutrino species.  The dominant reaction is charged-current
absorption of $\bar{\nu}_e$ neutrinos on free protons 
($\bar{\nu}_e p\rightarrow n e^+$).  The positron secondary emits \v{C}erenkov 
radiation, which is detected directly.  
Neutrino scattering on electrons, $\nu_i e^-\rightarrow\nu_i e^-$ also 
contributes for all neutrino species.  For this process, we use the
neutrino-electron scattering cross section (Sehgal 1974):
\beqa
\frac{d\sigma_i}{d\varepsilon_e}&=&\frac{1}{2}\frac{\sigma_o}{m_e c^2}
\left[A_i+\left(1-\frac{\varepsilon_e}{\varepsilon_\nu}\right)^2B_i\right] \nonumber \\
\Longrightarrow\;
\sigma_i&=&
\frac{1}{2}\sigma_o\Lambda_i
\left(\frac{\varepsilon_\nu}{m_e c^2}+\frac{1}{2}\right),
\label{neutrinoelectronscattercs}
\eeqa
where 
$$\Lambda_i=\frac{1}{4}\left(A_i+\frac{1}{3}B_i\right),\hspace{.5cm}
A_i=(C_V+C_A)^2, \hspace{.5cm} B_i=(C_V-C_A)^2,$$
and $C_V=\pm 1/2+2\sin^2 \theta_W$ for electron types and muon types, 
respectively. $C_A=+1/2$ for $\nu_e$ and $\aunu$, and $C_A=-1/2$ for $\aenu$ and $\unu$.
Note that the neutrino-electron scattering cross section is $6-7$ times
less for the $\mu$- and $\tau$-type neutrinos, owing to the fact that for them
the reaction can proceed via only the neutral current.
Other important reactions
include
\begin{equation}
\enu+^{16}{\rm O} \rightarrow ^{16}{\rm F}+e^- \;\;(\epsilon_{\rm th}=15.4 \,{\rm MeV}),
\label{enuo}
\end{equation}
\begin{equation}
\enu+^{18}{\rm O} \rightarrow ^{18}{\rm F}+e^- \;\;(\epsilon_{\rm th}=1.66 \,{\rm MeV}),
\label{enuo2}
\end{equation}
\begin{equation}
\aenu+^{16}{\rm O} \rightarrow ^{16}{\rm N}+e^+ \;\;(\epsilon_{\rm th}=11.4 \,{\rm MeV}),
\label{aenuo}
\end{equation}
and
\begin{equation}
\nu_i+^{16}{\rm O}\rightarrow \nu_i^\prime+^{16}{\rm O}^*\rightarrow\nu_i^\prime+\gamma+X,
\label{oexcite}
\end{equation}
where $\epsilon_{\rm th}$ is the energy threshold for the reaction.
For the neutral-current excitation of $^{16}$O (reaction \ref{oexcite}), the $\gamma$ secondary has
an energy in the range $5-10$ MeV and, hence, is detectable by SK.
Note that energy resolution for the secondary electron or positron in reactions
(\ref{enuo}), (\ref{enuo2}), and (\ref{aenuo})
is good to no better than $\sim$10$-$20\% (Burrows, Klein, \& Gandhi 1992; Totani et al.~1998).
For the study carried out here, we calculate event rates in SK
due to only $\bar{\nu}_e p\rightarrow n e^+$ and neutrino-electron scattering
for all neutrino species. We assume 100\% efficiency above the detector threshold.

Figure \ref{skn} shows the integrated number of $\nu_e$ and $\bar{\nu}_e$ 
neutrino events ($\log_{10}[N(<t)]$) detected in SK for our baseline 11\,M$_\odot$ progenitor
(thick lines) and for the 20\,M$_\odot$ progenitor (thin lines) 
described in \S\ref{section:progenitor}.  The $\nu_e$ neutrino signal 
from $\nu_e$-electron scattering (labeled, `$\nu_ee^-\rightarrow\nu_ee^-$') 
is the solid line for both progenitors.
The $\bar{\nu}_e$ neutrino signal from charged-current absorption on free protons 
(labeled, `$\bar{\nu}_ep\rightarrow n e^+$') is the
dashed line for both progenitors.  
Note that many hundreds more $\bar{\nu}_e$ events accumulate from the 20\,M$_\odot$
progenitor ($\sim1250$) than from the 11\,M$_\odot$ ($\sim650$) 
progenitor in the 250\,ms of post-bounce evolution shown here. 
This important difference results from the steeper density gradient outside
the iron core in the 11\,M$_\odot$ progenitor and the correspondingly
lower accretion luminosity (see Fig.~\ref{massfpro}). 
The inset in Fig.~\ref{skn} shows the first 55\,ms
of the signal and the initial rise of the number $\nu_e$ neutrinos
detected due to breakout.  The inset shows that approximately 5 $\nu_e$ events
can be expected from the collapse phase of the supernova, while $10-15$ $\nu_e$ events
might be observed solely from the breakout neutronization burst. This result seems to be independent
of the mass of the progenitor.  Although we do not include the results for
the 15\,M$_\odot$ progenitor (see \S\ref{section:progenitor}) in
Fig.~\ref{skn}, as indicated by Fig.~\ref{pro}, its $\nu_e$ breakout signature
is virtually identical to that for the 11\,M$_\odot$ 
model.  However, for the 15\,M$_\odot$ progenitor the luminosity 
of $\bar{\nu}_e$ neutrinos is slightly higher after 100\,ms.  As a result,  
250\,ms after bounce the integrated number of $\bar{\nu}_e$ events is $\sim$5\% higher than for 
the 11\,M$_\odot$ model. 

Note that the combined neutral-current $\nu_{\mu,\tau}$ and $\bar{\nu}_{\mu,\tau}$ signal 
due to scattering off of electrons is much smaller than that due to the process
$\nu_ee^-\rightarrow\nu_ee^-$.  This signal never contributes significantly to the total
neutrino event rate.  In particular, less than 1 event due to 
$\mu$- and $\tau$-neutrino scattering on electrons should be observed in the first 50\,ms.

Importantly for models of the neutrino oscillation phenomenon, the neutrino
signal through collapse ($\sim$5 $\nu_e$ events), breakout ($10-15$ $\nu_e$ events),
and just post-breakout ($\sim$5 $\nu_e$ events) should be composed of only $\nu_e$ 
neutrinos.  The break in the total number of events observed, which must occur
as the $\bar{\nu}_e$ neutrinos begin to dominate the signal just 20-30\,ms after bounce,
signals the end of this pure $\nu_e$ neutrino phase.  In addition, because
the reaction $\nu_ee^-\rightarrow\nu_ee^-$ is very forward-peaked and the process
$\bar{\nu}_ep\rightarrow n e^+$ is slightly backward-peaked for 
$\varepsilon_{\bar{\nu}_e}\lesssim15$\,MeV (Vogel \& Beacom 1999),
one might hope to distinguish the early $\bar{\nu}_e$ signal from
the post-breakout $\nu_e$ neutrino signal based on the directionality
of the \v{C}erenkov cones left by the positron and electron secondaries.
However, if there is large mixing between $\nu_e$ and $\nu_\mu$
or $\nu_\tau$ neutrinos, the early $\nu_e$ signal might be lessened significantly
in SK (e.g.~Takahashi et al.~2001).

\subsection{The Sudbury Neutrino Observatory}
\label{sno}

The Sudbury Neutrino Observatory (SNO) in Sudbury, Ontario,
is a light- and heavy-water neutrino detector.  It is a
spherical acrylic vessel 12 meters in diameter, containing 1 ktonne
of D$_2$O, that is surrounded by a 
cavity filled with light water.  The fiducial light-water mass is 
approximately 1.6 ktonne.  SNO is most sensitive to $\nu_\mu$ and
$\bar{\nu}_\mu$ neutrinos via breakup reactions on deuterons:
\begin{equation}
\nu_i + d \rightarrow n+p+\nu_i \;\;(\epsilon_{\rm th}=2.22 \,{\rm MeV})
\label{xnc1}
\end{equation}
\begin{equation}
\bar{\nu}_i + d \rightarrow n+p+\bar{\nu}_i \;\;(\epsilon_{\rm th}=2.22 \,{\rm MeV}).
\label{xnc2}
\end{equation}
The liberated neutron secondary is thermalized in the heavy water and is
then detected via capture on other nuclei within the detector volume.
Currently, the neutrons from neutral-current events are detected via
deuteron capture with an efficiency of $\sim$30\% (Ahmad et al.~2002; SNO Collaboration).
Several schemes are currently proposed and in development for improved 
neutron-capture efficiency.
The first is to lace the D$_2$O with two tonnes of NaCl.  The thermal
neutron absorption cross section on $^{35}$Cl is large ($\sim$83\% efficiency)
and would result in a $\gamma$ cascade peaked at 8\,MeV.  The $\gamma$s 
would be detected by the $\sim$9600 inward-looking photomultiplier tubes 
outside of the acrylic vessel.  The second proposed neutron detection method
involves hanging $^3$He proportional counters in the acrylic vessel and these would detect 
the neutrons directly (SNO collaboration webpage\hspace{-0.1cm}
\footnote{
http://www.sno.phy.queensu.ca/sno/sno2.html\#nc
}). 
Because the neutrons would have a $\sim$5\,ms
mean capture time if the $^{35}$Cl capture mechanism is employed
(which would spread the $\nu_\mu$ turn-on),
the neutral-current signal might be partially smeared with
the $^{35}$Cl approach.
The neutral-current deuteron breakup reactions (reactions \ref{xnc1} and \ref{xnc2})
also contribute to the $\nu_e$ and $\bar{\nu}_e$ signal in SNO, but for them the
charged-current deuteron capture processes,
\begin{equation}
\enu + d \rightarrow p+p+e^- \;\;(\epsilon_{\rm th}=1.44 \,{\rm MeV})
\end{equation}
\begin{equation}
\aenu + d \rightarrow n+n+e^+ \;\;(\epsilon_{\rm th}=4.03 \,{\rm MeV}),
\label{aenudeut}
\end{equation}
are more important above $\varepsilon_\nu\simeq10$\,MeV (Burrows, Klein, \& Gandhi 1992).  
Process (\ref{aenudeut})  can be used to observe
$\bar{\nu}_e$ neutrinos uniquely via the simultaneous detection of 
the secondary neutron and the
\v{C}erenkov emission from the final-state positron.
Although we include
the threshold energy in our calculations of the total signal, 
as with SK we assume a 100\% efficiency for SNO above the detector theshold.

In Fig.~\ref{skd} we show the integrated number of 
$\nu_e$ neutrinos detected in SNO via the processes
$\nu_ed\rightarrow ppe^-$ (dotted line), 
$\nu_ed\rightarrow np\nu_e$ (solid line), and 
$\nu_ee^-\rightarrow\nu_ee^-$ (dashed line) in the first 250\,ms after bounce
in our 11\,M$_\odot$ baseline model.  
The total $\nu_e$ signal is the thick solid line.  
For comparison, we provide the contribution to the total
number of detected events from all other neutrino species 
(long dashed line, labeled `Total(other)').
This signal includes the 
$\bar{\nu}_ed\rightarrow np\bar{\nu}_e$,
$\bar{\nu}_ed\rightarrow nne^+$, and
$\bar{\nu}_ep\rightarrow n e^+$ processes in the light-water portion of the detector,
as well as the processes $\bar{\nu}_{\mu,\tau}d\rightarrow np\bar{\nu}_{\mu,\tau}$,
${\nu}_{\mu,\tau}d\rightarrow np{\nu}_{\mu,\tau}$,
and neutrino-electron
scattering throughout the entire detector volume.
Note that it might be possible to exclude the charged-current
absorption process $\bar{\nu}_ed\rightarrow nne^+$ by cutting on events
that have simultaneous neutron and positron detection.  This
would lower somewhat the line labeled {\it Total(other)} 
and make the early $\nu_e$ breakout signal easier to see.
However, such a procedure requires accurate signal timing,
which would be a challenge if the $^{35}$Cl neutron capture
mechanism is employed.  As it stands, one might
expect to get $\sim$5 events in SNO from the $\nu_e$ breakout burst 
in the first $\sim$10\,ms after bounce.

\subsection{ICARUS}
\label{section:icarus}

Although it has not yet reached its goal mass,
ICARUS\footnote{http://www.aquila.infn.it/icarus/} 
is designed to be a 3.6-ktonne drift chamber of pure liquid $^{40}$Ar that 
will be sensitive primarily to electron neutrinos through  
electron-neutrino capture on Ar:
$$\nu_e+^{40}{\rm Ar}\rightarrow^{40}{\rm K}^*+e^-.$$
Recent shell-model calculations have shown that
this process can proceed through super-allowed Fermi transitions
to the 4.38 MeV excited isobaric analog state, as well as through several
Gamow-Teller (GT) transitions to other lower-lying states in $^{40}{\rm K}$ (Ormand et al.~1995).
This has recently been investigated
experimentally by studying the $\beta^+-$decay from $^{40}{\rm Ti}$ (Bhattacharya et al.~1998).
Employing isospin symmetry, Bhattacharya et al.~(1998) inferred the transition
strengths of $^{40}{\rm Ar}\rightarrow^{40}{\rm K}$ from $^{40}{\rm Ti}\rightarrow^{40}{\rm Sc}$.
These theoretical and experimental results indicate that the GT transitions enhance the total $\nu_e$
capture cross section by a factor of three over the pure Fermi transition cross section.
We take the super-allowed Fermi transition cross section from Raghavan (1986);
\begin{equation}
\sigma_{\rm F}\simeq1.02\times10^{-43}(\varepsilon_{\nu_e}-5.365)^2\,\,{\rm cm^2}
\;\;(\epsilon_{\rm th}=5.885 \,{\rm MeV})
\end{equation}
and we assume that the total $\nu_e$ absorption cross section is given by 
$\sigma^{\rm tot}_{\nu_e{\rm Ar}}=3\sigma_{\rm F}$.

The ICARUS detector is currently
at 600 tonnes and has recently been approved to expand to 3 ktonnes. 
Because of its sensitivity to $\nu_e$ neutrinos, ICARUS is of particular
importance in detecting the early breakout pulse so prominent
in Fig.~\ref{lts11}.  In the results presented below,
we assume that the detector efficiency is 100\% above threshold and
that the detector mass is 3\,ktonnes.

The $\nu_e$ breakout signal in ICARUS is very likely detectable.
Figure \ref{ski} shows $N(<t)$ for the ICARUS detector (with 3 ktonnes of Ar).  
We plot the two dominant
detection channels for $\nu_e$ neutrinos: $\nu_e$Ar absorption (thick solid line) 
and $\nu_ie^-\rightarrow\nu_ie^-$ (thin solid line). We also include the
event rate, $dN/dt$ (in units of $50$\,s$^{-1}$), so that the width of the $\nu_e$ breakout
spike can be compared with the detector signal in the very early phase.
Note that the other neutrino species will
contribute to the total signal through only the $\nu_ie^-\rightarrow\nu_ie^-$ process
and that the combined signal amounts to only two events in the first 250\,ms shown in 
Fig.~\ref{ski}.
Combining the $\nu_e$-electron scattering and $\nu_e$-Ar absorption rates, 
ICARUS should expect to detect approximately 10 $\nu_e$ events from the breakout pulse alone,
more than SNO and as many as Super K.

\subsection{The Influence on the Signal of Various Neutrino Processes and the EOS}
\label{section:microdetect}

\subsubsection{Weak Magnetism \& Recoil}

The weak-magnetism/recoil correction is relevant both in the supernova itself
and in the detection of $\bar{\nu}_e$ neutrinos in light-water detectors 
via the process $\bar{\nu}_ep\rightarrow ne^+$.  In \S\ref{weak:recoil}
we compared the resulting spectra from models of core-collapse with and without this
correction.  In all our detection calculations we retain the weak-magnetism/recoil 
correction in calculating the signal due to $\bar{\nu}_ep\rightarrow ne^+$.
Comparing the detector signal
with and without this correction we find differences
in events rates for neutrinos of all flavors of less than 1\%.  The
models that include the correction have slightly higher event rates.
For anti-neutrinos, in particular for $\bar{\nu}_e$ in SK, we find that
the event rate with the correction is larger by approximately 7\%.


\subsubsection{Inelastic Neutrino-Electron Scattering}

Inelastic neutrino-electron scattering in our core-collapse
models significantly affects the observed $\nu_\mu$ neutrino signal.  
The total $\nu_\mu$ event rate in SK via neutrino-electron scattering 200\,ms 
after bounce is $\sim$15\% higher in the model without inelastic 
neutrino-electron scattering in the supernova.  If this process
is neglected in the supernova models, the signal in SNO due to neutral-current 
deuteron breakup by $\nu_\mu$ neutrinos is nearly doubled.  
Approximately 200\,ms after bounce, the corresponding event rates are 
110\,s$^{-1}$ and 65\,s$^{-1}$, respectively.

Inelastic neutrino-electron scattering also affects the
observed early-time $\bar{\nu}_e$ signal in SK.  Figure \ref{ltnr} demonstrates that
the $\bar{\nu}_e$ luminosity very near breakout is larger in the model
without neutrino-electron scattering, though at later times models with and without
neutrino-electron scattering are quite similar.   
At 200\,ms after bounce, we obtain in SK 50 more $\bar{\nu}_e$ events 
in the model without neutrino-electron scattering (550 versus 500 events).
Just 20-30\,ms after bounce, the event rates differ by 10-15\%.

\subsubsection{Nucleon-Nucleon Bremsstrahlung}

Including nucleon-nucleon bremsstrahlung as a production
process for $\nu_\mu$ neutrinos softens and brightens the resulting spectrum
(see \S\ref{section:dyn_brem} and Fig.~\ref{aem}).  By including 
bremsstrahlung, the total event rate in SNO 
due to neutral-current deuteron breakup by $\nu_\mu$ neutrinos 200\,ms after bounce is increased  
by 13\%.  Similarly, the total number of neutrino-electron scattering
$\nu_\mu$ events in SK is increased  
by 20\% (10.5 versus 8.5 events after 200\,ms).
The event rates during this epoch of accretion-driven neutrino luminosity
are 62\,s$^{-1}$ and 55\,s$^{-1}$, respectively - a fractional difference of $\sim$11\%.

\subsubsection{LSEOS}

In \S\ref{section:comp}, we presented results from core-collapse
calculations employing different nuclear compressibilities in the
context of the LSEOS (see Appendix \ref{appendix.eos}). In the model with 
$\kappa=180$\,MeV we find that the $\nu_e$-electron scattering event rate in SK 
200\,ms after bounce is 160\,s$^{-1}$ and that the corresponding cumulative event total  
is 47.  The model with $\kappa=375$\,MeV gives 150\,s$^{-1}$ and 42,
respectively.  The event rates in SK due to $\bar{\nu}_e$ capture on free protons
200\,ms after bounce are 2870\,s$^{-1}$ and 3080\,s$^{-1}$ for $\kappa=375$\,MeV
and $\kappa=180$\,MeV, respectively - a fractional difference of just $\sim7$\%.  
The $\nu_\mu$ signal in both SK (due to
neutrino-electron scattering) and in SNO (due to deuteron breakup) are modified at the
10\% level, with the larger event rates corresponding to the lower
nuclear compressibility ($\kappa=180$\,MeV).

\subsection{Neutrino Oscillations}

Due to severe matter suppression of standard flavor neutrino oscillations 
at the high densities encountered in supernova cores, such oscillations do
not appreciably affect the dynamics, neutrino spectra, nor neutrino luminosities 
in the interior.  We have calculated a few models that approximately incorporate
flavor oscillations employing the oscillation parameters implied by the solar
neutrino
and atmospheric neutrino deficits and find almost no effect on core evolution.    

The standard Boltzmann equation involves intensities, not amplitudes and phases.
It is the evolution equations for the latter, or for a full density matrix, that 
should be calculated in order to take neutrino oscillations properly into account.
However, we have opted for a simpler approach that captures the basic effect.
For a freely propagating electron neutrino, the probability that it oscillates into 
a $\nu_{\mu}$ neutrino is:
\begin{equation}
P_{e\mu} = \sin^2 2\theta_m \sin^2 (\frac{\pi r}{L_m})\, ,
\label{pee}
\end{equation}
where $\theta_m$ and $L_m$ are the oscillation angle and length in matter, $r$ is
the distance propagated,
and $\theta$ is the oscillation angle in vacuum.
$L_m$ and $\theta_{m}$ are given by the expressions:
\begin{equation}
\sin^2 2\theta_m = \frac{\sin^2 2\theta}{\sin^2 2\theta + (\cos 2\theta - L_v/L_0)^2}\,,
\label{oscill}
\end{equation}
and
\begin{equation}
L_m = L_v\left[1-\frac{2 L_v}{L_0} \cos 2\theta +
(\frac{L_v}{L_0})^2\right]^{-1/2} = L_v \frac{\sin 2\theta_m}{\sin 2 \theta} \, ,
\label{ell}
\end{equation}
where 
\begin{equation}
L_v = \frac{4\pi\hbar \varepsilon_{\nu}}{\delta m^2 c} \sim 500\, {\rm km}\,\,
\left(\frac{\varepsilon_{\nu}}{10\,\,{\rm MeV}}\right)
\left(\frac{5\times 10^{-5}\,\,{\rm eV}^2}{\delta m^2}\right) 
\label{ell2}
\end{equation}
is the vacuum oscillation length, $L_0 \sim 3.3\times 10^{-3}\,{\rm cm}/\rho_{12}$, 
$\rho_{12}=\rho/10^{12}$\,g\,\,cm$^{-3}$, and $\delta m^2$
is the standard mass-squared-difference.   The matter suppression of neutrino
flavor oscillations
is quite severe.  At resonance, $\theta_m = \pi/4$ and $L_v/L_0 = \cos
2\theta$,
but for the range of densities that obtain in protoneutron stars and the
values of $\delta m^2$ ($\sim 5\times 10^{-5}\,{\rm eV}^2$) and $\sin^2 2\theta$ 
($> 0.8$) derived from the KamLAND (Eguchi et al.~2003) and solar neutrino data, 
there are no MSW resonances in supernova interiors.  We have not taken into
consideration the possibility of density fluctuations induced neutrino flavor
depolarization (Loreti et al.~1995).  If we assume that a hard scattering or 
absorption decoheres the state (``resets the clock''),
then a rough measure of the probability of oscillation from $\nu_e$ to $\nu_{\mu}$
in a radius interval $dr$
is $P_{e\mu} e^{-r/\Lambda} dr/\Lambda$, where $\Lambda$ is the extinction
mean-free path.
The appropriate integral gives us the approximate transition rate
\begin{equation}
\Gamma_{e\mu} = \frac{2\pi^2c}{\Lambda_e} \sin^2\theta_m
\bigl(\frac{\Lambda_e}{L_m}\bigr)^2\, .
\label{trans}
\end{equation}
There are corresponding expressions for $\Gamma_{\mu e}$, etc.  These rates are
plugged into
rate equations that conserve total neutrino number:
\begin{eqnarray}
\frac{d J_{\nu_e}}{d t} = -\Gamma_{e\mu} \,J_{\nu_e} + \Gamma_{\mu e} \,J_{\nu_\mu}/4
\nonumber
\\
\frac{d J_{\nu_\mu}}{d t} = \Gamma_{e\mu} \,J_{\nu_e} - \Gamma_{\mu e} \,J_{\nu_\mu}/4 \, ,
\label{rates}
\end{eqnarray}
where $J_i$ is the energy dependent zeroth moment of the specific 
intensity for neutrino species $i$ ($\times$ 4 for the $\nu_{\mu}$s).
Hence, using eqs.~(\ref{rates}), we can incorporate the effects of neutrino
oscillations on the internal supernova dynamics and spectra.  As stated, they are small.

The edge of our computational grid resides 
deep inside the supernova progenitor.  The spectra we have calculated and
folded with terrestrial neutrino detectors will undoubtedly be modified
by oscillation effects in propagating from the edge of our computational
domain to the detector through the low-density progenitor envelope.  
Hence, our neutrino number spectra must be multiplied by
effective survival probabilities taking into account the density profile of 
the overlying progenitor and, depending upon the position of the detector, the 
strata of the Earth itself. Because our focus is here on producing reliable
emergent spectra from core-collapse and the diagnosis of the basic effects
on this signal of changes in the supernova microphysics, and because modifications
to the signal in detectors depends on the wide range of progenitor envelope
profiles, oscillation parameters, and Earth day-night effects, we save a detailed 
investigation of neutrino oscillations for a future work.  The reader
is referred to Ando \& Sato (2002), Dighe \& Smirnov (2000), Dutta et al.~(1999),
and Takahashi et al.~(2001) for recent discussions of possible neutrino oscillation
effects in the detected supernova neutrino signal.

\section{Summary and Discussion}
\label{summary}

We have constructed fully dynamical models of core collapse
in spherical symmetry, focusing on the shock breakout phenomenon
and the first 100's of milliseconds after bounce.  
We employed a newly-developed algorithm
for radiation-hydrodynamics, which gives a Boltzmann
solution to the neutrino transport problem using the Feautrier
technique, the tangent-ray method, and Accelerated Lambda Iteration.
The code provides good angular and spectral resolution of the radiation field and incorporates 
realistic neutrino microphysics.  We find that at least 20 energy groups are
required to adequately resolve the neutrino spectrum and the dynamics;
10 energy groups are not enough.  Furthermore, the tangent-ray algorithm provides
very good and adaptive angular resolution that is well-suited to the
spherically-symmetric supernova problem in which the neutrino fields transition
from a diffusive interior to a free-streaming exterior that nevertheless still resides within the
relevant computational domain.  Using this scheme, we have presented some of the first high-resolution
angular distribution functions for neutrinos in stellar collapse.  

We have explored the effects on 
dynamical simulations of core collapse and its neutrino signatures of
nucleon-nucleon bremsstrahlung, inelastic neutrino-electron scattering, weak-magnetism/recoil,
and artificial decreases in the neutrino-nucleon scattering rates.
Our new algorithm for incorporating inelastic scattering explicitly as a source and
a sink in the collision term of the Boltzmann equation is both
efficient and robust (\S\ref{section:inelastic}).  
Furthermore, we have studied the dependence of the collapse dynamics
and neutrino signal on variations in the equation of state, as parametrized
by Lattimer and Swesty (1991).   
Surprisingly, the dependence of the salient quantities of collapse
on the EOS (again, as parametrized by Lattimer and Swesty) is weak.

We have computed the collapse, bounce, and post-bounce evolution
and neutrino signatures for three progenitors: 11, 15, and 20\,M$_\odot$.
In the process, we have compared key quantities
such as core entropy and electron fraction at bounce, peak neutrino breakout
luminosity, and the evolution of the neutrino spectra over hundreds of
milliseconds with the corresponding quantities in other recent supernova 
simulations that employ comparably sophisticated algorithms.
We evolve most models to several hundred milliseconds after bounce and
although all of our models form regions of net neutrino heating
behind the shock, none yields an explosion in the first 250 milliseconds.  

We have folded our neutrino spectra with the sensitivities and
thresholds of several underground neutrino detectors 
and have focused on those detectors most likely to observe and uniquely 
identify the electron-neutrino breakout burst.  We show that SuperKamiokande,
the Sudbury Neutrino Observatory, and the ICARUS detector might all 
see this important signature of core collapse.  
We find that the collapse and breakout electron-neutrino signal
in these detectors is remarkably similar (see Fig.~\ref{skn}) for
all three progenitors considered in this work.  The pre-collapse
profiles are different in entropy and composition, but the breakout
peak is roughly the same in each model.  These results are further 
confirmed in preliminary runs with a 40\,M$_\odot$ progenitor.

Figure \ref{skn} shows that the event rate due to charged-current capture of $\bar{\nu}_e$ neutrinos
on free protons for the 20\,M$_\odot$ progenitor (5850\,s$^{-1}$) is more than double that
for the 11\,M$_\odot$ model (2920\,s$^{-1}$) 175\,ms after bounce.
The nearly linear rise in $N(<t)$ reflects the fact that the 
accretion luminosity is roughly constant in this 
early post-bounce epoch.  The difference between 
the 11 and 20\,M$_\odot$ models can be attributed primarily to 
a difference in $\dot{M}$ (see Fig.~\ref{massfpro}), which, in turn,  
is a consequence of the relatively steeper envelope density profile  
of the lighter progenitor.  The behaviors of the density profiles of the 
11\,M$_\odot$ and 20\,M$_\odot$ models roughly bracket those found 
in the full progenitor model suite of Woosley \& Weaver (1995).  
Hence, we expect that most other progenitors should yield accretion luminosities
that produce $\bar{\nu}_e$ signals between those of the 11\,M$_\odot$ and 20\,M$_\odot$
models in Fig.~\ref{skn}.

\acknowledgments

We would like to acknowledge discussions with Stan Woosley, Eli Livne, Itamar Lichtenstadt,
Ron Eastman, Thomas Janka, Markus Rampp, Bronson Messer, Tony Mezzacappa, Georg Raffelt,
Chuck Horowitz, Jorge Horvath, Rolf Walder, Christian Ott, Casey Meakin, and Jeremiah Murphy.
Support for this work is provided in part by 
the Scientific Discovery through Advanced Computing (SciDAC) program
of the DOE, grant number DE-FC02-01ER41184, a NASA GSRP program fellowship
and by NASA through Hubble Fellowship
grant \#HST-HF-01157.01-A awarded by the Space Telescope Science
Institute, which is operated by the Association of Universities for Research in Astronomy,
Inc., for NASA, under contract NAS 5-26555.

\begin{appendix}


\section{Tabular Nuclear Equation of State}
\label{appendix.eos}

We employ the equation of
state due to Lattimer \& Swesty (1991) (the LSEOS), based on the finite-temperature liquid drop
model of nuclei developed in Lattimer et al.~(1985).  We have
constructed an efficient three-dimensional tabular version of this EOS,
taking the matter temperature ($T$), mass density ($\rho$), and
electron fraction ($Y_e$) as variables.  The table consists of 180 equally spaced
points in $\log_{10}[\rho\,\,({\rm g \,\,cm^{-3}})]$ with
$6.4\leq\log_{10}[\rho\,\,({\rm g \,\,cm^{-3}})]\leq15.1$, 50 equally spaced  
$Y_e$ planes with $0.05\leq Y_e \leq 0.51$, and 180 equally spaced zones in 
$\log_{10}[T\,\,({\rm MeV})]$.
The table is not cubic.  For $\log_{10}[\rho\,\,({\rm g \,\,cm^{-3}})]=6.4$,
$-0.8\leq\log_{10}[T\,\,({\rm MeV})]\leq1$.
At $\log_{10}[\rho\,\,({\rm g \,\,cm^{-3}})]=15.1$,
$0\leq\log_{10}[T\,\,({\rm MeV})]\leq1.6$.  We find this
range of temperatures and densities sufficient for all calculations we
perform.  At each point in the table we save the specific internal energy ($E$),
the pressure ($P$), the entropy per baryon ($s$), the specific heat at constant volume ($C_V$),
the neutron, proton,
alpha particle, and heavy nucleus fractions ($X_n$, $X_p$, $X_\alpha$, $X_H$, respectively),
$A$ and $Z/A$ for the heavy nucleus,
the electron chemical potential ($\mu_e$),
$\Gamma_s=\p\ln P/\p\ln\rho|_s$, $\p P/\p T|_\rho$, $\hat{\mu}=\mu_n-\mu_p$,
and the derivatives $\p\hat{\mu}/\p T$ and $\p\hat{\mu}/\p Y_e$.
In total, our tabular LSEOS takes up $\sim$200 Megabytes of memory.
Given $T$, $\rho$, and $Y_e$, the EOS performs three six-point bivariant
interpolations (Abromowitz \& Stegun 1972) in the $T-\rho$ planes nearest to and
bracketing the given $Y_e$ point.  A quadratic interpolation is then executed between
$Y_e$ points to obtain
the desired thermodynamic quantity.  This procedure is employed for all quantities
in the table except the mass fractions of neutrons, protons, alpha particles, and the
heavy nucleus.  For these quantities, two four-point bivariant interpolations in the $T-\rho$ plane
are combined with a linear interpolation between $Y_e$ planes.
The table uses integer arithmetic to find nearest neighbor points, thus alleviating
the need for time-intensive search algorithms.
Because most hydrodynamics
routines update specific internal energy, we include a Newton-Raphson/bisection scheme
which iterates on temperature, given a fixed internal energy, until the root is found
to within a part in $10^8$.  Similar iteration routines are employed if one wishes to
iterate on entropy or pressure.

The LSEOS extends down to only $\sim5\times10^6$ g cm$^{-3}$ and its validity
in this density regime is guaranteed only for fairly high temperatures - the
assumption of NSE being thereby satisfied.  For calculations in which an explosion occurs,
the shock will quickly evolve down the progenitor density gradient to regions
where the LSEOS breaks down.  For this reason we have 
coupled to our tabular version of the LSEOS the Helmholtz EOS
(Timmes \& Arnett 1999; Timmes \& Swesty 2000), which contains electrons and
positrons at arbitrary degeneracy and relativity, photons, nuclei 
and nucleons as non-relativistic ideal gases, and Coulomb corrections.
At $\rho=6\times10^7$ g cm$^{-3}$ we assume the LSEOS is valid.  At $\rho=4\times10^7$ g cm$^{-3}$
we employ only the Helmholtz EOS.  For densities between the two, we quadratically interpolate
all relevant thermodynamic quantities.  Because the LSEOS assumes NSE and the Helmholtz EOS
takes non-NSE abundances, there is no thermodynamic consistency between the
two EOSs.  Fortunately, in thermodynamical regimes relevant for core collapse
at these densities, the electron-positron/photon component of the matter dominates the
pressure. Therefore, one expects few hydrodynamic artifacts in piecing together these two
equations of state.  In fact, given a realistic composition for
the Helmholtz EOS, at a density of $\rho=5\times10^7$ g cm$^{-3}$,
pressure differences between the LSEOS and Helmholtz EOS are $\sim1$\%
in the stellar profiles we employ in this study.
Differences in entropy, however, are of order 5-10\%.

We have performed an extensive set of tests of our tabular implementation
of the LSEOS to ensure that thermodynamical consistency is maintained
during dynamical simulations.  These tests include the adiabatic compression
of a single fluid element over eight orders of magnitude (from an initial density
of 10$^7$ g cm$^{-3}$) and the purely hydrodynamical core collapse of realistic 
supernova progenitors.  For the single-element tests, we use the fractional
change in the element entropy ($\Delta s/s_{\rm i}$, where $s_{\rm i}$ is
the initial entropy at low density) to assess the consistency 
of the LSEOS as the element is compressed and the density 
increases to $\sim5\times10^{14}$ g cm$^{-3}$.
For fractional changes in the density (or specific volume) at each step of
1\%, our table gives $\Delta s/s_{\rm i}\sim0.5$\%. This is to be compared with 
the results using the actual analytic LSEOS, from which the table was composed, 
which yield $\Delta s/s_{\rm i}\sim0.25$\%. 
The number of temperature and
density points used in the table does effect $\Delta s/s_{\rm i}$.
For example, using just 100 $T$, 100 $\rho$, and 50 $Y_e$ points
over the same ranges as the larger table raises $\Delta s/s_{\rm i}$ to about $\sim1.25$\%.
In all our calculations we have opted for the larger table.
For the adiabatic collapse tests, we coupled the LSEOS to a variety 
of one-dimensional hydrodynamics schemes.  With radiation turned off and
before any shocks form, each mass zone should remain isentropic during the 
collapse of a massive progenitor star.
Using a variety of zoning schemes with both artificial-viscosity and Riemann solvers for
shock capturing and resolution, we found that higher-order EOS interpolation schemes
as described in Swesty (1996) are not needed to maintain the degree of adiabaticity (1\%)
during collapse and bounce that we were able to achieve with our more
straightforward tabular and interpolation routines.

\section{Inelastic Neutrino Scattering}
\label{section:inelastic}

A number of authors have addressed the issue of inelastic neutrino-electron scattering as
an important energy redistribution process, which helps thermalize the neutrinos and
increases their energetic coupling to the matter in supernova explosions (Bruenn 1985;
Mezzacappa \& Bruenn 1993).  
Here, we review the Legendre expansion formalism 
for approximating the angular dependence of the scattering kernel, and detail our own
implementation of scattering terms in the Boltzmann equation.

The general collision integral for inelastic scattering may be written (Bruenn 1985) as
\beqa
{\cal L}^{\rm scatt}_\nu[f_\nu]&=&(1-f_\nu)\int\frac{d^3p_\nu\pr}{c(2\pi\hbar c)^3}f_\nu\pr\,
R^{\rm in}(\varepsilon_\nu,\varepsilon_\nu\pr,\cos\theta) \nonumber
-f_\nu\int\frac{d^3p_\nu\pr}{c(2\pi\hbar c)^3}(1-f_\nu\pr)\,
R^{\rm out}(\varepsilon_\nu,\varepsilon_\nu\pr,\cos\theta) \nonumber \\
&=&\tilde{\eta}_\nu^{\rm scatt}-\tilde{\chi}_\nu^{\rm scatt}f_\nu
\label{gencoll}
\eeqa
where $\cos\theta$ is the cosine of the scattering angle, $\varepsilon_\nu$ is the 
incident neutrino energy,
and $\varepsilon_\nu\pr$ is the scattered neutrino energy.   
Although we suppress it here, the incident
and scattered neutrino phase space distribution functions ($f_\nu$ and $f_\nu\pr$, respectively) 
have the following dependencies: $f_\nu=f_\nu(r,t,\mu,\varepsilon_\nu)$ and 
$f_\nu\pr=f_\nu\pr(r,t,\mu\pr,\varepsilon_\nu\pr)$.  $\mu$ and $\mu\pr$ are 
the cosines of the angular
coordinate of the zenith angle in spherical symmetry and are related to $\cos\theta$ through
\begin{equation}
\cos\theta=\mu\mu\pr+[(1-\mu^2)(1-\mu^{\prime\,2})]^{1/2}\cos(\phi-\phi\pr).
\label{costheta}
\end{equation}
$R^{\rm in}$ is the scattering kernel for scattering {\bf in}to the 
bin ($\varepsilon_\nu$, $\mu$) from any bin ($\varepsilon_\nu\pr$, $\mu\pr$)
and $R^{\rm out}$ is the scattering kernel for scattering {\bf out} of the 
bin ($\varepsilon_\nu$, $\mu$) to any bin ($\varepsilon_\nu\pr$, $\mu\pr$).
The kernels are Green functions, correlation functions that connect points
in energy and momentum space.  In fact, one may also write 
$R(\varepsilon_\nu,\varepsilon_\nu\pr,\cos\theta)$ as $R(q,\omega)$, where
$\omega(=\varepsilon_\nu-\varepsilon_\nu\pr)$ is the energy transfer and
$q(=[\varepsilon_\nu^2+\varepsilon_\nu^{\prime\,2}-2\varepsilon_\nu
\varepsilon_\nu\pr\cos\theta]^{1/2})$
is the momentum transfer, so that the kernel explicitly reflects these dependencies.

It is clear from eq.~(\ref{gencoll}) that the source, or the net scattering into the
beam ($\varepsilon_\nu$, $\mu$) is a function of the occupancy ($f_\nu\pr$) in all other beams.
The sink term also depends on $f_\nu\pr$ through the blocking term ($1-f_\nu\pr$), 
reflecting the Fermi-Dirac statistics of the neutrino.  
A solution to the Boltzmann equation yields $f_\nu$ (and, hence, $f_\nu\pr$)
at all times, radii, energies, and angles.  $f_\nu$ is not known a priori and cannot
be assumed to be Fermi-Dirac.  Only in equilibrium should $f_\nu$ approach a Fermi-Dirac
distribution, characterized by the local temperature and with a chemical potential 
that reflects the local neutrino number density.  The transport problem is difficult
enough without the added complication of non-linear integral source terms.
The full implicit solution, including energy and angular redistribution, is
numerically cumbersome. Instead, we make several simplifications 
that make the problem tractable, efficiently solved, and explicit.

An important simplification comes from detailed balance, a consequence
of the fact that these scattering rates must drive the distribution to equilibrium;
$R^{\rm in}=e^{-\beta\omega}R^{\rm out}$, where $\beta=1/T$.  We may therefore deal
only with $R^{\rm out}$.  The scattering kernels 
generally have complicated dependencies on scattering angle and the angular integration
over scattered neutrino phase space, implicit in eq.~(\ref{gencoll}), cannot be 
computed analytically.  For this reason, we approximate the angular dependence
of the scattering kernel with a truncated Legendre series (Bruenn 1985).  That is, we take
\begin{equation}
R^{\rm out}(\varepsilon_\nu,\varepsilon_\nu\pr,\cos\theta)
=\sum_{l=0}^\infty\frac{2l+1}{2}\Phi(\varepsilon_\nu,\varepsilon_\nu\pr,\cos\theta)
P_l(\cos\theta),
\end{equation}
where 
\begin{equation}
\Phi_l(\varepsilon_\nu,\varepsilon_\nu^\prime)=\int_{-1}^{+1}d(\cos\theta)\, 
R^{\rm out}(\varepsilon_\nu,\varepsilon_\nu^\prime,\cos\theta)P_l(\cos\theta).
\label{momentkernel}
\end{equation}
In practice, we expand only to first order so that
\begin{equation}
R^{\rm out}(\varepsilon_\nu,\varepsilon_\nu\pr,\cos\theta)\sim
\frac{1}{2}\Phi_0(\varepsilon_\nu,\varepsilon_\nu\pr)+
\frac{3}{2}\Phi_1(\varepsilon_\nu,\varepsilon_\nu\pr)\cos\theta.
\label{kernelapprox}
\end{equation}
Substituting into the first term on the right-hand-side of eq.~(\ref{gencoll})
(the source) gives
\begin{equation}
\tilde{\eta}_\nu^{\rm scatt}=(1-f_\nu)
\int_0^\infty \frac{d\varepsilon_\nu\pr\varepsilon_\nu^{\prime\,2}}{c(2\pi\hbar c)^3}\,e^{-\beta\omega}
\int_{-1}^{+1}d\mu\pr f_\nu\pr\int_0^{2\pi}d\phi\pr
\left[\frac{1}{2}\Phi_0+
\frac{3}{2}\Phi_1\cos\theta\right]
\end{equation}
Substituting for $\cos\theta$ using eq.~(\ref{costheta}) and using the definitions
\begin{equation}
\tilde{J}_\nu=\frac{1}{2}\int_{-1}^{+1}d\mu f_\nu\hspace*{1 cm}{\rm and}\hspace*{1 cm}
\tilde{H}_\nu=\frac{1}{2}\int_{-1}^{+1}d\mu \mu f_\nu
\end{equation}
we have that 
\begin{equation}
\tilde{\eta}_\nu^{\rm scatt}=(1-f_\nu)\frac{4\pi}{c(2\pi\hbar c)^3}
\int_0^\infty d\varepsilon_\nu\pr \varepsilon_\nu^{\prime\,2} e^{-\beta\omega}
\left[\frac{1}{2}\Phi_0\,\tilde{J}_\nu\pr+\frac{3}{2}\Phi_1\mu\tilde{H}_\nu\pr\right].
\end{equation}
Integrating over $\mu$ to get the source for the zeroth-moment equation,
\begin{equation}
\frac{1}{2}\int_{-1}^{+1}d\mu\,\tilde{\eta}_\nu^{\rm scatt}=
\frac{4\pi}{c(2\pi\hbar c)^3}
\int_0^\infty d\varepsilon_\nu\pr \varepsilon_\nu^{\prime\,2} e^{-\beta\omega}
\left[\frac{1}{2}\Phi_0\,\tilde{J}_\nu\pr(1-\tilde{J}_\nu)-
\frac{3}{2}\Phi_1\,\tilde{H}_\nu\tilde{H}_\nu\pr\right].
\label{jtildeeta}
\end{equation}
Similarly, employing the Legendre expansion, 
we may write the sink term of the Boltzmann collision integral
in terms of the angular moments of $f_\nu^{\prime}$:
\begin{equation}
\tilde{\chi}_\nu^{\rm scatt}=\frac{4\pi}{c(2\pi\hbar c)^3}
\int_0^\infty d\varepsilon_\nu\pr \varepsilon_\nu^{\prime\,2} 
\left[\frac{1}{2}\Phi_0(1-\tilde{J}_\nu\pr)-\frac{3}{2}\Phi_1\mu\tilde{H}_\nu\pr\right].
\end{equation}
The contribution to the zeroth-moment equation is then
\begin{equation}
\frac{1}{2}\int_{-1}^{+1}d\mu(-\tilde{\chi}_\nu^{\rm scatt}f_\nu)=-\frac{4\pi}{c(2\pi\hbar c)^3}
\int_0^\infty d\varepsilon_\nu\pr \varepsilon_\nu^{\prime\,2} 
\left[\frac{1}{2}\Phi_0(1-\tilde{J}_\nu\pr)\tilde{J}_\nu-
\frac{3}{2}\Phi_1\tilde{H}_\nu\tilde{H}_\nu\pr\right].
\label{jtildechi}
\end{equation}
Combining these equations, we find that
\begin{equation}
\frac{1}{2}\int_{-1}^{+1}d\mu\,{\cal L}^{\rm scatt}_\nu[f_\nu]=
\frac{4\pi}{c(2\pi\hbar c)^3}\int_0^\infty d\varepsilon_\nu\pr 
\varepsilon_\nu^{\prime\,2} 
\left\{\frac{1}{2}\Phi_0\left[\tilde{J}_\nu\pr(1-\tilde{J}_\nu)e^{-\beta\omega}-
(1-\tilde{J}_\nu\pr)\tilde{J}_\nu \right]
-\frac{3}{2}\Phi_1\tilde{H}_\nu\tilde{H}_\nu\pr(e^{-\beta\omega}-1)\right\}.
\end{equation}
One can see immediately that including another term in the Legendre expansion,
i.e.~taking 
$$R^{\rm out}\sim(1/2)\Phi_0+(3/2)\Phi_1\cos\theta+
(5/2)\Phi_2(1/2)(3\cos^2\theta-1),$$ 
necessitates including $\tilde{P}_\nu$ and 
$\tilde{P}_\nu\pr$, the second angular moment of the neutrino phase-space distribution 
function, in the source and sink terms.  Our transport scheme porvides both the zeroth- and first-moment equations
with the spectral Eddington factors $p_\nu=\tilde{P}_\nu/\tilde{J}_\nu$ and 
$g_\nu=\tilde{N}_\nu/\tilde{J}_\nu$,
where
\begin{equation}
\tilde{P}_\nu=\frac{1}{2}\int_{-1}^{+1}d\mu \mu^2 f_\nu \,\,\,\,\,\,\,
{\rm and} \,\,\,\,\,\,\,
\tilde{N}_\nu=\frac{1}{2}\int_{-1}^{+1}d\mu \mu^3 f_\nu.
\label{ptilde}
\end{equation}
It is therefore straightforward for us to include terms up to $R^{\rm out}\propto\cos^3\theta$
in our Legendre expansion of the scattering kernel.  In practice, however, we retain only 
the linear term.

\subsection{Neutrino-Electron Scattering}
\label{app:escatt}

The scattering kernel $R(\varepsilon_\nu,\varepsilon_\nu\pr,\cos\theta)$ in \S\ref{section:inelastic}
is related
to the fully relativistic structure function for neutrino-electron scattering;
\begin{equation}
R^{\rm out}(\varepsilon_\nu,\varepsilon_\nu\pr,\cos\theta)=
2G^2\frac{q_\mu^2}{\varepsilon_\nu\pr\varepsilon_\nu}
[A{\cal S}_1(q,\omega)+{\cal S}_2(q,\omega)+B{\cal S}_3(q,\omega)](1-e^{-\beta\omega})^{-1}.
\end{equation}
The relativistic structure functions (${\cal S}_i$) are given in terms of the imaginary
part of the polarization functions ($\Pi$).
Each of the retarded polarization functions 
can be written in terms of one-dimensional integrals over
electron energy ($\varepsilon_e$) (see Reddy, Prakash, \& Lattimer 1998).

Even though we have made the simplifying assumption that the scattering kernel can be
approximated by a Legendre series truncated at first order (eq.~\ref{kernelapprox}),
in a fully dynamical calculation it is numerically costly to compute
the Legendre moments of the scattering kernel ($\Phi_0(\varepsilon_\nu,\varepsilon_\nu\pr)$ and 
$\Phi_1(\varepsilon_\nu,\varepsilon_\nu\pr)$)
via eq.~(\ref{momentkernel}) at each point on the radial grid (at each temperature, density, 
and composition point) at 
each time step.  In order to decrease the computation time, we 
tabulate $\Phi_0(\varepsilon_\nu,\varepsilon_\nu\pr)$ 
and $\Phi_1(\varepsilon_\nu,\varepsilon_\nu\pr)$ for each $\varepsilon_\nu$ 
and $\varepsilon_\nu\pr$, for each
neutrino species, on a grid in temperature and $\eta_e$.  At each $T$-$\eta_e$ point, 
we calculate both kernels for each $\varepsilon_\nu$-$\varepsilon_\nu\pr$ point, given 
the energy grouping for the particular
calculation.  The angular integrals over $\cos\theta$ for $l=0$ and $l=1$ 
in eq.~(\ref{momentkernel}) 
are carried out using 16-point Gauss-Legendre
quadratures.  During an actual supernova calculation, we use simple trapezoidal rule quadrature to 
calculate the integral over $\varepsilon_\nu\pr$ for a given $\varepsilon_\nu$.  Each term in this
integral contains the kernels, which must be evaluated for each 
$\varepsilon_\nu$-$\varepsilon_\nu\pr$
pair at the temperature/density/composition point currently being addressed.  $\eta_e$ is
calculated by the equation of state and we do a six-point bivariant interpolation 
in $T$-$\eta_e$ space, for the given 
$\varepsilon_\nu$-$\varepsilon_\nu\pr$ combination.  In practice, we use 40 energy groups ($n_f$),
30 temperature points ($N_T$), and 30 $\eta_e$ points ($N_\eta$).  
The tables are then $l\times n_f\times n_f\pr\times N_T\times N_\eta$
in size, with $l=2$ ($\Phi_0$ and $\Phi_1$), or approximately 50 Megabytes.  Since
the vector and axial-vector couplings for electron scattering are different for each 
neutrino species,
we compute tables for $\nu_e$, $\bar{\nu}_e$, and $\nu_\mu$ for each dynamical calculation.
$\tilde{\eta}_\nu^{\rm scatt}$ and $\tilde{\chi}_\nu^{\rm scatt}$
are then included explicitly as a source and sink, respectively, in a manner analogous to 
any of the absorption and emission processes detailed in Burrows (2001).
Using this method, our calculations including neutrino-electron scattering,
are just 10-15\% slower than our calculations ignoring this important equilibration process.

Because we do not evaluate the scattering source and sink implicitly, we introduce an
explicit timescale into the energy and electron fraction updates returned by our 
transport algorithm.  In effect, if the scattering timescale $(c\tilde{\chi}_\nu^{\rm scatt})^{-1}$
is shorter than our timestep, we may encounter a numerical instability.  For this reason,
at high densities, where the neutrino phase-space distribution function is in local thermodynamic
equilibrium, we divide the source and sink by a factor 
(typically 100 above $\rho=10^{14}$ g cm$^{-3}$).  
This reduces the rate artificially, thus increasing the timescale for scattering.  
As this process is totally
negligible in this regime, particularly considering the fact that $f_\nu=f_\nu^{\rm eq}$ at these
high densities, this approximation is acceptable.

\end{appendix}

\pagebreak

\clearpage

\clearpage

\clearpage

\clearpage

\clearpage

\clearpage

\clearpage

\clearpage

\clearpage

\clearpage

\clearpage

\clearpage

\clearpage

\clearpage

\clearpage

\clearpage

\clearpage

\clearpage

\clearpage

\clearpage

\begin{figure} 
\vspace*{6.0in}
\hbox to\hsize{\hfill\includegraphics{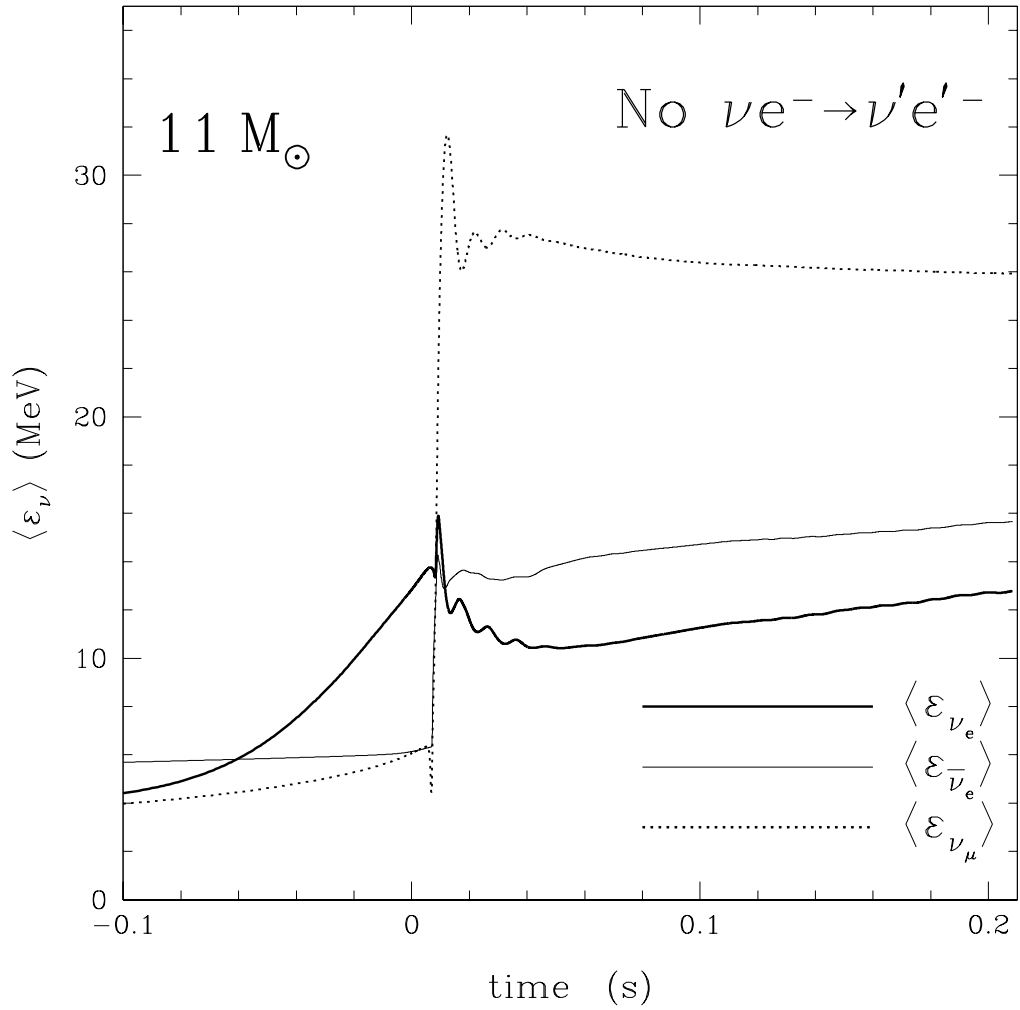}\kern+6in\hfill}
\caption[$\langle\varepsilon_{\nu_i}\rangle$
without inelastic $\nu_i e^-$ scattering]
{$\langle\varepsilon_{\nu_e}\rangle$ (thick solid line), 
$\langle\varepsilon_{\bar{\nu}_e}\rangle$ (thin solid line), 
and $\langle\varepsilon_{\nu_\mu}\rangle$  (dotted line)
at infinity for the 11\,M$_\odot$ progenitor without
inelastic neutrino-electron scattering.
Compare with Fig.~\ref{ltnr}, which shows
the corresponding luminosities and Fig.~\ref{ets11}, 
which shows the average energy evolution in our fiducial model
(including inelastic neutrino-electron scattering).
}
\label{etnr}
\end{figure}

\clearpage

\begin{figure} 
\vspace*{6.0in}
\hbox to\hsize{\hfill\includegraphics{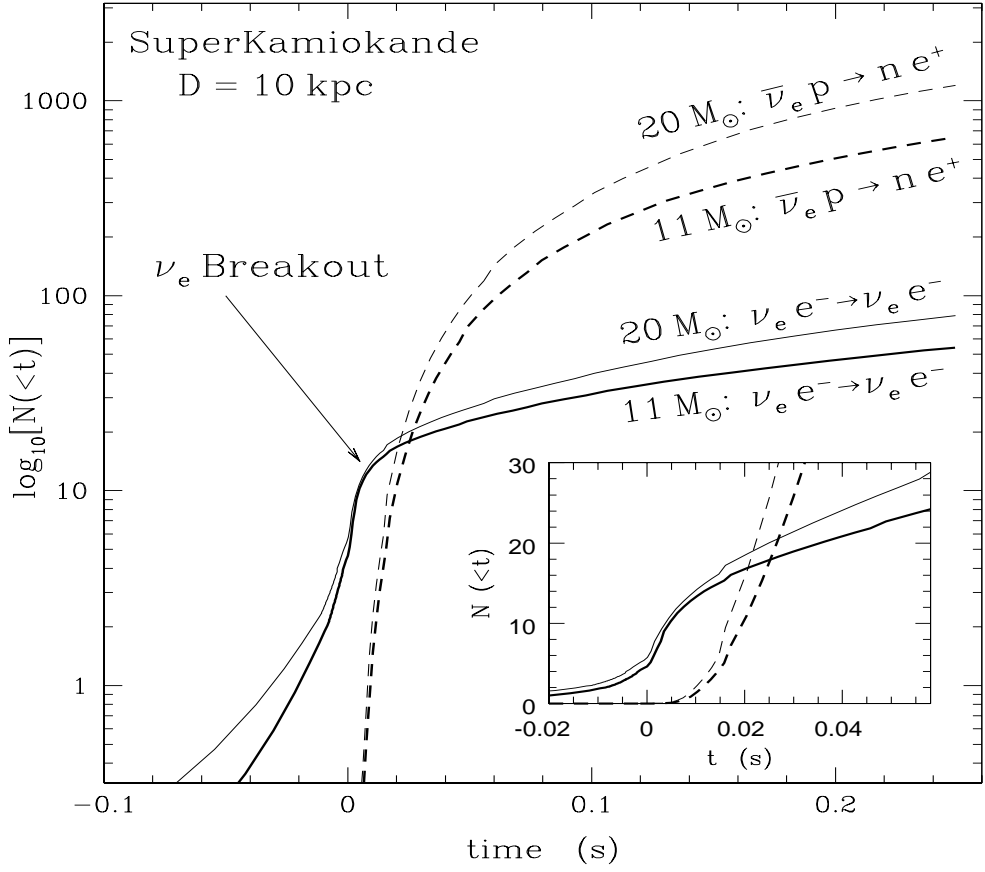}\kern+6in\hfill}
\caption[]
{The integrated number of neutrino detection events ($\log_{10}[N(<t)]$) 
in SK for our  11\,M$_\odot$ (thick lines) and 20\,M$_\odot$ (thin lines) 
models via $\nu_e-$electron scattering ($\nu_ee^-\rightarrow\nu_ee^-$, 
solid lines) and $\bar{\nu}_e$ capture on free 
protons ($\bar{\nu}_e p\rightarrow n e^+$, 
dashed lines) for supernovae at 10\,kpc.
The inset is an expanded view of the $\nu_e$ breakout signal.
Note that both models imply that SK should see a distinct and 
observable $\nu_e$ signature in the
first $\sim30$\,ms. Nearly 20 $\nu_e$ events accumulate before being
swamped by the dominant $\bar{\nu}_e$ signal.
For clarity, we do not include the corresponding lines for the 
15\,M$_\odot$ progenitor
in this figure.  The $\nu_e$ neutrino signal in the 15\,M$_\odot$ model
is identical to that of the 11\,M$_\odot$ in the first 50\,ms.  The 
$\bar{\nu}_e$ event rate is slightly larger than that for the 
11\,M$_\odot$ model;
the 15\,M$_\odot$ model accumulates 680 events whereas the 11\,M$_\odot$ model
yields 650 events in the 250\,ms after bounce shown here.}
\label{skn}
\end{figure}

\clearpage

\begin{figure} 
\vspace*{6.0in}
\hbox to\hsize{\hfill\includegraphics{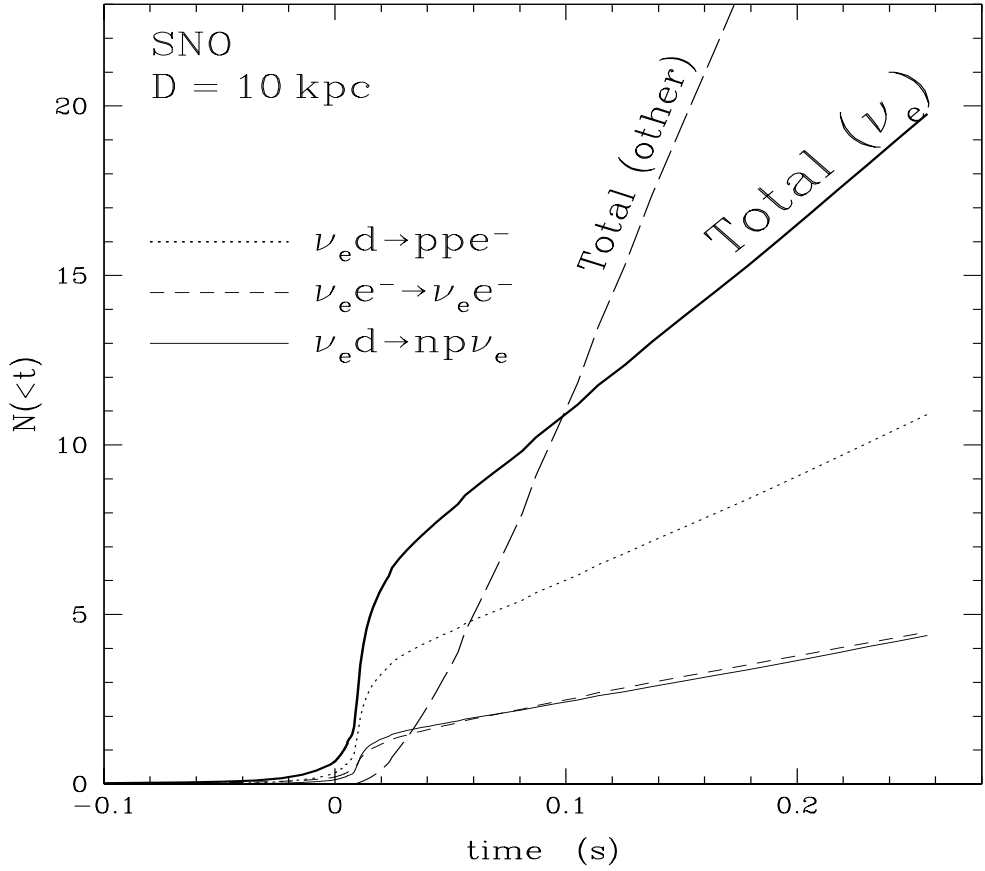}\kern+6in\hfill}
\caption[The $\nu_e$ breakout signal in SNO]
{Integrated number of $\nu_e$ neutrinos detected in SNO 
for a supernova at 10\,kpc as a function of time.
The individual contributions to the detected signal via the  
$\nu_ed\rightarrow ppe^-$ (dotted line), 
$\nu_ed\rightarrow np\nu_e$ (solid line), and 
$\nu_ee^-\rightarrow\nu_ee^-$ (dashed line) processes are shown.  
The sum 
is the thick solid line.  The sum of all other
processes, including $\bar{\nu}_ed\rightarrow np\bar{\nu}_e$,
$\bar{\nu}_ed\rightarrow nne^+$,
$\bar{\nu}_ep\rightarrow n e^+$ in the light-water portion of the detector,
as well as
$\bar{\nu}_{\mu,\tau}d\rightarrow np\bar{\nu}_{\mu,\tau}$,
${\nu}_{\mu,\tau}d\rightarrow np{\nu}_{\mu,\tau}$,
and $\bar{\nu}_e$-, ${\nu}_{\mu,\tau}$-, and $\bar{\nu}_{\mu,\tau}$-electron
scattering throughout the entire detector volume,
is shown as the long dashed line.
This signal corresponds
to our fiducial 11\,M$_\odot$ core-collapse model. }
\label{skd}
\end{figure}

\clearpage

\begin{figure} 
\vspace*{6.0in}
\hbox to\hsize{\hfill\includegraphics{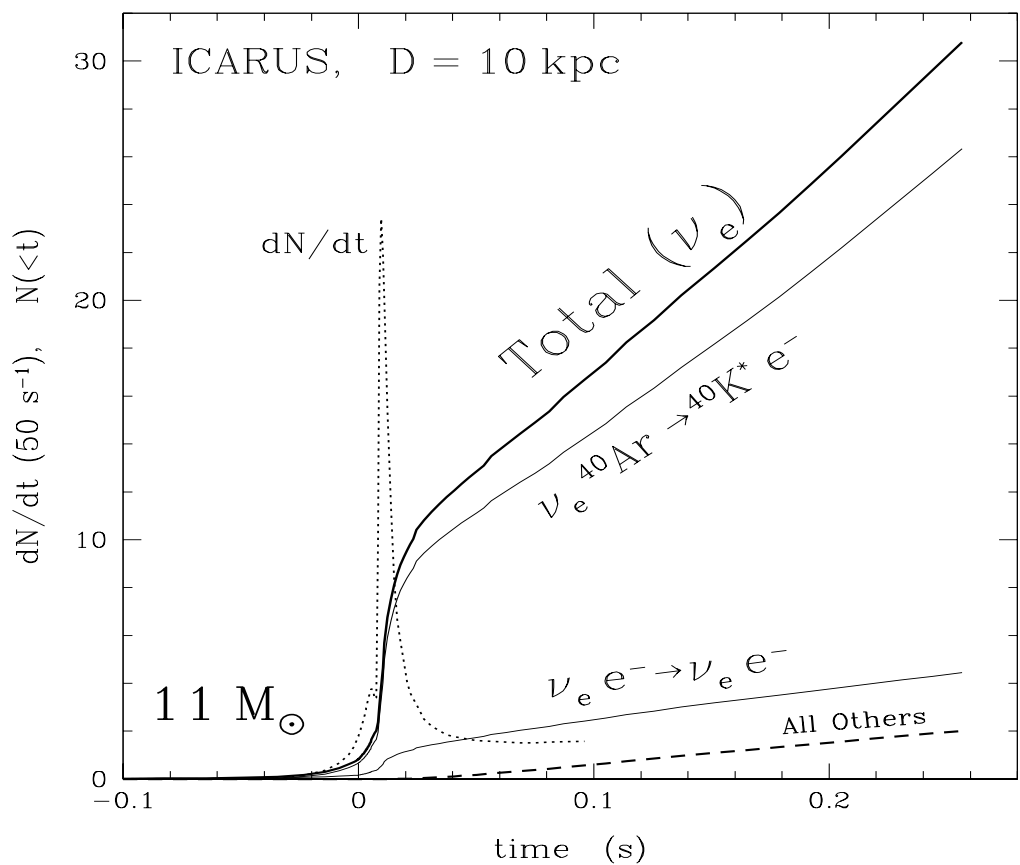}\kern+6in\hfill}
\caption[The $\nu_e$ breakout signal in ICARUS]
{Integrated number of $\nu_e$ events as a function of time via
absorption on Argon 
(labeled `$\nu_e {\rm Ar}\rightarrow {\rm K}^*+e^-$', solid line),
including the Fermi and Gamow-Teller transitions to $^{40}{\rm K}^*$ (see \S\ref{section:icarus}),
and via $\nu_e$-electron scattering, labeled `$\nu_ee^-\rightarrow\nu_ee^-$' (solid line).
The total number of $\nu_e$ neutrinos detected via these channels is the thick solid line.
The thick dashed line (labeled `All Others') shows the contribution to the total neutrino 
signal from $\bar{\nu}_e$, 
$\nu_\mu$, $\bar{\nu}_\mu$, $\nu_\tau$, and $\bar{\nu}_\tau$ neutrinos via scattering
on free electrons.  For reference, we also show the $\nu_e$ detection frequency ($dN/dt$, 
dotted line) during $\nu_e$ breakout and the first 100\,ms after bounce.
}
\label{ski}
\end{figure}

\end{document}